\documentclass[11pt]{article}

\usepackage{subfig}

\usepackage{amsfonts}
\usepackage{amssymb}
\usepackage{epsfig}
\usepackage{dcolumn}
\usepackage{color}

\advance \textheight by \topskip
\usepackage{amsmath,dsfont}
\usepackage[english]{babel}

\makeatletter
 \@addtoreset{equation}{section}
\makeatother

\usepackage[utf8]{inputenc}
\usepackage{amsmath,amssymb}
\makeatletter \@addtoreset{equation}{section}
\makeatother
\def\be{\begin{equation}}

\def\be{\begin{equation}}
\def\ee{\end{equation}}

\def\cW{{\mathcal{W}}}

\def\cH{{\mathcal{H}}}

\def\A{\mathbb A}
\def\B{\mathbb B}
\def\Z{\mathbb Z}

\def\W{\mathbb W}

\usepackage{amsmath,amssymb}
\def\bea{\begin{eqnarray}}
\def\eea{\end{eqnarray}}
\def\barray{\begin{array}}
\def\earray{\end{array}}

\def\sn{\mathrm{sn}}
\def\cn{\mathrm{cn}}
\def\dn{\mathrm{dn}}
\def\ns{\mathrm{ns}}

\def\sech{\mathrm{sech}}

\begin{document}

\title{
{\bf  Soliton defects in one-gap periodic system \\
and exotic supersymmetry}
}

\author{ {\small \textrm{\textup{\textsf{Adri\'an Arancibia${}^{a}$,  
Francisco Correa${}^{b,c}$, 
V\'{\i}t Jakubsk\'y${}^{d}$,
Juan Mateos Guilarte${}^{e}$,
Mikhail S. Plyushchay${}^{a}$}}} } \\
[15pt]
{\small \textit{
${}^{a}$Departamento de F\'{\i}sica,
Universidad de Santiago de Chile, Casilla 307, Santiago 2,
Chile  }}\\
{\small \textit{ ${}^{b}$Leibniz Universit\"at Hannover, Appelstra\ss{}e 2, 30167 Hannover, Germany }}\\
{\small \textit{ ${}^{c}$Centro de Estudios Cient\'{\i}ficos (CECs), Arturo Prat 514, Valdivia, Chile}}\\
{\small \textit{${}^{d}$Department of Theoretical Physics, 
Nuclear Physics Institute, 25068  Re\v z, 
Czech Republic}}\\
{\small \textit{ ${}^{e}$Departamento de 
F\'{\i}sica Fundamental and 
IUFFyM, Universidad de Salamanca, }}\\
{\small \textit{Salamanca E-37008, Spain }}\\
[10pt]
 \sl{\small{E-mails: 
 adaran.phi@gmail.com, 
correa@cecs.cl, jakub@ujf.cas.cz,}}\\ 
 \sl{\small{guilarte@usal.es,
mikhail.plyushchay@usach.cl
}}
}

\date{}


\maketitle

\begin{abstract}
By applying Darboux-Crum transformations
to the quantum one-gap Lam\'e system, 
we  introduce 
an arbitrary countable number of  bound states
into forbidden bands. 
The perturbed  potentials 
are reflectionless and 
contain two types of 
soliton  defects in the periodic 
background. The 
bound states 
with finite number of nodes  
are supported in the lower forbidden band
by the  periodicity defects 
of the potential well type, while the 
pulse type bound states 
in the gap have infinite number of nodes and 
are trapped
by defects of 
the compression modulations 
nature. 
We investigate the exotic  nonlinear $\mathcal{N}=4$ 
supersymmetric structure  in 
such  paired  Schr\"odinger systems, which 
extends an ordinary
$\mathcal{N}=2$ supersymmetry and 
involves two bosonic 
generators  composed  from  
Lax-Novikov integrals of the subsystems.
One of the bosonic integrals 
has a nature of a central charge,
and allows us to liaise the obtained
systems with the stationary equations 
of the Korteweg-de Vries and 
modified Korteweg-de Vries 
hierarchies.  
This exotic supersymmetry
opens the way for the construction 
of self-consistent condensates based on
the Bogoliubov-de Gennes equations
and associated with them 
new solutions to the Gross-Neveu model.
They 
correspond to the kink or
kink-antikink defects of the 
crystalline background
in dependence on whether the exotic supersymmetry
is unbroken or spontaneously broken.
\end{abstract}

\newpage

\section{Introduction}

Quantum  periodic finite-gap systems find 
many interesting applications in physics 
\cite{Lam1}--\cite{PAN}.
They can be  related via the algebro-geometric 
approach with the integrable 
Korteweg-de Vries (KdV) and modified Korteweg-de Vries 
(mKdV) equations \cite{solitons,AlgGeo}. The 
potentials of  finite-gap Schr\"odinger systems 
correspond to the `snapshots'
of the evolving 
in time  generalisations of
 cnoidal waves  solutions to the KdV
 equation \cite{Draz}.
In a similar way, via the Miura transformation,
the scalar Dirac finite-gap potentials can be associated 
with solutions to the mKdV equation.
The  infinite-period limit of such
potentials corresponds to reflectionless systems 
\cite{KayMos}
and the solitary waves solutions to the KdV and mKdV equations.

Reflectionless second and first order
quantum systems  can be constructed 
via the Darboux-Crum transformations 
\cite{MatSal}
from the quantum  free  particle Schr\"odinger 
and Dirac systems. The same transformations
provide an effective dressing method
for construction of 
Lax-Novikov integrals for 
these systems.
The condition of conservation of them
generates the higher order nonlinear stationary
equations for the KdV and mKdV hierarchies
\cite{PlNi,AGP1,AGP2,SolScat}.
This picture also applies for a more general case of
Zakharov-Shabat -- Ablowitz-Kaup-Newell-Segur
hierarchy  \cite{CJ}.

It was shown recently in  \cite{SolScat} that 
the Darboux-Crum transformations
yield a possibility 
 to relate reflectionless systems
with different number of bound states in their spectra
via a soliton scattering  picture.
It was also
demonstrated that the pairs of reflectionless 
Schr\"odinger systems are described not by
the ordinary linear or nonlinear 
$\mathcal{N}=2$  supersymmetry, as this happens
in the case of ordinary, non-transparent 
quantum  systems related by a
Darboux-Crum transformation. Instead,
they are characterised  by exotic  nonlinear 
$\mathcal{N}=4$ supersymmetric structure.
It is generated by two pairs of the 
supercharges, which are the $2\times 2$ 
matrix differential operators of the odd and even 
orders. In addition, the 
exotic supersymmetric structure includes 
two bosonic generators composed
from the Lax-Novikov integrals of 
subsystems, which are 
differential operators of higher odd-order
\cite{AGP1,AGP2}. 

Among all such paired reflectionless 
Schr\"odinger systems,
there is a  special class, in which  
two lower-order supercharges have
the differential order one.
 In this case, one of the 
two bosonic  integrals transmutes into the
central charge of the exotic nonlinear 
$\mathcal{N}=4$ superalgebra, while the second
bosonic integral generates rotations 
between the first-order and even-order 
supercharges.
One of the first-order supercharges 
can be reinterpreted as the Dirac Hamiltonian,
which is 
characterised by its own exotic supersymmetry
associated with the central charge of the 
initial extended Schr\"odinger system.
It is, in fact, the Bogoliubov-de Gennes Hamiltonian,
whose potential, being superpotential of the 
initial extended Schr\"odinger system, provides us with
self-consistent condensates. The latter
supply us, particularly, with kink and kink-antikink type
solutions for the Gross-Neveu model
\cite{AGP2}.    
A similar picture related to the exotic supersymmetry 
was also revealed in the pairs
of mutually displaced one-gap Lam\'e systems
\cite{PAN}.

A natural question that appears here 
is whether the Darboux-Crum transformations 
can be employed to unify the reflectionless
and finite-gap properties  in the same quantum
system. Such a quantum system
could  be associated with  
the KdV and mKdV equations, 
and its potential would correspond to
solitary wave solutions propagating in a  background of 
finite-gap, cnoidal wave type solutions. The related question then
is what happens with the exotic nonlinear 
supersymmetric  structure in such quantum systems.

In this article we answer the posed questions. 
To this aim we  apply the Darboux-Crum transformations 
to the quantum one-gap periodic Lam\'e system
to introduce into its spectrum an arbitrary countable  number
of bound states in its two, the lowest and the intermediate, 
forbidden  bands. This procedure will provide us 
the reflectionless non-periodic one-gap potentials, 
which will contain two essentially different 
types of  soliton  defects in the periodic background
in dependence on in which of the two 
forbidden bands they support the  bound states. 
Coherently with this, as it will be shown,
the corresponding  two types of the 
bound states  possess  essentially  different properties.  
We also investigate the exotic nonlinear 
supersymmetric structure associated with 
such quantum systems. 

Some general 
mathematical aspects of  the theory of the 
class of the systems we investigate   here 
 were discussed 
in   
\cite{Ges}.  The simplest particular examples 
were considered  in \cite{Sam}. 
For the discussion of 
the problem of defects in  a more general context 
of integrable  classical and quantum field theoretical systems, 
see Refs. \cite{DefQFT1,DefQFT2,DefQFT3}.  
\vskip0.1cm

The  article is organised as follows.
In next section, generic properties of the 
quantum one-gap periodic Lam\'e  
system are summarised,  and
its  infinite period limit corresponding to
the simplest reflectionless P\"oschl-Teller model 
with one bound state
is discussed in the light
of Darboux-Crum transformations.
In Section 3, we consider 
Darboux translations for Lam\'e system.
We apply Darboux-Crum transformations 
in Section 4 to introduce soliton defects 
into the one-gap Lam\'e system. 
The  procedure is developed  first to generate  
arbitrary number of periodicity 
defects supporting bound states in 
the lower forbidden band.  Then 
we do the same for  the gap separating  
the allowed valence and conduction bands.
As we shall see,
the cases of the even and odd numbers 
of the bound states  in the intermediate forbidden 
band are characterised by different 
Darboux-Crum schemes.
Finally,  we show how to generalise the 
construction to introduce 
the  bound states in both 
forbidden bands. 
We discuss also the application of Darboux-Crum 
dressing procedure  for the construction 
of the irreducible Lax-Novikov integrals.
Section 5 is devoted to investigation
of the exotic nonlinear $\mathcal{N}=4$ 
supersymmetric structure  which 
appears in the extended Shr\"odinger systems  
composed from two arbitrary 
one-gap systems with periodicity defects.
Special attention is given there for the most
interesting from the viewpoint of  physical applications case,
in which two of the four supercharges 
are given by the matrix differential operators of the 
first order. We consider the cases 
of the unbroken and spontaneously broken 
exotic supersymmetries, and 
indicate  the relation of the obtained
systems with the KdV and mKdV 
equations. The results are summarised in 
Section 6.  We point out   there
further possible research directions 
for the  development of  the obtained results
and some interesting 
applications. Appendix is devoted
to a more technical demonstration 
of a non-singular nature  of the 
constructed 
one-gap potentials of a generic 
form with arbitrary number
of the periodicity defects. 

\section{One-gap Lam\'e system and its infinite period limit}

In this section we summarise   generic properties of the 
quantum  one-gap periodic Lam\'e system, and  
discuss its infinite period limit corresponding
to the reflectionless P\"oschl-Teller model. The Darboux transformations 
associate  the latter
system with a free particle and 
allow us, particularly, 
to identify its nontrivial Lax-Novikov integral
via the dressing procedure.
All this will  form the basis 
for application of the method  
of the Darboux-Crum transformations 
to introduce  two different  
types of non-periodic soliton defects
into the Lam\'e system.

\subsection{Spectral properties of one-gap Lam\'e system}

\begin{figure}[h!]\begin{center}
    \includegraphics[scale=1.0]{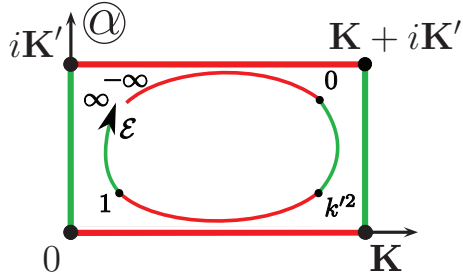}
    \caption{Spectrum of the one-gap Lam\'e system (\ref{HLame}) 
    as a function of complex parameter $\alpha$.}
    \label{fig1}
\end{center}
\end{figure}

The quantum one-gap Lam\'e system is described by the 
Hamiltonian operator
\begin{equation}\label{HLame}
    H_{0,0}=-\frac{d^2}{dx^2}+V_{0,0}(x),\qquad
    V_{0,0}(x)= 2k^2\sn^2x-k^2=-2\dn^2 x+1+k'^2\,,
\end{equation}
with a  periodic potential
$V_{0,0}(x)=
V_{0,0}(x+2{\rm {\bf K}})$~\footnote{
${\rm {\bf K}}={\rm {\bf K}}(k)$
is a complete elliptic integral of the first kind 
corresponding to the modular parameter $k$,
$0<k<1$. We also denote
${\rm {\bf K}}'={\rm {\bf K}}(k')$, 
where $k'$, $0<k'<1$,
$k^2+k'^2=1$, is the complementary modular parameter.
For the 
 properties of Jacobi elliptic and related 
 functions see \cite{WW}.
 For  a short summary of the properties we use here,
 see Appendix in \cite{PAN}. 
 The dependence of these functions on $k$ 
 is not shown explicitly.  In the case when
 they depend on  $k'$
 instead of $k$,  we indicate such a dependence 
explicitly.}. 
The sense of 
the  
lower indexes introduced here  
will be clarified in what follows.
The eigenstates of $H_{0,0}$ can be found
in a closed analytic form for any complex eigenvalue 
$\mathcal{E}$. Parametrising the latter in terms of
Jacobi's elliptic $\dn$-function,
$\mathcal{E}(\alpha)=
\dn^2\alpha$,  we obtain  the solutions of the stationary 
Schr\"odinger equation $H_{0,0}\Psi^\alpha_\pm= 
\mathcal{E}(\alpha)\Psi^\alpha_\pm$,
\begin{equation}\label{Psi+-}
    \Psi_\pm^{\alpha}\left(x\right)=
    \frac{{\rm H}\left(x\pm \alpha\right)}{\Theta\left(x\right)}
    \exp\left[
    \mp x {\rm Z}\left(\alpha\right)\right]\,.
\end{equation}
Here $\Theta$, ${\rm H}$,  and ${\rm Z}$ are
Jacobi's  Theta, Eta and Zeta functions, while
 parameter $\alpha$ can take arbitrary complex 
 values. 
Since the periods of  
the doubly periodic elliptic function
$\dn^2 \alpha$ are $2{\rm {\bf K}}$ and 
$2i{\rm {\bf K}}'$, and it is an even function,
without any loss of generality one can restrict 
a consideration to a rectangular domain with
vertices in $0$, $ {\rm {\bf K}}$,
$ {\rm {\bf K}}+i{\rm {\bf K}}'$ and $i{\rm {\bf K}}'$.
Hamiltonian (\ref{HLame}) is a Hermitian operator, and 
we are interested in the real eigenvalues 
$\mathcal{E}(\alpha)$ \footnote{
The $PT$-symmetric 
generalisation \cite{PTsym1,PTsym2} 
of  (\ref{HLame}) 
can  also be associated 
with real  values of $\mathcal{E}(\alpha)$, see below.}.
These are provided 
by further restriction of  the  values of the 
parameter $\alpha$ to  the borders
of the indicated 
rectangle, see Fig. \ref{fig1}. 
The horizontal edges correspond to
the \emph{lower} and  \emph{upper} 
forbidden  zones (lacunas) in the spectrum.
The vertical edges correspond, respectively, to the 
\emph{valence} and \emph{conduction} bands.
\begin{table}[ht]
\caption{\underline{Bands and their characteristics.}\quad
 Here 
${\rm z}(\beta^-)={\rm Z}(\beta^-)+\cn\,\beta^-{\rm ds}\,\beta^-$,
$\kappa(\gamma^\pm\vert k')=\frac{\pi}{2{\rm{\bf K}}}
   (1-\frac{\gamma^\pm}{{\rm{\bf K}}'})-{\rm Z}(\gamma^\pm\vert k')
   + f_\pm$, $f_-=k'^2\sn(\gamma^-\vert k'){\rm cd}(\gamma^-\vert k')$, and
 $f_+=\sn(\gamma^+\vert k'){\rm dc}(\gamma^+\vert k')$.  
} \label{T1}
\begin{center}
\begin{tabular}{|c|c|c|c|c|c|c|}\hline
Band  &  $\alpha=\beta+i\gamma$ & $\mathcal{E}(\alpha)$ & $\kappa(\alpha)$ \\\hline\hline
 {\small lower forbidden}&  {\small $ \beta\equiv\beta^-\in (0,{\rm{\bf K}})$, \, \, $\gamma={\rm{\bf K}}'$ }  
 &  {\small $(-\infty,0) \ni\mathcal{E}=-{\rm cs}^2\beta^-$}& 
 {\small $-i{\rm z}(\beta^-)$}
   \\[1pt]\hline
  {\small valence} & {\small $\beta={\rm{\bf K}}$,\, \,  
  $\gamma\equiv \gamma^-\in[0,{\rm{\bf K}}']$} & 
   {\small $[0,k'^2] \ni\mathcal{E}=k'^2{\rm cd}^2(\gamma^-\vert k')$ }& 
   {\small $\kappa(\gamma^-\vert k')$}
 \\[1pt]\hline
 {\small upper forbidden (gap)}   &  
 {\small $\beta\equiv \beta^+\in (0, {\rm{\bf K}})$,\, \,  $\gamma=0$} &  
{\small  $(k'^2,1) \ni \mathcal{E}={\rm dn}^2 \beta^+$}
& 
{\small $\frac{\pi}{2{\rm{\bf K}}}- i{\rm Z}(\beta^+)$} 
  \\[1pt]\hline
{\small conduction}  & {\small $\beta=0$,\, \,  $\gamma\equiv \gamma^+\in 
[0,  {\rm{\bf K}}')$} & 
{\small $ [0,+\infty)\ni  \mathcal{E}={\rm dc}^2(\gamma^+\vert k')$}
 & 
 {\small $\kappa(\gamma^+\vert k')$} \\[1pt]\hline
\end{tabular}
\end{center}
\end{table}
The necessary information on the bands' structure, 
including the values of quasi-momentum 
$\kappa(\alpha)$, see below,  is summarised in Table \ref{T1}.
We supply the parameters $\beta$ and $\gamma$,
corresponding to real and imaginary parts of the 
complex parameter $\alpha$,
with upper index $-$/$+$ to distinguish whether 
they correspond to the lower/upper 
forbidden and  allowed bands, respectively.

While  the real parameter $\beta^-$ increases
 in the open 
interval $(0, {\rm {\bf K}})$, 
the energy increases in the lower,
semi-infinite forbidden  band, but decreases in the
finite gap  separating the allowed bands
when $\beta^+$ varies in the same interval.
In the valence band 
the energy increases when the
 parameter $\gamma^-$ 
decreases from ${\rm {\bf K}}'$ to $0$;
the variation of the parameter $\gamma^+$ in the
semi-open interval 
$[0,{\rm {\bf K}}')$ gives 
the energy monotonically 
increasing in the semi-infinite conduction 
band.
 
Under the shift for the real period  $2{\rm {\bf K}}$
of the potential, 
the eigenstates (\ref{Psi+-}) undergo the transformation
\begin{equation}\label{kappal}
 \Psi_\pm^{\alpha}\left(x+2{\rm {\bf K}}\right)=
    \exp\left(\mp i2{\rm {\bf K}}\,
    \kappa(\alpha)\right)\Psi_\pm^{\alpha}\left(x\right)\,,
 \quad
{\rm where}\quad 
	\kappa(\alpha)=\frac{\pi}{2{\rm {\bf K}}}-
	i{\rm Z}(\alpha)\,
\end{equation}
 is the \emph{quasi-momentum}, in which the first term 
 is associated with the $2{\rm {\bf K}}$-anti-periodicity  of the 
 Eta function, ${\rm H}(x+2{\rm {\bf K}})=-
 {\rm H}(x)$. 
 The analytical form of the quasi-momentum
 $\kappa(\alpha)$  allows us to determine explicitly
when it takes real or complex values,
 and therefore to locate the allowed and forbidden
 bands.
Thus, making use of the properties of the 
 Jacobi's Zeta function, one finds that in the 
lower forbidden band, the quasi-momentum 
takes pure imaginary values,
$
	\kappa(\beta^-+i{\rm {\bf K}}')=-i{\rm z}(\beta^-),
$
$
	{\rm z}(\beta^-)=
	\frac{d}{d\beta^-}
	\log{\rm H}(\beta^-)
$.
In accordance with this, 
the quasi-momentum varies 
in the complex plane 
along  the imaginary axis so that 
$\kappa\rightarrow -i\infty$ 
for $\beta^-\rightarrow 0$, ${\mathcal E}\rightarrow -\infty$,
and  
$\kappa\rightarrow 0$ 
when $\beta^-\rightarrow {\rm {\bf K}}$, 
${\mathcal E}\rightarrow 0$.
The amplitude of the wave functions
 (\ref{Psi+-}) in this band 
increases exponentially in one of the 
two directions on the real axis $x$, 
and  eigenfunctions $\Psi^{\alpha=\beta^- +i{\rm {\bf K}}'}
_\pm(x)$
correspond therefore
to non-physical states.
In the valence band, the quasi-momentum takes
real values,
$	\kappa({\rm {\bf K}}+i\gamma^-)=\frac{\pi}{
	2{\rm {\bf K}}}\big(1-\frac{\gamma^-}{{\rm {\bf K}}'}\big)
	-\frac{d}{d\gamma^-}\log \Theta(\gamma^-+{\rm {\bf K}}'\vert k'),$
where it  increases monotonically   from
$\kappa=0$ (${\mathcal E}=0$)
to  $\kappa=\frac{\pi}{2{\rm {\bf K}}}$ 
(${\mathcal E}=k'^2$).  The wave functions 
 (\ref{Psi+-}) inside the valence band 
 correspond to the two 
 linearly independent Bloch states.
In the intermediate energy gap, the 
quasi-momentum is complex-valued,
$
\kappa(\beta^+)=\frac{\pi}{
	2{\rm {\bf K}}}- i{\rm Z}(\beta^+)
$.
In accordance with the relation 
$\frac{d}{d\beta}{\rm Z}(\beta)=
\dn^2\beta-\frac{{\rm {\bf E}}}{{\rm {\bf K}}}$,
where ${\rm {\bf E}}$ is the complete
elliptic integral of the second kind,
and $k'^2<\frac{{\rm {\bf E}}}{{\rm {\bf K}}}<1$,
the imaginary part in $\kappa(\beta^+)$ 
varies monotonically in the interval $\beta^+\in(0,\beta_*]$, 
$0<{\rm Z}\leq {\rm Z}(\beta_*)$, where  $\beta_*$
corresponds to the equality 
$\dn^2\beta_*=\frac{{\rm {\bf E}}}{{\rm {\bf K}}}$,
and then decreases monotonically 
approaching to the
zero value
in the interval $\beta^+\in(\beta_*,{\rm {\bf K}})$.
 In the conduction band,
 like in the valence band, the quasi-momentum
takes real values,
$	\kappa(i\gamma^+)=\frac{\pi}{
	2{\rm {\bf K}}}\big(1-\frac{\gamma^+}{{\rm {\bf K}}'}\big)
	-\frac{d}{d\gamma^+}\log {\rm H}(\gamma^+
	+{\rm {\bf K}}'\vert k').
$
 It  increases here monotonically from
$\frac{\pi}{
	2{\rm {\bf K}}}$ (${\mathcal E}=1$) to
	$+\infty$  (${\mathcal E}\rightarrow\infty$).
Inside this band, for any value of the energy
the two wave functions  (\ref{Psi+-}) correspond to
the two linearly independent 
physical Bloch states.

 \begin{figure}[h!]\begin{center}
    \includegraphics[scale=0.7]{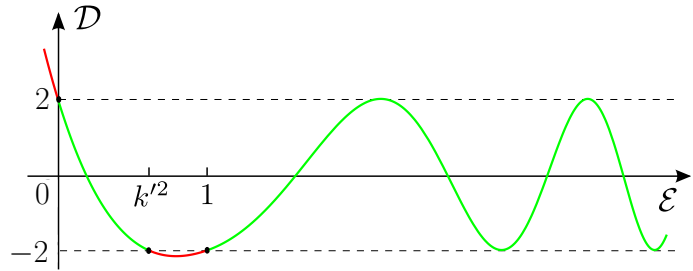}
    \caption{ The discriminant $\mathcal{D}(\mathcal{E})$ of the one-gap Lam\'e system.
    The scale is linear in energy for 
    $\mathcal{E}<1$, while for  $\mathcal{E}>1$ a logarithmic scale is used
    here. The parts shown in red correspond to 
    the lower ($\mathcal{E}<0$) and to upper
    ($k'^2<\mathcal{E}<1$) forbidden bands.
    } 
    \label{fig2}
\end{center}
\end{figure}

The properties of a periodic quantum system
are effectively reflected by
the discriminant 
$\mathcal{D}(\mathcal{E})$
(Lyapunov function)
of the corresponding stationary Schr\"odinger 
equation,  which
is defined as a trace of the monodromy 
matrix representing the operator 
of the translation for the period of the
potential \cite{solitons,Hill,CJP,Rosu}.
 Its form $\mathcal{D}(\mathcal{E})=2\cos \left(2{\rm {\bf K}}
\kappa(\mathcal{E})\right)$
 for the one-gap Lam\'e system
(\ref{HLame}) is shown on Fig.  \ref{fig2}.
In the lower prohibited zone 
and in the valence band
the explicit analytic  form is given, respectively,   by 
$\mathcal{D}(\mathcal{E}(\beta^-+i{{\rm {\bf K}}'}))=
2\cosh \left(2{\rm {\bf K}}{\rm z}(\beta^-)\right)$,
and 
$\mathcal{D}(\mathcal{E}({\rm {\bf K}}+i\gamma^-))=
2\cos\left(2{\rm {\bf K}}\kappa(\gamma^-\vert k')\right)$.
In the energy gap
separating 
the valence and conduction bands, 
it reduces to $\mathcal{D}(\mathcal{E}(\beta^+))=
-2\cosh \left(2{\rm {\bf K}}{\rm Z}(\beta^+)\right)$.
The minimum of the curve at
$\mathcal{E}=\dn^2\beta_*=
\frac{{\rm {\bf E}}}{{\rm {\bf K}}}$
corresponds to the 
maximum value 
${\rm Z}(\beta_*)>0$
of the Zeta function.
In the conduction band  we have
$\mathcal{D}(\mathcal{E}(\gamma^+))=
2\cos\left(2{\rm {\bf K}}\kappa(\gamma^+\vert k')\right)$.
The infinite number of oscillations of the curve 
between $-2$ and $+2$ extrema values 
of the $\mathcal{D}(\mathcal{E})$
is associated in this band 
with the zero of $\cn\,(\gamma^+\vert k')$
at $\gamma^+={\rm {\bf K}}'$
appearing in the denominator of  the 
function $f_+$ in the structure of 
$\kappa(\gamma^+\vert k')$, see
Table \ref{T1}.

At the edges of the
valence and conduction bands,
where $\vert\mathcal{D}\vert=2$, $\frac{d\mathcal{D}}{d\mathcal{E}}\neq 0$,
 the two  wave functions (\ref{Psi+-}) reduce, up to numerical 
factors,  to the same periodic,
$\psi_1=\dn\, x$ (${\mathcal E}=0$), and anti-periodic,
$\psi_2=\cn\, x$ (${\mathcal E}=k'^2$) and $\psi_3=\sn\, x$ 
(${\mathcal E}=1$),
 eigenstates. The second, linear 
 independent eigenfunctions at the edges of the  
valence and conduction bands are given by 
$\Psi_i(x)=\psi_i(x)\mathcal{I}_i$, $i=1,2,3$, where
 $\mathcal{I}_i(x)=\int dx/\psi_i^2(x)$
are expressed in terms of 
the incomplete elliptic integral of the 
second kind, 
${\rm E}(x)=\int_0^x \dn^2x\, dx$\,:
$\mathcal{I}_1(x)=\frac{1}{k'^2}
{\rm E}(x+{\rm {\bf K}})$, 
$\mathcal{I}_2(x)=x-\frac{1}{k'^2}
{\rm E}(x+{\rm {\bf K}}+i{\rm {\bf K}}')$,
$\mathcal{I}_3(x)=x-
{\rm E}(x+i{\rm {\bf K}}')$.
The functions $\Psi_i(x)$ are not bounded 
on the real line and correspond to non-physical 
eigenstates of the Lam\'e Hamiltonian operator.
They also can be obtained from the states 
(\ref{Psi+-}) by differentiation in $\alpha$. 
Namely, derivatives of the 
functions $\Psi^\alpha_+(x)$ in $\alpha$ at 
$\alpha=0$ and $\alpha={\rm {\bf K}}$ 
give some linear combinations of the 
functions $\psi_i(x)$ and $\Psi_i(x)$ with  $i=3$ and $i=2$,
respectively,
while the derivative of the function (\ref{Fdef})
in parameter $\beta^-$ at $\beta^-={\rm {\bf K}}$
gives a linear combination of
$\psi_1(x)$ and $\Psi_1(x)$.

\vskip0.1cm

For any value of the parameter $\alpha$,
under the parity reflection, $Pf(x)=f(-x)$, 
the states (\ref{Psi+-}) satisfy
the relation
\begin{equation}\label{Psireflec}
	P \Psi_\pm^\alpha(x)=-\Psi_\mp^\alpha(x)\,.
\end{equation}
The properties of the wave functions 
(\ref{Psi+-}) in corresponding bands 
under  
 the $T$, $Tf(x)=f^*(x)$, and the composed 
 $PT$ 
operations \cite{PTsym1,PTsym2}
are shown in Table \ref{Table2}.

\begin{table}[ht]
\caption{\underline{Properties of the eigenfunctions
under the $T$ and $PT$ operations.}  Here 
 $c=\exp\left(i\frac{\pi\beta^-}{{\rm {\bf K}}}\right)$.
} \label{Table2}
\begin{center}
\begin{tabular}{|c|c|c|c|c|c|c|}\hline
Band  & $\Psi_\pm^\alpha(x)$ & 
${T} \Psi_\pm^\alpha(x) $ & ${P T} \Psi_\pm^\alpha(x) $  
\\\hline\hline
 {\small lower forbidden}&  {\small $\Psi_\pm^{\beta^-+iK'}(x)$}
 &     {\small $-c \Psi_\pm^{\beta^-+iK'}(x)$ }& 
 {\small $ c \Psi_\mp^{\beta^-+iK'}(x)$} 
   \\[1pt]\hline
  {\small valence} &  {\small $\Psi_\pm^{K+i\gamma^-}(x)$} 
  &{\small $-\Psi_\mp^{K+i\gamma^-}(x)$} &
  {\small $\Psi_\pm^{K+i\gamma^-}(x)$} 
 \\[1pt]\hline
 {\small upper forbidden (gap)}   &   
 {\small  $\Psi_\pm^{\beta^+}(x)$ }& 
 {\small  $\Psi_\pm^{\beta^+}(x)$ } &  
 {\small  $-\Psi_\mp^{\beta^+}(x)$ }
  \\[1pt]\hline
{\small conduction}  &  {\small $ \Psi_\pm^{i\gamma^+}(x)$ } &   
 {\small $ \Psi_\mp^{i\gamma^+}(x)$ } 
  &  {\small $ -\Psi_\pm^{i\gamma^+}(x)$ } 
 \\[1pt]\hline
\end{tabular}
\end{center}
\end{table}

Notice that in the lower forbidden band 
\begin{equation}\label{Psi*gap0}
	\Psi^{\beta^-+i{\rm {\bf K}}'}_\pm(x)=
	\pm iq^{-1/4}
	\exp\left(- i\frac{\pi\beta^-}{2{\rm {\bf K}}}\right)
	{F}(\pm x;\beta^-)\,,
\end{equation}
where 
\begin{equation}\label{Fdef}
	{F}(x;\beta^-)=\frac{\Theta(x+\beta^-)}{\Theta(x)}
	\exp(-x{\rm z}(\beta^-))\,
\end{equation}	
is a real-valued function of $x$,
which takes positive values, ${F}(x;\beta^-)>0$.
Here $q=\exp(-\pi {\rm {\bf K}}'/{\rm {\bf K}})$ is the 
Jacobi's nome, and we used the relation
${\rm H}(x+i{\rm {\bf K}}')=iq^{-1/4}\exp(-i\frac{\pi x}{2{\rm {\bf K}}})
\Theta(x)$. 
In this band one can employ 
alternatively  the real functions 
$F(x;\beta^-)$ and $F(-x;\beta^-)=P F(x;\beta^-)$
as 
two linear independent
solutions.  

The  operator PT distinguishes whether the function 
(\ref{Psi+-})  
belongs to the forbidden or allowed band. 
When it corresponds to 
the physical Bloch state, it is also 
the eigenfunction of 
the $PT$. In contrast, 
the functions (\ref{Psi+-})  from the forbidden bands cease 
to be eigenstates of the $PT$  operator.
Instead, certain linear combinations of the 
 two states (\ref{Psi+-})
with the opposite sign of the 
complex-valued there quasi-momentum
are  the eigenstates of the $PT$  in those bands.

\subsection{Infinite period limit: reflectionless 
P\"oschl-Teller system
and  Darboux transformations}\label{PT1Dar}

Before we pass to the discussion of the
introduction of the periodicity defects,
 corresponding to solitons,
into the spectrum of the the one-gap Lam\'e system,
we consider briefly the analogous procedure for the infinite 
period limit case. The picture in such a limit case is more simple and 
transparent, and is useful to bear  it  in mind when
we generalise the method  to the very
Lam\'e system. 

In the infinite period limit ${\rm {\bf K}}\rightarrow
\infty$, that is   equivalent to any 
of the three limits $k\rightarrow 1$, 
$k'\rightarrow 0$, or
 ${\rm {\bf K}}'\rightarrow
\pi/2$,  operator
(\ref{HLame}) transforms into the 
Hamiltonian of the reflectionless P\"ochl-Teller system
\begin{equation}\label{HPT}
    H_{1}=-\frac{d^2}{dx^2}+V_1(x)\,,\qquad
    V_1(x)=-\frac{2}{\cosh^2 x}+1\,.
\end{equation}
In this limit the valence band shrinks into 
one discrete energy level $\mathcal{E}=0$.
The wave functions (\ref{Psi+-}) of the
valence band with $\alpha={\rm {\bf K}}+i\gamma^-$,
$\gamma^-\in[0,{\rm {\bf K}}']$ transform into 
the unique bound state described by the 
normalisable wave 
function
$\Psi_{\mathcal{E}=0}(x)=\sech\,x$. 
The conduction band, parametrised by 
$\alpha=i\gamma^+$, $\gamma^+\in [0,{\rm {\bf K}}')$, 
transforms into the 
scattering part of the spectrum of the system
(\ref{HPT}). In the limit, we have 
$\gamma^+\in [0,\frac{\pi}{2})$. Introducing the notation
$\tan\gamma^+={\rm k}$, $0\leq {\rm k}<\infty$,
we find that the rescaled wave functions 
$q^{-1/4}\Psi^{\alpha=i\gamma^+}_{\mp}(x)$
of the conduction band 
transform, up to inessential constant multiplicative factor,
into the wave functions
\begin{equation}\label{Psipmk}
	\Psi^{\rm k}_{\pm}(x)=(\pm i{\rm k}-
	\tanh x)e^{\pm i{\rm k}x}\,.
\end{equation}
Corresponding  energy $\mathcal{E}=\dn^2(i\gamma^+\vert k)=
{\rm dc^2(\gamma^+\vert k')}$
 transforms in the limit $k'\rightarrow 0$
 into
${1}/{\cos^2\gamma^+}=1+{\rm k}^2$,
which is the  eigenvalue
of the eigenstates (\ref{Psipmk}) of
the P\"oschl-Teller Hamiltonian (\ref{HPT}).
The non-degenerate state $	\Psi^{\rm 0}=\tanh x$
 (${\rm k}=0$)
corresponds here to the state of energy 
$\mathcal{E}=1$ described by 
$\sn\, x$  at the edge of the conduction band 
of the Lam\'e system (\ref{HLame}).

The scattering states  (\ref{Psipmk})
can be presented in the form
$\Psi^{\rm k}_{\pm}(x)=A_{\varphi} e^{\pm i{\rm k}x}$ in terms 
of the first order differential operator 
\begin{equation}\label{Adef}
A_{\varphi}=\varphi(x)\frac{d}{dx}\frac{1}{\varphi(x)}
=\frac{d}{dx}-
	\tanh x\,,\qquad
	\varphi(x)=\cosh x\,.
\end{equation} 
	Operator $A_{\varphi}$ together with the Hermitian conjugate
	$A_{\varphi}^\dagger$ 
	intertwine the reflectionless system (\ref{HPT}) with the
	free particle  Hamiltonian shifted for an additive  constant,
	\begin{equation}\label{H0}
	H_0=-\frac{d^2}{dx^2}+1\,,
	\end{equation}
	 and 
	provide the factorisation of both:
	\begin{equation}\label{Darboux0PT}
	A_{\varphi}A_{\varphi}^\dagger=H_{1}\,,\qquad
	 A_{\varphi}^\dagger A_{\varphi}=H_{0}\,,\qquad
	A_{\varphi}H_0=H_{1}A_{\varphi}\,,\qquad  
	A_{\varphi}^\dagger H_{1}=H_0 A_{\varphi}^\dagger\,.
\end{equation}
Relations (\ref{Darboux0PT}) correspond to the Darboux transformations
which relate the free particle system with the 
reflectionless P\"oschl-Teller system.
The alternative form
to express the same relation between the 
systems corresponds to the equality
\begin{equation}\label{HphiH}
	H_{1}=H_0-2\frac{d^2}{dx^2}\log \varphi(x)\,.
\end{equation}
The wave function $\varphi(x)=\cosh x$ is a nodeless 
non-physical eigenstate of the free particle $H_0$,
and the operator $A_{\varphi}$ produces an almost isospectral
mapping of all the physical and non-physical
states of $H_0$, except $\varphi(x)$,
$A_{\varphi}\varphi(x)=0$,
 into corresponding states of 
the system $H_1$. The only physical
bound state
$\Psi_{\mathcal{E}=0}(x)=\sech\,x$ of $H_{1}$ of 
zero energy, for which there is no bound state analog 
in the physical spectrum of
$H_0$,  is obtained by applying  the operator 
$A_{\varphi}$ 
to the wave function
 $\tilde{\varphi}(x)=
\varphi(x)\int \frac{dx}{\varphi^2(x)}$.
This is the non-physical eigenstate of 
(\ref{H0}) of the same zero eigenvalue as 
$\varphi(x)$. It reduces here
just to the derivative of the latter,
$ \tilde{\varphi}(x)=\sinh x=\varphi'(x)$. 
Analogously, the application of the operator $A_{\varphi}^\dagger$ to the 
eigenstates of $H_1$ in correspondence 
with the last relation in (\ref{Darboux0PT})
produces the eigenstates of $H_0$.
The unique bound state 
$\Psi_{\mathcal{E}=0}(x)=\sech\,x$ of $H_{1}$
is the zero mode of the first order operator
$A_{\varphi}^\dagger$. 

The free particle system (\ref{H0}) 
has a nontrivial integral $p=-i\frac{d}{dx}$. It
distinguishes the plane waves $e^{\pm i{\rm k}x}$,
which are 
the eigenstates of $H_0$ of the same 
energy, and detects a unique non-degenerate 
state $\Psi_{\mathcal{E}=1}(x)=1$ corresponding to ${\rm k}=0$ 
by annihilating it. 
In correspondence with the two
last relations in (\ref{Darboux0PT}) and 
the described picture
of the mapping associated with the Darboux transformations,
one finds that the operator 
\begin{equation}\label{P1PT1}
	\mathcal{P}=-iA_{\varphi}\frac{d}{dx}A_{\varphi}^\dagger
\end{equation}
is the Hermitian integral for the reflectionless system 
$H_{1}$.  
We refer to this as the dressing procedure.
Similarly to $p$, this operator
distinguishes the eigenstates 
(\ref{Psipmk}) being analogs of the plane wave 
states for the free
particle, $\mathcal{P}\Psi^{\rm k}_{\pm}(x)=
\pm {\rm k}(1+{\rm k}^2)\Psi^{\rm k}_{\pm}(x)$. 
It annihilates  the lowest non-degenerate state 
$	\Psi^{\rm 0}(x)=\tanh x$
in the scattering sector, and  the bound 
state$\,$\footnote{Being the third order 
differential operator, (\ref{P1PT1}) also  turns  into zero 
the state $\varphi(x)=\cosh x$, which is a  
a non-physical eigenstate of the free particle Hamiltonian
(\ref{H0}) 
\cite{CorPly}.}  
$\Psi_{\mathcal{E}=0}(x)=\sech\,x$. 
Integral (\ref{P1PT1}) satisfies the
Burchnall-Chaundy  relation \cite{BurCha}
\begin{equation}\label{P2BC}
	\mathcal{P}^2=H_{1}^2(H_{1}-1)\,.
\end{equation}
Since the free particle has the integral
$p=-i\frac{d}{dx}$, the $H_0$  and the 
P\"oschl-Teller Hamiltonian (\ref{HPT})
can be intertwined not only by the first order 
operator (\ref{Adef}) and its conjugate $A_{\varphi}^\dagger$,
but also by the second order operators 
\begin{equation}\label{Bphi}
	B_{\varphi}=A_{\varphi}\frac{d}{dx}\quad 
	{\rm and}  \quad
	B_{\varphi}^\dagger\,.
\end{equation}	
The first and second order intertwining 
operators together with the integrals $p$ and 
$\mathcal{P}$  of the systems $H_0$ and 
$H_{1}$ constitute the building blocks of 
the exotic  centrally extended $\mathcal{N}=4$
nonlinear supersymmetry of the  system described
by the $2\times 2$ matrix 
Hamiltonian $\mathcal{H}={\rm diag}\,(H_0,H_{1})$ 
\cite{SolScat}.

Suppose now that we want to construct another 
reflectionless  system proceeding from the P\"oschl-Teller
system (\ref{HPT}) by means of a new 
Darboux transformation, or a composition of them,
that corresponds to the Darboux-Crum transformation.
There are three different ways to do this.
First, one can construct a reflectionless system 
with an additional, second bound state lying below
the unique, zero energy bound state of the system 
(\ref{HPT}). Another case corresponds to the situation 
when we want to introduce a bound state with the energy
level lying between the zero energy level of the 
already existing bound state and the edge of the scattering 
sector of energy $\mathcal{E}=1$. 
At last, one can construct a reflectionless system completely
isospectral to the system (\ref{HPT})
but with the displaced potential (`soliton center').
Having at hands the building blocks corresponding 
to the described three possibilities, by the appropriate
generalisation of
the procedure, we can construct reflectionless system 
with arbitrary number of bound states and arbitrary positions 
of the corresponding soliton centres \cite{AGP2,AGP1}.

The first situation is realised by the 
construction in a way similar 
to  (\ref{Adef}) of the 
Darboux generator on the basis of the 
nodeless function 
\be\label{phi1state}
\varphi_1(x;\kappa_1,\tau_1)
=A_{\varphi}\sinh \kappa_1(x+\tau_1), 
\ee
where $\kappa_1>1$
and $\tau_1$ is an arbitrary real parameter. 
The function $\varphi_1(x;\kappa_1,\tau_1)$ is 
the non-physical eigenstate of  (\ref{HPT})
with energy $1-\kappa_1^2$, and $\tau_1$ 
is associated with the center 
(phase) of the second 
soliton (the first soliton is characterised by  
 $\tau_0=0$ and the  amplitude $\kappa_0=1$)
 in the potential of the system 
 \be\label{H2refless}
 H_2=H_{1}-2\frac{d^2}{dx^2}\log
 \varphi_1(x)
 \ee
 with two bound states, cf. (\ref{HphiH}).
 Note that alternatively $H_2$ can be 
 presented in terms of the second order 
 Darboux-Crum transformation applied  to 
 the free particle,
 $H_2=H_0-2\frac{d^2}{dx^2}\log
 \W(x)$, where $\W(x)$ is the Wronskian
 of the two non-physical states of the 
 free particle, $\varphi=\cosh x$ and $\phi=\sinh \kappa_1(x+\tau_1)$,
 $\W(x)=W(\varphi,\phi)=\varphi \phi'-\varphi'
\phi$. 

 To obtain reflectionless system 
 with an additional  bound state inside the 
 energy interval $(0,1)$,
 which separates the bound state level
 of the system (\ref{HPT}) with the
  continuous part of the spectrum,
 one can  apply  to (\ref{HPT}) 
 the Darboux-Crum transformation 
generated by the two non-physical states
 $\phi_1(x;\kappa_1,\tau_1)
=A_{\varphi}\cosh \kappa_1(x+\tau_1)$
and   $\phi_2(x;\kappa_2,\tau_2)
=A_{\varphi}\sinh \kappa_2(x+\tau_2)$. 
If we restrict the parameters $\kappa_{1,2}$
by the 
condition $0<\kappa_1<\kappa_2<1$,
the corresponding Wronskian 
$\W(x)=W(\phi_1,\phi_2)$ has no zeros.
This produces a system with a regular reflectionless
potential
 \be\label{V3V1}
 	V_3(x)=V_1(x)-2\frac{d^2}{dx^2}\log \W(x),
 \ee
which has three bound states with energies $1-\kappa_1^2$,
$1- \kappa_2^2$ and $0$.
Sending then one of the two translation parameters, $\tau_2$ or $\tau_1$,
to any of the limits $+\infty$ or $-\infty$, we get 
a reflectionless system with two bound states of energies 
$1-\kappa_1^2$ and $0$ when we send 
$\vert \tau_2 \vert\rightarrow
\infty$, or with energies $1-\kappa_2^2$ and $0$
when $\vert\tau_1\vert\rightarrow
\infty$.
 The indicated limit changes the translation parameters 
 of the remaining added soliton as well as of the initial 
 one with $\kappa_0=1$ and $\tau_0=0$ in correspondence 
 with the picture of soliton scattering, see \cite{SolScat}.

There is another possibility to  introduce one additional bound 
state into the spectrum of the system (\ref{HPT})
with the energy inside  the interval $(0,1)$.
One can apply to  (\ref{HPT}) a Darboux transformation 
constructed on the basis of its non-physical state 
 $\phi(x;\kappa,\tau)
=A_{\varphi}\sinh \kappa(x+\tau)$, $0<\kappa<1$. 
This will produce a singular system.
Shifting  then
$\tau\rightarrow \tau +i\frac{\pi}{2\kappa}(1-\kappa)$
and  $x\rightarrow x+i\frac{\pi}{2}$, 
we get a regular reflectionless system with two bound states 
with energies $1-\kappa^2$ and~$0$.

Finally, to produce a system completely isospectral to the system
(\ref{HPT}), one can apply to the latter the Darboux transformation
based on the function \cite{SolScat}
$f(x;\kappa)
=A_{\varphi}\exp (\kappa x)$, where $\kappa>1$.
In the present simplest case of $H_1$ 
this will give us the shifted  system (\ref{HPT}), 
in which the argument of the potential $x$ changes  
for~\footnote{In the case of a reflectionless system  
with $n>1$  bound states, the  isospectral deformation of the potential,
which can be generated by applying the appropriate Darboux-Crum 
transformation, 
corresponds to a  `snapshot' of the evolved $n$-soliton
solution of the Korteweg-de Vries equation, see Refs. 
\cite{SolScat,AGP2,AGP1}. In that case,
like in the case of Lam\'e system with periodicity defects 
we consider below, the form of 
isospectrally deformed potential is
different from the original one.}
 $x+\lambda$, where 
 $\lambda=\frac{1}{2}\log \frac{\kappa -1}{\kappa+1}$.

In all three indicated  cases, the corresponding 
extended system $\mathcal{H}=
{\rm diag}\,(H_1,\tilde{H})$  
will be described by the exotic centrally extended 
nonlinear $\mathcal{N}=4$ supersymmetry \cite{AGP1,AGP2,SolScat}.
Such reflectionless systems will correspond 
to the $k\rightarrow 1$ limit of the systems 
obtained from the one-gap Lam\'e system 
by introducing into it the periodicity defects 
by means of the appropriate Darboux(-Crum)
transformation.

In the subsequent sections we describe how to 
introduce such periodicity defects, and discuss 
the associated exotic nonlinear supersymmetric 
structure. 

\section{Darboux translations of the  Lam\'e system}\label{Darbtrans}

Assume that we have a system described by a Hamiltonian operator
of the most general form 
$H=-\frac{d^2}{dx^2}+U(x)$, and that $\psi(x)$ is its arbitrary 
 physical, or non-physical eigenstate, $H\psi=\mathcal{E}\psi$.
 As in (\ref{Adef}), we define the first order operators
 \begin{equation}\label{Apsidef}
	A_\psi=\psi\frac{d}{dx}\frac{1}{\psi}=\frac{d}{dx}
	+\Delta(x)\,,\quad
	 \Delta(x)=-\frac{d}{dx}\log \psi(x), 
\end{equation}
and 
\begin{equation}
	A^{\sharp}_\psi=-\frac{1}{\psi}\frac{d}{dx} \psi=
	-\frac{d}{dx}+\Delta(x).
\end{equation}
If  $\psi(x)$ is a real valued function modulo a possible 
complex multiplicative constant, then 
the operators $A_\psi$ and $A^{\sharp}_\psi$ are mutually conjugate,
$A^{\sharp}_\psi=A^{\dagger}_\psi$.
Another, linear independent  eigenstate of $H$ of the same eigenvalue 
$\mathcal{E}$ is given by $\tilde{\psi}(x)=\psi(x)\int dx/\psi^2(x)$.
The action of  the operator $A_\psi$ on this eigenstate produces 
a kernel of the operator $A^{\sharp}_\psi$,
$A_\psi\tilde{\psi}(x)=1/\psi(x)$. 
The second order operator 
$A^{\sharp}_\psi A_\psi=-\frac{d^2}{dx^2}+\Delta^2(x)-\Delta'(x)$
has exactly the same kernel, spanned by $\psi(x)$
and $\tilde{\psi}(x)$, 
as the second order differential operator
$H-\mathcal{E}$,  and therefore,  $A^{\sharp}_\psi A_\psi=H-\mathcal{E}$,
and $\Delta^2(x)-\Delta'(x)=U(x)-\mathcal{E}$.

Consider now the operator 
$A _\psi A^{\sharp}_\psi=-\frac{d^2}{dx^2}+\Delta^2(x)+\Delta'(x)=
A^{\sharp}_\psi A_{\psi}+2\Delta'(x)\equiv \tilde{H}-\mathcal{E}$.
The wave function $1/\psi(x)$ is the eigenstate 
of the Schr\"odinger Hamiltonian operator $\tilde{H}$ of eigenvalue 
$\mathcal{E}$.
Another, linear independent  eigenstate of $\tilde{H}$
of the same eigenvalue $\mathcal{E}$
is $\frac{1}{\psi(x)}\int \psi^2(x) dx$. The latter
is mapped by the operator $A^{\sharp}_\psi$
into the state $\psi(x)$ being the zero mode of $A_\psi$.

Let us return now to the Lam\'e system 
(\ref{HLame}). Its eigenstates $\Psi^\alpha_+(x)$ 
obey the following property
\begin{equation}\label{Psi-Psi}
	\Psi^\alpha_+(-x-\alpha-i{\rm {\bf K}}')=
	-\Psi^\alpha_-(x+\alpha+i{\rm {\bf K}}')=
	\frac{\mathcal{C}(\alpha)}{\Psi^\alpha_+(x)}\,,
	\end{equation}
where $\mathcal{C}(\alpha)=-\exp\left(\alpha\left({\rm Z}(\alpha)+i
	\frac{\pi}{2{\rm {\bf K}}}\right)+i{\rm {\bf K}}'{\rm Z}(\alpha)\right)$.
Taking $\psi(x)=\Psi^\alpha_+(x)$ in  (\ref{Apsidef}),
we obtain the factorisation for the one-gap Lam\'e Hamiltonian,
\begin{equation}\label{H(x+alpha}
	A^{\sharp}_{\Psi^\alpha_+} A_{\Psi^\alpha_+} 
	=H_{0,0}(x)-\mathcal{E}(\alpha)\,.
\end{equation}
Making use of  the relation (\ref{Psi-Psi}) we find then 
that 
\begin{equation}\label{AA+H}
	A_{\Psi^\alpha_+}A^{\sharp}_{\Psi^\alpha_+} 
	 =H_{0,0}(x+\alpha+i{\rm {\bf K}}')-\mathcal{E}(\alpha)\,.
\end{equation}
As the Darboux-partner of the Lam\'e Hamiltonian 
$H_{0,0}(x)$ we obtain  therefore the translated 
Hamiltonian operator
$H_{0,0}(x+\alpha+i{\rm {\bf K}}')$. 

In the case of the lower prohibited band,  
the wave function $\Psi^{\beta^-+i{\rm {\bf K}}'}_+(x)$ reduces 
to the real function $F(x;\beta)$ modulo 
a constant multiplier, see Eqs.  (\ref{Psi*gap0}),
(\ref{Fdef}), 
and we have $A_{\Psi^\alpha_+}=A_{F}$,
$A^{\sharp}_{\Psi^\alpha_+}=A_{F}^{\dagger}$. 
The property $\dn\,(x+2i{\rm {\bf K}}')=-\dn\, x$
gives us then in (\ref{AA+H}) 
the same Hermitian Lam\'e Hamiltonian operator
but shifted for the real distance $\beta^-$, $0<\beta^-<{\rm {\bf K}}$,
$H_{0,0}(x+\alpha+i{\rm {\bf K}}')=
H_{0,0}(x+\beta^-)$.
 The
obtained Darboux  transformations,
supersymmetry and physics associated with them
were studied in diverse aspects in \cite{PAN}.
Note here that the real function ${F}(x;\beta^-)$, 
shown on Figure \ref{fig3},
 takes  positive values  for all $x$, 
blows up exponentially  when $x\rightarrow -\infty$,
and tends to zero for  $x\rightarrow +\infty$.
The limit case $\beta^-={\rm {\bf K}}$ 
corresponds to a translation for the half of the
period of Lam\'e Hamiltonian. It is produced on the 
basis of the ground state $\psi(x)=\dn\,x$ \cite{CJNP}. 
The obtained 
 Darboux transformations are analogous to the 
translation transformations  in the case of  
the P\"oschl-Teller system 
(\ref{HPT})
with one bound state, 
which are constructed on the basis of 
the exponent-like non-physical 
eigenstates 
$\psi=A_{\varphi} \exp \kappa x$, $\kappa >1$,
of $H_1$.

In the forbidden  band separating the allowed bands,
the eigenfunction $\Psi_+^{\beta^+}(x)$ 
takes real values, but it has infinite number of zeroes 
at the points $-\beta^++2n{\rm {\bf K}}$,
$n\in \Z$.  In this case relation (\ref{H(x+alpha}) 
gives us the factorisation of the Lam\'e Hamiltonian
$H_{0,0}(x)$ in terms of the singular 
mutually conjugate Darboux generators.
The alternative product (\ref{AA+H}) of these  first order 
differential operators produces the Hermitian operator
 $H_{0,0}(x+\beta^++i{\rm {\bf K}}')$ with the 
singular Treibich-Verdier potential \cite{TV}
\begin{equation}\label{TVpot}
	V_{0,0}(x+\beta^++i{\rm {\bf K}}')=
	\frac{2}{\sn^2(x+\beta^+)}-k^2\,,
\end{equation}
where we have taken into account the
identity $\sn\,(x+i{\rm {\bf K}}')={1}/{k}\,{\sn\,x}\,$.
The limiting case $\beta^+=0$ corresponds to the
singular Darboux transformation constructed on the 
basis of the eigenfunction $\psi(x)=\sn\,x$
at the edge of the conduction band.
Another limit case $\beta^+={\rm {\bf K}}$ 
gives rise to
the singular transformation based on
 the eigenfunction $\psi(x)=\cn\,x$
 at  the edge of the valence band, for which the Treibich-Verdier 
potential reduces to 
\begin{equation}\label{V00dc}
V_{0,0}(x+{\rm {\bf K}}+i{\rm {\bf K}}')=
	2\,{\rm dc}^2 x-k^2\,,
\end{equation} 
	where we have employed the identity
$\sn\,(x+{\rm {\bf K}}+i{\rm {\bf K}}')
={\dn\,x}/{k\,\cn\,x}\,$.	

Inside the valence band, 
the eigenstate $\Psi_+^{{\rm {\bf K}}+i\gamma^-}(x)$
takes nonzero but complex values.
The Darboux partner (\ref{AA+H})
reduces in this case to the nonsingular $PT$-symmetric 
Hamiltonian with the potential
\begin{equation}\label{V00valence}
	V_{0,0}(x+\alpha+i{\rm {\bf K}}')=
	2\,{\rm dc}^2(x+i\gamma^-)
	-k^2\,.
\end{equation}
The edge value $\gamma^-={\rm {\bf K}}'$ 
corresponds here to the regular Hermitian 
Lam\'e Hamiltonian operator 
shifted for the half-period, 
$H_{0,0}(x+{\rm {\bf K}})$.  Another edge 
value $\gamma^-=0$ gives
the singular Hermitian Treibich-Verdier 
Hamiltonian (\ref{V00dc}) obtained  on the  basis 
of the edge state $\psi(x)=\cn\,x$.

At last, inside the conduction band, the  Hamiltonian
in (\ref{AA+H}) reduces to the
regular $PT$-symmetric operator 
with the potential 
\begin{equation}\label{V00valence}
	V_{0,0}(x+i\gamma^++i{\rm {\bf K}}')=\frac{2}{\sn^2(x+i\gamma^+)}
	-k^2\,.
\end{equation}
The edge case $\gamma^+=0$ reduces 
to the singular Treibich-Verdier 
potential generated via 
the choice $\psi(x)=\sn\,x$.

The 
described first order Darboux 
transformations  can
also be considered 
for  the values of the 
parameter $\alpha$ lying inside the 
rectangular on Fig. \ref{fig1}. In this case
the partner Hamiltonian
will be  nonsingular with the potential taking 
complex values, which, however, will
be neither Hermitian nor $PT$-symmetric
operator. 
Indeed, under Hermitian conjugation
the shifted Hamiltonian operator from (\ref{AA+H})
transforms as 
$(H_{0,0}(x+\alpha+i{\rm {\bf K}}'))^\dagger
=H_{0,0}(x+\alpha^*+i{\rm {\bf K}}')$,
where we have taken into account the 
pure imaginary period 
$2i{\rm {\bf K}}'$ of the potential $V_{0,0}(x)$.
Analogously, we have $PT(H_{0,0}(x+\alpha+i{\rm {\bf K}}'))=
H_{0,0}(x-\alpha^*+i{\rm {\bf K}}')$, where 
the even nature of the potential have 
additionally been taken into account.
The shifted Hamiltonian is therefore Hermitian 
if $\alpha-\alpha^*=2n{\rm {\bf K}}
+2im{\rm {\bf K}}'$, $n,m\in\Z$,
while it is $PT$-symmetric when 
$\alpha+\alpha^*=2n{\rm {\bf K}}
+2im{\rm {\bf K}}'$.
For the $\alpha$-region shown on Figure~\ref{fig1}
the first condition is satisfied only on the 
upper and lower horizontal
edges of the rectangular, which correspond to the 
prohibited zones in the spectrum,
while the second relation takes place only
on the vertical edges corresponding 
to the allowed valence  and conduction bands.

Below we shall see that 
the higher order Darboux-Crum transformation
corresponding to a  composition of the 
Darboux transformations, each of which 
generates translated Lam\'e system 
of the form (\ref{AA+H}),
produces the Lam\'e system with 
a shift of the argument equal to the sum 
of individual translations.

\section{Lam\'e system deformed by 
non-periodic, soliton  defects}

In this section we show how to introduce the
reflectionless, soliton (non-periodic) defects 
into the one-gap Lam\'e system.

\subsection{Lower forbidden band}

\begin{figure}[!h]
  \centering
  \includegraphics[scale=0.7]{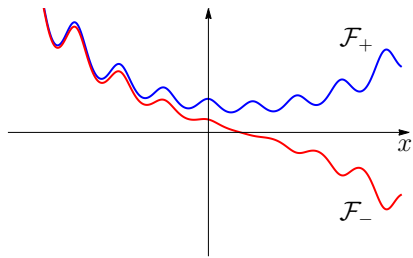}\\
  \caption{At $C = 1$,  
  $ \mathcal{F}_+ (x; \beta ^ -, C) $ is an even function,
while $ \mathcal {F}_- (x; \beta ^ -, C)$ is odd.
The symmetry of non-physical eigenfunctions 
$\mathcal{F}_\pm (x; \beta ^ -, C) $ of $H_{0,0}$ is broken 
for $C \neq 1$. Here,
the case $C> 1$ is shown.  With $C$ increasing, the minimum
of $ \mathcal{F}_+ (x; \beta ^ -, C) >0$  and zero of 
$ \mathcal{F}_- (x; \beta ^ -, C) $ are displaced to the the right.
A similar situation occurs when $ 0 <C < 1$ but 
with a displacement to the negative coordinate axis. In fact, the
form of the functions  for $ 0 <C < 1$ is obtained from that for $C>1$
via the relation $\mathcal{F}_\pm (x; \beta ^ -, C)=\pm
\mathcal {F}_\pm (-x; \beta ^ -, C^{-1})$.}
  \label{fig3}
\end{figure}

The  real-valued eigenfunction
 $F(x;\beta^-)$ in the lower prohibited band 
 has the modulated exponent-like behaviour.  
Let us take  a linear combination of the 
two eigenfunctions  of the same eigenvalue, 
\begin{equation}\label{Fsi+-}
\mathcal{F}_\pm(x;\beta^-,C)=C {F}(x;\beta^-)\pm 
\frac{1}{C}{F}(-x;\beta^-)\,,
\end{equation}
where 
${\rm {\bf K}}>\beta^->0$, 
and a real parameter $C$ is restricted by the condition 
$C>0$.
These states have the properties 
$\mathcal{F}_\pm(-x;\beta^-,C^{-1})=
\pm \mathcal{F}_\pm(x;\beta^-,C)$.
The function $\mathcal{F}_+(x;\beta^-,C)$
takes strictly positive values, 
and blows up exponentially in the 
limits $x\rightarrow\pm\infty$.
The function $\mathcal{F}_-(x;\beta^-,C)$, on the other hand,
tends  exponentially to $+\infty$ and $-\infty$
when $x$ tends to $-\infty$ and $+\infty$,
respectively, and has a unique zero  whose position 
 depends on the 
values of the parameters $\beta^-$ and $C$. 
The form of the functions 
$\mathcal{F}_\pm(x;\beta^-,C)$ is shown
on Figure \ref{fig3}.

Construct now the first order  operator 
\begin{equation}\label{A1def}
	A_{0,1}=\mathcal{F}_{+}(1)\frac{d}{dx}
	\frac{1}{\mathcal{F}_{+}(1)}=\frac{d}{dx}-\frac{d}{dx}
	\log\, \mathcal{F}_{+}(1),
\end{equation}
where $\mathcal{F}_{+}(1)=\mathcal{F}_+(x;\beta^-_1,C_1)$.
We have 
$A_{0,1}^\dagger A_{0,1}=H_{0,0}
-\varepsilon^-_1$,
and 
$ A_{0,1}A_{0,1}^\dagger=
H_{0,1} -\varepsilon^-_1$, where 
$\varepsilon^-_1\equiv
\mathcal{E}(\beta^-_1+i{\rm {\bf K}}')
=-\cn^2\beta^-_1/\sn^2\beta^-_1<0$,
\begin{equation}\label{H01}
H_{0,1}=H_{0,0}-2\frac{d^2}{dx^2}
(\log \mathcal{F}_+(1))=-\frac{d^2}{dx^2}
+V_{0,1}(x)\,,
\end{equation}
\begin{equation}\label{V01}
V_{0,1}(x)=1+k'^2-2\frac{{\rm {\bf E}}}{{\rm {\bf K}}}
-2\frac{d^2}{dx^2}
\Big(\log \chi^{\beta^-_1}_{0,1}(x;C_1)\Big)\,,
\end{equation}
\begin{equation}\label{var01}
\chi^{\beta^-_1}_{0,1}(x;C_1)=C_1\Theta(x+\beta^-_1)
\exp(-x{\rm z}(\beta^-_1))
+
\frac{1}{C_1}\Theta(x-\beta^-_1)\exp(x{\rm z}(\beta^-_1))\,.
\end{equation}

\begin{figure}
  \centering
  \includegraphics[scale=0.7]{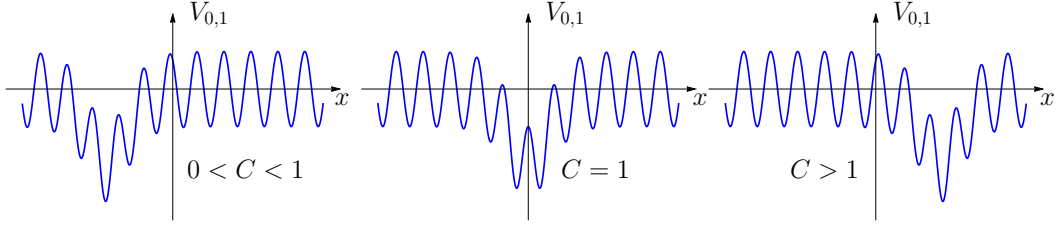}\\
  \caption{Potential 
  with a one-soliton defect which
  supports a bound state in the lower forbidden band.
  The soliton is broader when the energy of the bound state is closer to zero, 
  and a greater number of oscillations are observable within it. 
  The depth (amplitude) of the soliton, on the other hand, increases 
  when the negative energy of the bound state is deeper. 
  The sequence of the pictures illustrates  the propagation of the
  soliton in the periodic background of the Lam\'e potential.}\label{fig4}
\end{figure}

\begin{figure}[!h]
  \centering
  \includegraphics[scale=0.6]{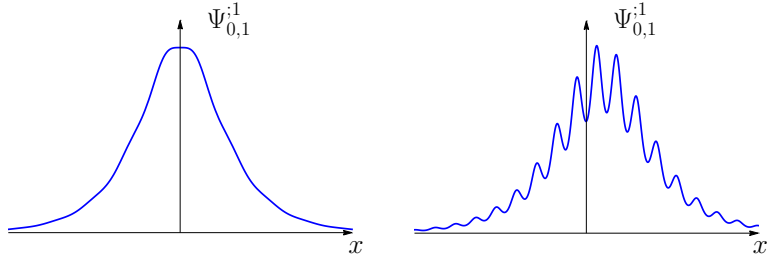}\\
  \caption{The bound state eigenfunction of the system 
  $H_{0,1}$. The state on the left corresponds to the 
  potential $V_{0,1}$ with $C=1$
   in the central picture on 
  Figure \ref{fig4}.
 The state on the right,
 with $C>1$, has
 energy closer to zero: 
 when the energy modulus
 is lower, the state is broader and 
 the oscillations in it are well notable. 
 By varying the parameter $C$, 
 the soliton defect in the potential is displaced 
 as well as the position of the bound state
 supported by it. 
 In correspondence with this,
 in the case of  $0<C<1$ not shown here,
 a localisation of the wave function
of the bound state  is shifted to the $x<0$ 
 region in comparison with 
 the case $C>1$.}
\label{fig5}
\end{figure}

The $\Theta(x)$ function 
appearing in the denominator  of $\mathcal{F}_+(x)$, see eq. 
(\ref{Fdef}),  
cancels the nontrivial potential term $-2\dn^2x$
in the Lam\'e Hamiltonian $H_{0,0}$ 
via the equality
$\frac{d^2}{dx^2}(\log\Theta(x))=
\dn^2x-\frac{{\rm {\bf E}}}{{\rm {\bf K}}}$,
that results in the non-periodic potential (\ref{H01}), (\ref{V01}),
see Figure \ref{fig4}.
By the Darboux construction, 
the system $H_{0,1}$ has
the same spectrum as  the one-gap
Lam\'e system  except that it possesses an additional
 discrete level of energy
$\varepsilon^-_1$. This is the eigenvalue 
of the bound state described by the normalisable 
nodeless
wave function
\begin{equation}\label{bound1}
	\Psi_{0,1}^{;1}(x;\beta^-_1,C_1)=
	\frac{1}{\mathcal{F}_+(x;\beta^-_1,C_1)}
\end{equation}
shown on Figure \ref{fig5}, which is a zero mode of the operator
$A^\dagger_{0,1}$.
The nonzero lower index in the Hamiltonian and potential 
reflects here the property that the system possesses one bound state 
in the \emph{lower} forbidden band. The upper index 
in notation for the wave function of the bound state 
is introduced 
bearing in mind a generalisation for the case
of a perturbed Lam\'e system with  various
 bound states supported both 
in lower and upper forbidden bands.

Other physical  and non-physical 
eigenfunctions  of $H_{0,1}$ are
given by 
$A_{0,1}\Psi_\pm^{\alpha}\left(x\right)$. 
They correspond to the same permitted and prohibited 
values of energy
as the eigenstates $\Psi_\pm^{\alpha}\left(x\right)$ 
of the periodic Lam\'e Hamiltonian.
This shows that the introduced non-periodic defect 
is reflectionless: physical Bloch states are transformed 
into the Bloch states.

Asymptotically, in the limit $x\rightarrow -\infty$ the potential 
has a form of the one-gap periodic Lam\'e potential,
$V_{0,1}(x)\rightarrow V_{0,1}^{-\infty}(x)= V_{0,0}(x+\beta^-_1)$. 
In another limit $x\rightarrow +\infty$, we have
$V_{0,1}(x)\rightarrow V_{0,1}^{+\infty}(x)=
V_{0,0}(x-\beta^-_1)$.
So, the defect produces a phase shift  
between the asymptotically periodic 
one-gap potentials which is equal to
$-2\beta^-_1$.
This observation follows also directly  from (\ref{Fsi+-}).
Asymptotically we have 
$\mathcal{F}_+(x;\beta^-_1,C_1)\rightarrow
C_1F(x;\beta^-_1) $ when $x\rightarrow -\infty$, and 
$\mathcal{F}_+(x;\beta^-_1,C_1)\rightarrow
C^{-1}_1F(-x;\beta^-_1) $ for $x\rightarrow \infty$.
Employing the results discussed below
(\ref{AA+H}), we can write
\begin{equation}\label{Hasympt}
A_{0,1}A^\dagger_{0,1} 
\rightarrow H_{0,0}(x\pm \beta^-_1)
-\varepsilon^-_1\quad
{\rm for}\quad
x\rightarrow \mp\infty\,.
\end{equation}
We get the phase displacement 
\begin{equation}\label{phase_1}
\Delta\phi(\beta^-_1)=-2\beta^-_1,\,\qquad
\varepsilon^-_1=
-{\rm cd}^2\beta^-_1
<0\,,
\end{equation}
where we indicate  the discrete energy level 
of the bound state of  $H_{0,1}$. 
The potential $V_{0,1}(x)$ may be treated 
as a soliton defect in the 
background of  the  one-gap 
periodic Lam\'e system.

Notice that in the limit 
$C_1\rightarrow \infty$  (or, $C_1\rightarrow 0$),
the soliton `goes' to infinity, and in correspondence with 
Eq. (\ref{H01}),
$H_{0,1}$ transforms into the shifted 
Lam\'e Hamiltonian $H_{0,0}(x+\beta^-_1)$
(or, $H_{0,0}(x-\beta^-_1)$).
\vskip0.1cm 

Before we proceed further, 
let us show that the infinite period limit 
of the obtained system with 
a periodicity defect
corresponds to a
reflectionless system  
of a  generic form (\ref{H2refless})
with two bound states
of energies $\mathcal{E}_0=
0$ and $\mathcal{E}_1=1
-\kappa_1^2<0$.
To this aim we apply the 
limit $k\rightarrow 1$ to the operator 
(\ref{A1def}). 
The non-physical eigenfunction $\mathcal{F}_+(1)$ 
of the Lam\'e system in this 
limit transforms into the eigenfunction 
(\ref{phi1state}), whose explicit form is
\be\label{Psixx+}
\varphi_1(x;\kappa_1,\tau_1)=
\frac{1}{\cosh x}W(\cosh x,\sinh\kappa_1(x+\tau_1)).
\ee
Indeed, in  the indicated limit 
${\rm Z}(\beta\vert 1)=\tanh \beta$, and 
 ${\rm z}(\beta^-_1)$, defined in Table \ref{T1},
reduces to
${\rm z}(\beta^-_1)\rightarrow 
\tanh \beta^-_1+\frac{1}{
\sinh \beta^-_1\cosh \beta^-_1}=
{\rm cotanh}\,\beta^-_1\equiv \kappa_1$,
where $1<\kappa_1<\infty$ 
since  ${\rm {\bf K}}\rightarrow \infty$
and  then  $\beta^-_1\in (0,\infty)$. 
We have also 
$\frac{\Theta(x\pm\beta\vert 1)}{\Theta(x\vert 1)}=
\frac{\cosh (x\pm\beta)}{\cosh x}$.
Introducing the notation $C_1\equiv \exp\kappa_1\tau_1$, 
where $\tau_1$ is an arbitrary real parameter,
we find that 
$\mathcal{F}_+(1)$ transforms into 
$\frac{1}{\cosh x}\left(
\cosh (x+\beta^-_1)\exp(-\kappa_1(x+\tau_1))
+\cosh (x-\beta^-_1)\exp(\kappa_1(x+\tau_1))\right)$.
This function reduces, up to inessential 
nonzero multiplicative
constant $\sinh \beta^-_1$, to (\ref{Psixx+}).
Then in correspondence with the discussion of Section
\ref{PT1Dar}, the limit of the operator 
(\ref{A1def}) is the Darboux generator, 
which intertwines the reflectionless
P\"oschl-Teller Hamiltonian (\ref{HPT}) 
with the  Hamiltonian operator 
(\ref{H2refless}).  Thus, we conclude 
that the infinite period limit of (\ref{H01})
corresponds to the reflectionless system
(\ref{H2refless}). 
\vskip0.1cm

To introduce several discrete energies into the  spectrum of 
the one-gap 
Lam\'e system by making use of its non-physical states 
from the lower prohibited band, consider first the case of the two 
bound states. It is not difficult to show that 
the Wronskian 
$W(\mathcal{F}_+(1),\mathcal{F}_-(2))=\mathcal{F}_+(1)
\mathcal{F}'_-(2)-\mathcal{F}'_+(1)\mathcal{F}_-(2)$, where  
$\mathcal{F}_+(1)=\mathcal{F}_+(x;\beta^-_1,C_1)$,
$\mathcal{F}_-(2)=\mathcal{F}_-(x;\beta^-_2,C_2)$,
takes strictly negative values,  $W(x)<0$, if
 ${\rm {\bf K}}>\beta^-_1>\beta^-_2>0$, see Appendix.
The corresponding energies of the non-physical
eigenstates of $H_{0,0}$ are ordered  then as 
$0>\mathcal{E}( \beta^-_1+i{\rm {\bf K}}')>
\mathcal{E}(\beta^-_2+i{\rm {\bf K}}')$.  
With such a choice of the states,
we can construct the Darboux-Crum transformation 
producing a non-periodic deformation of  Lam\'e system,
which in addition to the one-gap spectrum
of $H_{0,0}(x)$  has 
two discrete energy values $\varepsilon^-_j=\mathcal{E}( \beta^-_j
+i{\rm {\bf K}}')$,
$j=1,2$,  
\begin{equation}\label{H02}
H_{0,2}=-\frac{d^2}{dx^2} +V_{0,2}(x)\,,\quad
V_{0,2}(x)=V_{0,0}(x)-2\frac{d^2}{dx^2}\big(\log 
W(\mathcal{F}_+(1),\mathcal{F}_-(2))\big)\,.
\end{equation}
The discrete energy levels $\varepsilon^-_1$ and $\varepsilon^-_2$ 
correspond, respectively,  to the two
bound states 
\begin{equation}\label{2statesE1}
	\Psi_{0,2}^{;1}(x;\beta^-_1,C_1,\beta^-_2,C_2)=
	\frac{W(\mathcal{F}_+(1),\mathcal{F}_-(2),\mathcal{F}_-(1))}
	{W(\mathcal{F}_+(1),\mathcal{F}_-(2))}\,,
\end{equation}
\begin{equation}\label{2statesE2}
	\Psi_{0,2}^{;2}(x;\beta^-_1,C_1,\beta_2,C_2)=
	\frac{W(\mathcal{F}_+(1),\mathcal{F}_-(2),\mathcal{F}_+(2))}
	{W(\mathcal{F}_+(1),\mathcal{F}_-(2))}\,.
\end{equation}
Other physical and non-physical eigenstates of 
the system (\ref{H02}) are given by
\begin{equation}\label{2statesEalpha}
	\Psi^\alpha_{0,2;\pm}(x;\beta^-_1,C_1,\beta^-_2,C_2)=
	\frac{W(\mathcal{F}_+(1),\mathcal{F}_-(2),\Psi^\alpha_\pm)}
{W(\mathcal{F}_+(1),\mathcal{F}_-(2))}\,,
\end{equation}
and correspond to  the Darboux-Crum mapping
of the eigenstates (\ref{Psi+-}) of the initial
Lam\'e system. The energies of these states
are defined by the values of the parameter
$\alpha$  exactly in the same way as 
for the system (\ref{HLame}).
In accordance with (\ref{Fsi+-}),
expressions  (\ref{2statesE1}) and (\ref{2statesE2})
for the bound states correspond
to linear combinations of the eigenstates 
(\ref{2statesEalpha}) with 
$\alpha=\beta^-_1+i{\rm {\bf K}}'$ 
and $\alpha=\beta^-_2+i{\rm {\bf K}}'$, respectively.

Let us take now $n$ states
\be\label{Fsjn}
\mathcal{F}_{s_{j}}(j)=\mathcal{F}_{s_{j}}(x;\beta^-_j,C_j)
\quad {\rm with} \quad
{\rm {\bf K}}>\beta^-_1>\beta^-_2>\ldots>\beta^-_n>0\,,
\ee
where $s_{j}$ corresponds to a linear combination of the form 
(\ref{Fsi+-})
with index  $+$ ($-$)  for $j$ odd (even). Then by
applying  the Darboux-Crum construction on the basis of
these eigenstates, we obtain a non-periodic  
deformation  $H_{0,n}$ of the 
Lam\'e system $H_{0,0}$ with  $n$ bound states 
with energies
$0>\varepsilon^-_1>
\varepsilon^-_2>\ldots \varepsilon^-_n>-\infty$.

The potential of this system is given by a generalisation 
of Eq. (\ref{H02}), in which the Wronskian 
has to be changed for 
\begin{equation}\label{W_n}
\W_{0,n}(x)=W(\mathcal{F}_{+}(1),
\mathcal{F}_{-}(2),\ldots,\mathcal{F}_{s_{n}}(n))\,.
\end{equation}
The $n$ bound states of energies 
$\varepsilon^-_j$
are described by the normalisable 
wave functions 
\begin{equation}\label{nstatesEj}
	\Psi_{0,n}^{;j}(x;\beta^-_1,C_1,\ldots,\beta^-_n,C_n)=
	\frac{W(\mathcal{F}_+(1),\mathcal{F}_-(2),
	\ldots,\mathcal{F}_{s_{n}}(n),
	\mathcal{F}_{-s_{j}}(j))}
	{\W_{0,n}}\,,\qquad j=1,\ldots, n\,,
\end{equation}
while other corresponding eigenstates of $H_{0,n}$
are given by the generalisation of Eq. 
(\ref{2statesEalpha}),
\begin{equation}\label{nstatesEalpha}
	\Psi^\alpha_{0,n;\pm}(x;\beta^-_1,C_1,\ldots,\beta^-_n,C_n)=
	\frac{W(\mathcal{F}_+(1),\mathcal{F}_-(2),\ldots,\mathcal{F}_{s_{n}}(n),
	\Psi^\alpha_\pm)}
	{\W_{0,n}}\,.
\end{equation}
As in the case (\ref{H02}), 
bound states (\ref{nstatesEj}) may be obtained 
from (\ref{nstatesEalpha})
by putting there 
$\alpha=\beta^-_j+i{\rm {\bf K}}'$, $j=1,\ldots, n$,
and changing  the wave functions  $\Psi^\alpha_\pm$
on the r.h.s. for the corresponding linear combinations
of them.

Applying then the limit $x\rightarrow -\infty$ to the Wronskian
$\W_{0,n}(x)$, we find that it transforms, up to a 
multiplicative constant, into 
$\W_{0,n}(x)=W(F(x;\beta^-_1)(x),\ldots,F(x;\beta^-_n))$.
Asymptotically we get a potential 
$V_{0,n}^{-\infty}(x)=\lim_{x\rightarrow-\infty}
(-2\frac{d^2}{dx^2}\log \W_{0,n}(x))=
V_{0,0}(x+b)$, where $b=\sum_{j=1}^n\beta_j$.
Analogously, in another limit $x\rightarrow +\infty$ we get
the asymptotic form of the potential
$V_{0,n}^{+\infty}(x)=
V_{0,0}(x-b)$. The phase displacement produced by the 
$n$ solitons (defects) is 
\begin{equation}\label{phase_n}
\Delta\phi(\beta^-)=-2\sum_{j=1}^n\beta^-_j\,,
\end{equation}
that generalises the one-soliton effect (\ref{phase_1}).

The eigenstates of the system $H_{0,n}$
(\ref{nstatesEj}) and (\ref{nstatesEalpha})
can be presented in an alternative form \cite{AGP2}
\begin{equation}\label{An}
\Psi(x;\beta^-_1,C_1,\ldots,\beta^-_n,C_n)=
\A_{0,n}\Psi(x)\,,\qquad
\A_{0,n}=A_{0,n} A_{0,n-1}\ldots A_{0,1}\,,
\end{equation}
where the wave function on the l.h.s. corresponds to 
(\ref{nstatesEj}) for the choice 
$\Psi=\mathcal{F}_{-s_{j}}(j)$ on the r.h.s.,
while it corresponds to the eigenfunctions
 (\ref{nstatesEalpha})
for the choice 
$\Psi=\Psi^\alpha_\pm$ on the r.h.s.
The operator $\A_{0,n}$ is a differential operator 
of  order $n$, which is constructed 
in terms of the recursively defined first order 
differential operators (\ref{A1def}) and 
\begin{equation}\label{Aj}
A_{0,j}=(\A_{0,j-1}\mathcal{F}_{s_{j}}(j))\frac{d}{dx}
\frac{1}{(\A_{0,j-1}\mathcal{F}_{s_{j}}(j))}=
\frac{d}{dx}+\mathcal{W}_{0,j}\,,\quad
j=2,\ldots,
\end{equation}
where 
\begin{equation}\label{WjOmega}
\mathcal{W}_{0,j}=\Omega_{0,j}-\Omega_{0,j-1},\,\quad
\Omega_{0,j}=-\left(\log \W_{0,j}\right)_x\,,
\end{equation}
and  $\W_{0,1}\equiv \mathcal{F}_+(1)$. Eqs. (\ref{Aj}) and 
(\ref{WjOmega}) can also be used for $j=1$ by 
putting $\W_{0,0}=1$.
Note here that making use of Eqs. (\ref{An}),
it is easy  to see that  in the case 
of the two-soliton defect, particularly,  the bound states
(\ref{2statesE2}) and (\ref{2statesE1}) are reduced
modulo multiplicative constants to 
the functions $\mathcal{F}_+(1)/\W_{0,2}$, and 
$\mathcal{F}_-(2)/\W_{0,2}$,
respectively. This  shows explicitly that the first function describing
the discrete ground state is nodeless, while the 
second wave function corresponding to 
the first excited bound state
has exactly one zero as it should be 
for the lowest bound states in the spectrum.

Relation (\ref{An}) means that the operator  
$\A_{0,n}$ maps the eigenstates of the Lam\'e system 
(\ref{HLame}) into the corresponding eigenstates 
of $H_{0,n}$. Its $n$-dimensional kernel is spanned 
by the eigenstates $\mathcal{F}_{s_{j}}(j)$, $j=1,\ldots,n$.
These relations reflect the fact that 
the Darboux-Crum transformation of order $n$ corresponds 
to a composition of $n$ subsequent  Darboux 
maps $H_{0,0}\rightarrow H_{0,1}\rightarrow \ldots
\rightarrow H_{0,n}$. In accordance with this,
the operators $\A_{0,n}$ and   $\A_{0,n}^\dagger$
intertwine the Hamiltonian operator  $H_{0,n}(x)$
with the Lam\'e Hamiltonian $H_{0,0}(x)$,
\begin{equation}\label{HA0nH}
\A_{0,n}H_{0,0}=H_{0,n}\A_{0,n}\,,\qquad
\A_{0,n}^{\dagger}H_{0,n}=H_{0,0}\A_{0,n}^{\dagger}\,.
\end{equation}
The products of the operator $\A_{0,n}$ and its conjugate are
\begin{equation}\label{HA0nH}
\A_{0,n}\A_{0,n}^{\dagger}=\prod_{j=1}^n(H_{0,n}-\varepsilon^-_j)\,,\qquad
\A_{0,n}^{\dagger}\A_{0,n}=\prod_{j=1}^n(H_{0,0}-\varepsilon^-_j)\,.
\end{equation}

Alternative representation 
given by Eqs. (\ref{An}), (\ref{Aj})  
is valid for arbitrary   Darboux-Crum transformations
generated on the basis of $n$ eigenstates of a generic 
Schr\"odinger  Hamiltonian 
 \cite{AGP2}. In particular case of the 
one-gap Lam\'e system $H=H_{0,0}$
and the choice of eigenstates $\psi_j(x)=\Psi^{\alpha_j}_+(x)$,
each of which, as we saw in the previous section,
generates the translation of the Lam\'e system for
$\alpha_j+i{\rm{\bf K}}'$,   
we obtain the Darboux-Crum transformation
producing  the translation of $H_{0,0}(x)$ for
$\sum_{j=1}^n \alpha_j +in{\rm{\bf K}}'$.
Taking into account that the system 
(\ref{HLame}) besides the real period 
$2{\rm{\bf K}}$ possesses also the imaginary
period $2i{\rm{\bf K}}'$, the shift 
produced by the Darboux-Crum transformation
reduces to $\sum_{j=1}^{2r} \alpha_j$ in the case of 
even $n=2r$, and to  $\sum_{j=1}^{2r+1} \alpha_j
+i{\rm{\bf K}}'$ when $n=2r+1$ is odd.
Making use of this observation, it is 
obvious that when the total shift produced by the Crum-Darboux 
transformation  reduces to a nontrivial period 
$2{\rm{\bf K}}n_1+2i{\rm{\bf K}}'n_2$
of the system  (\ref{HLame}) with $n_1^2+n_2^2\geq 2$, 
the corresponding higher order 
generator $\A_n$ gives us the integral
(multiplied in a generic case by
a polynomial in $H_{0,0}$ \cite{AP2})  of the
one-gap Lam\'e system.
This  is  the analog of  the  integral
(\ref{P1PT1}) of the reflectionless
P\"oschl-Teller system (\ref{HPT}),  
which is  the 
Lax-Novikov integral $\mathcal{P}_{0,0}$ 
for the system (\ref{HLame}),
\begin{equation}\label{LN00}
i\mathcal{P}_{0,0}=\frac{d^3}{dx^3}
+(1+k^2-3k^2\sn^2x)\frac{d}{dx}
-3k^2\sn\,x\,\cn\,x\,\dn\,x\,.
\end{equation}
In the limit $k\rightarrow 1$ it transforms into 
(\ref{P1PT1}). 
The kernel of this third order differential operator is spanned
by eigenfunctions $\dn\,x$, $\cn\,x$ and $\sn\,x$,
which correspond to the edges of the allowed bands. 
In correspondence with this, it admits an infinite number
of factorisations. Particularly, it can be presented in the form
\begin{equation}\label{P3factor}
	i\mathcal{P}_{0,0}=A_{1/\cn\,x}A_{{\cn\,x}/{\dn\,x}}A_{\dn\,x},
\end{equation}
where $A_{\dn\,x}$ is defined by 
relation of the form (\ref{Apsidef}) with 
$\psi(x)=\dn\,x$, etc.

The sense of the factorisation (\ref{P3factor}) is the following.
The first factor on the right, $A_{\dn\,x}$,  in accordance with its 
definition, annihilates $\dn\,x$, which is the state at the lower edge
of the valence band, or, that is the same
 up to inessential multiplicative factor,
is the limit case of the state
$F(x;\beta^-)$ with $\beta^-={\rm {\bf K}}$. 
Acting on the wave function $\sn\,x$, which 
corresponds to the lower edge of the conduction band,
the operator $A_{\dn\,x}$ translates it, as well as 
all other eigenstates of the Lam\'e system,
for the half-period ${\rm {\bf K}}$,
$\sn\,(x+{\rm {\bf K}})=\cn\,x/\dn\,x$, and then this  
$\sn$-function with a shifted argument 
 is annihilated by the operator $A_{{\cn\,x}/{\dn\,x}}$.
Acting on the wave function $\cn\,x$, which 
describes  the upper edge state of the valence band,
the $A_{\dn\,x}$  transforms it into 
$\cn\,(x+{\rm {\bf K}})$, while the subsequent action of the 
$A_{{\cn\,x}/{\dn\,x}}$  transforms this into 
$\cn\,(x+{\rm {\bf K}}+i{\rm {\bf K}}')=-i{k'}/k\cn\,x$,
that is annihilated finally by the 
first order operator $A_{1/\cn\,x}$.
In a similar way, one can construct five other 
factorisations of $\mathcal{P}_{0,0}$ having a simple 
interpretation in terms  of the Darboux transformations 
(translations)
generated by the edge states.
Relation (\ref{P3factor}) corresponds here 
to the Darboux-Crum transformation which generates 
the total shift for the nontrivial period 
$2{\rm{\bf K}}n_1+2i{\rm{\bf K}}'n_2$ 
with $n_1=n_2=1$ in correspondence 
with the discussion presented above.

The Lam\'e system's integral $\mathcal{P}_{0,0}$
satisfies the Burchnall-Chaundy relation 
\begin{equation}\label{BCLame}
	\mathcal{P}_{0,0}^2=H_{0,0}(H_{0,0}-k'^2)(H_{0,0}-1)\,,
\end{equation}
which lies in the basis of the hidden  bosonized 
nonlinear supersymmetry of the one-gap Lam\'e system 
\cite{CNP}.
The zeros of the third order polynomials in $H_{0,0}$  
correspond to the energies of the edges 
of the allowed bands of (\ref{HLame}).
In the limit $k\rightarrow 1$, (\ref{BCLame})
transforms into relation (\ref{P2BC}), in which the double
factor $H_1^2$ originates from the first two factors 
in (\ref{BCLame}), and 
roots  in  the  shrinking of the  valence band.

By analogy with the Lax-Novikov integral (\ref{P1PT1})
for the reflectionless P\"oschl-Teller system 
with one bound state, we can find the 
analogous integral for the $H_{0,n}$ system,
\begin{equation}\label{P3factor}
	\mathcal{P}_{0,n}=\A_{0,n}\mathcal{P}_{0,0}\A_{0,n}^\dagger\,,
	\qquad
	[\mathcal{P}_{0,n},H_{0,n}]=0\,,
\end{equation}
which is the differential operator of the order $2n+3$.
In correspondence with (\ref{BCLame})
and (\ref{HA0nH}),
it satisfies the Burchnal-Chaundy relation 
\begin{equation}
	\mathcal{P}_{0,n}^2=H_{0,n}(H_{0,n}-k'^2)(H_{0,n}-1)
	\prod_{j=1}^n(H_{0,n}-\varepsilon^-_j)^2\,.
\end{equation}

The systems 
$H_{0,0}$ and $H_{0,n}$ can be intertwined  not only by the 
operators  $\A_{0,n}$ and $\A_{0,n}^\dagger$,
but also by the operators 
\begin{equation}\label{B0n}
\B_{0,n}=\A_{0,n} \mathcal{P}_{0,n}\,,\quad
{\rm and}\quad
\B_{0,n}^\dagger\,.
\end{equation}

\subsection{Intermediate forbidden band}

Let us consider the intermediate prohibited band (gap) 
and the following linear combinations of eigenstates
(\ref{Psi+-})  in it,
\begin{equation}\label{Phi+}
	\Phi_+(1)\equiv \Phi_+(x;\beta^+_1,C_1)=C_1\Psi^{\beta^+_1}_+(x)+
	\frac{1}{C_1}\Psi^{\beta^+_1}_-(x),	
\end{equation}
\begin{equation}\label{Phi-}
	\Phi_-(2)\equiv \Phi_-(x;\beta^+_2,C_2)=C_2\Psi^{\beta^+_2}_+(x)-
	\frac{1}{C_2}\Psi^{\beta^+_2}_-(x),	
\end{equation}
where 
$0<\beta^+_l<{\rm {\bf K}}$, and 
$C_l$, $l=1,2$, are arbitrary real constants restricted by the condition
$C_l>0$. Taking into account relation  (\ref{Psireflec}),
the linear combinations used here differ
effectively in sign in comparison with those employed in (\ref{Fsi+-}).
This is related with the fact that the 
eigenvalue $\mathcal{E}(\beta^-+i{\rm {\bf K}}')$ 
is increasing function of the
real parameter $\beta^-$ in the lower prohibited band, while 
$d\mathcal{E}(\beta^+)/{d\beta^+}<0$ in the intermediate, 
upper forbidden band.
Both these functions have infinite number of zeros on the real line.
The choice of any of these two functions as the
function $\psi$ in operator (\ref{Apsidef}) produces by means 
of the first order Darboux transformation 
a singular partner for  the system $H_{0,0}(x)$.

Our next goal is to show  how by appropriate use 
of  the second 
order Darboux-Crum transformation
applied to $H_{0,0}$, one can generate 
a regular system with
two bound states in the gap.

Zeros of the non-physical 
eigenfunctions $\Psi^{\beta^+}_+(x)$ are
$-\beta^++2n{\rm {\bf K}}$, while the infinite set of zeros of 
the eigenstates $\Psi^{\beta^+}_-(x)$ is 
$\beta^+ +2n{\rm {\bf K}}$, $n\in\Z$.
On the open intervals $(-\beta^+,\beta^+)+2n{\rm {\bf K}}$
functions $\Psi^{\beta^+}_+(x)$ and $\Psi^{\beta^+}_-(x)$
take nonzero values of the opposite sign,
whereas on the open intervals 
$(\beta^+,2{\rm {\bf K}}-\beta^+)+2n{\rm {\bf K}}$
they  take values of the same sign.
Therefore, zeros of the linear combination (\ref{Phi+})
of $\Psi^{\beta^+}_+(x)$ and $\Psi^{\beta^+}_-(x)$ with 
$\beta^+_1=\beta^+$
are inside the first of the indicated set of the open  intervals,
and zeros of (\ref{Phi-}) with $\beta^+_2=\beta^+$ are inside the second 
set of the intervals. 
Since $\Phi_+$ and $\Phi_-$ 
are linearly independent eigenstates of the same eigenvalue 
$\mathcal{E}(\beta^+)$,
in correspondence with the oscillation theorem, 
each of the indicated open intervals contains exactly 
one zero of the respective function. 

\begin{figure}[!h]
  \centering
  \includegraphics[scale=0.9]{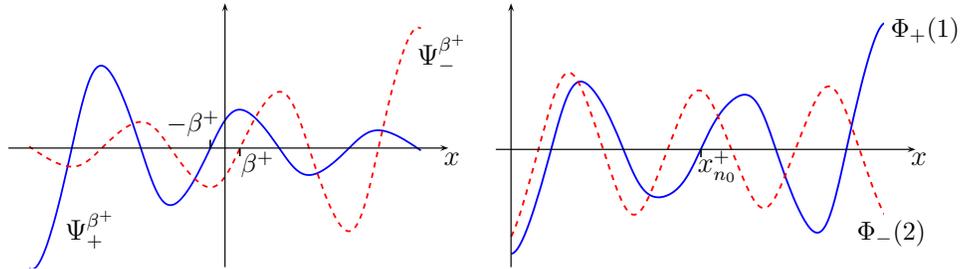}\\
  \caption{Zeros of 
  $\Psi^{\beta^+}_\pm(x)$ are in the equidistant points
  $2n{\rm {\bf K}}\mp\beta^+$, the 
  amplitudes of these two functions 
   increase exponentially in opposite directions. 
  The amplitudes of the oscillating  
  states $\Phi_\pm$ increase exponentially in both directions.
  The graphic on the right corresponds to the case 
  $\beta^+_1<\beta^+_2$.}
  \label{fig6}
\end{figure}

We want to generate a nontrivial non-singular Darboux-Crum transformation
based on the pair of the eigenfunctions  (\ref{Phi+}) and   (\ref{Phi-}).
For this the Wronskian of these functions should take nonzero 
non-constant values.
The choice  
\begin{equation}\label{beta+beta-}
0<\beta^+_1<\beta^+_2<{\rm{\bf K}} \quad 
\Leftrightarrow\quad 
\mathcal{E}(\beta^+_1)>\mathcal{E}(\beta^+_2) 
\end{equation}
guarantees then that the  intervals containing zeroes 
of the functions (\ref{Phi+}) and (\ref{Phi-}) do not intersect, and 
between each two neighbour zeros $x^+_n$ and $x^+_{n+1}$
of the $\Phi_+(x;\beta^+_1,C_1)$
there will appear exactly one zero $x^-_n$ of the 
$\Phi_-(x;\beta^+_2,C_2)$,
\begin{equation}\label{x+x-zeros}
	x^+_n\in \mathcal{I}^+_n(1)\,,\quad
	x^-_n\in \mathcal{I}^-_n(2)\,,\quad 
	\mathcal{I}^+_n(1)\cap \mathcal{I}^-_{n'}(2)=\emptyset\,,
\end{equation}
where
\begin{equation}\label{defI+I-}
	 \mathcal{I}^+_n(1)=(-\beta^+_1, \beta^+_1)+2n
 {\rm {\bf K}},\quad
 \mathcal{I}^-_n(2)=
	 (\beta^+_2, 2{\rm {\bf K}}-\beta^+_2)+2n
{\rm {\bf K}}\,.
\end{equation}
The amplitudes of the oscillating
functions  $\Psi^{\beta^+}_+(x)$ and $\Psi^{\beta^+}_-(x)$ 
increase exponentially for $x\rightarrow -\infty$ 
and $x\rightarrow +\infty$, respectively.
As a consequence,  
in the limit $x\rightarrow  +\infty$ the zeros 
$x^+_n$  tend to the right edges of the 
 intervals $\mathcal{I}^+_n(1)$, 
 while  $x^-_n$ tend to the left edges of 
the intervals $\mathcal{I}^-_n(2)$.
In another limit $x\rightarrow  -\infty$,  the corresponding 
zeros  tend to the opposite edges of the indicated intervals.

The Wronskian of the eigenfunctions  (\ref{Phi+}) and (\ref{Phi-})
obeys the relation 
\begin{equation}\label{Wy1y2}
	\frac{d}{dx}W(y_1,y_2)=
	(\mathcal{E}(\beta^+_1)-
	\mathcal{E}(\beta^+_2))y_1(x)y_2(x)\,,
\end{equation}
where $y_1=\Phi_+(1)$  $y_2(x)= \Phi_-(2)$.
{}From (\ref{Wy1y2}) it follows that zeros $x^\pm_n$ 
correspond exactly to the  local extrema of the Wronskian.
Let us  choose a zero 
$x^+_{n_0}$ of $y_1$, $y_1(x^+_{n_0})=0$,  such that
$y'_1(x^+_{n_0})>0$. 
Then in principle we have two possibilities: either (i) $y_2(x^+_{n_0})>0$,
or (ii) $y_2(x^+_{n_0})<0$.
In the case (i) we find that $W(x^\pm_n)<0$ 
while in the case (ii)  we would have $W(x^\pm_n)>0$ for
any $n\in \Z$. 
Differentiation of (\ref{Wy1y2}) in $x$ shows that 
in the case (i) the zeros $x^-_n$ and $x^+_n$ correspond 
to the local maxima and minima of the Wronskian, respectively.
In the case (ii) the role of these zeros as local maxima and minima would be
interchanged. Then in the case (i) we  conclude 
that the Wronskian takes strictly  negative values for all $x$,
while  in the case (ii) it would be strictly positive  function.
Though in both cases we would have a nodeless Wronskian,
let us show that the  case (i) is realised here.
In the limits $x\rightarrow\pm\infty$,
in correspondence with definition (\ref{Phi+}), (\ref{Phi-}),
 we have 
 \begin{equation}\label{W2limit+}
	\lim_{x\rightarrow +\infty}W(\Phi_+(1),\Phi_-(2))=-\frac{1}{C_1C_2}
	W(\Psi^{\beta^+_1}_-(x),\Psi^{\beta^+_2}_-(x))\,,
\end{equation}
\begin{equation}\label{W2limit-}
	\lim_{x\rightarrow -\infty}W(\Phi_+(1),\Phi_-(2))=
	C_1C_2 W(\Psi^{\beta^+_1}_+(x),\Psi^{\beta^+_2}_+(x))\,.
\end{equation}
Using these relations and the described above 
behaviour of the zeros of the functions 
$\Phi_+(1)$ and $\Phi _-(2)$ in the limit $x\rightarrow +\infty$,
the corresponding local extrema values 
of $W$ are given by 
\begin{equation}\label{W2limitextrema}
	\lim_{x^\pm_n\rightarrow +\infty}W(x^\pm_n)=
	-\frac{1}{C_1C_2}\frac{{\rm H}'(0){\rm H}(\beta^+_2-\beta^+_1)}{\Theta^2(\beta_j)}
	\exp\Big((\beta^+_j+2n{\rm{\bf K}})({\rm Z}(\beta^+_1)
	+{\rm Z}(\beta^+_2))\Big),\quad
	n\gg1,
\end{equation}
where $j=1,2$, and $\beta^+_1$ ($\beta^+_2$) corresponds 
here to $x^+_n$
 ($x^-_n$).
 For the limits $x^\pm_n\rightarrow -\infty$ we have a similar 
 expression with a unique change of the coefficient ${1}/({C_1C_2})$
 for $C_1C_2$. Taking into account that ${\rm H}'(0)=\sqrt{2kk'{\rm {\bf K}}/\pi}>0$,
 and that  ${\rm H}(\beta^+_2-\beta^+_1)>0$ because
 $0<\beta^+_2-\beta^+_1<{\rm {\bf K}}$, we conclude
 finally  that $\W_{2,0}(x)=W(\Phi_+(1),\Phi_-(2))$ takes strictly negative values on all the real line. 
 Additionally, we conclude  that $-\W_{2,0}(x) $ blows up exponentially 
 in both limits  $x\rightarrow \pm\infty$.

Similarly to (\ref{H02}), 
we construct now the Hamiltonian
\begin{equation}\label{H20}
H_{2,0}=-\frac{d^2}{dx^2} +V_{2,0}(x)\,,\quad
V_{2,0}(x)=V_{0,0}(x)-2\dfrac{d^2}{dx^2}\log 
W({\Phi}_+(1),{\Phi}_-(2))\,.
\end{equation}
This quantum system has the same spectrum as the Lam\'e
system except two additional discrete energy levels
$\varepsilon^+_l\equiv\mathcal{E}(\beta^+_l)$,
$l=1,2$. These are described 
by the wave functions given by relations
of the form (\ref{2statesE1}), (\ref{2statesE2})
with $\mathcal{F}_\pm(j)$ there changed for 
corresponding functions
$\Phi_\pm(l)$. With some algebraic manipulations,
the  wave eigenfunctions can be presented in the form
\begin{equation}\label{boundinter1}
	\Psi_{2,0}^{1;}(x)={\rm const} \,
	\frac{\Phi_-(2)}
	{\W_{2,0}}\,,\qquad H_{2,0}\Psi_{2,0}^{1;}(x)=\varepsilon^+_1
	\Psi_{2,0}^{1;}(x)\,,
\end{equation}
\begin{equation}\label{boundinter2}
	\Psi_{2,0}^{2;}(x)={\rm const} \,
	\frac{\Phi_+(1)}
	{\W_{2,0}}\,,\qquad
	H_{2,0}\Psi_{2,0}^{2;}(x)=
	\varepsilon^+_2\Psi_{2,0}^{2;}(x)\,.
\end{equation}
\begin{figure}[!h]
  \centering
  \includegraphics[scale=0.8]{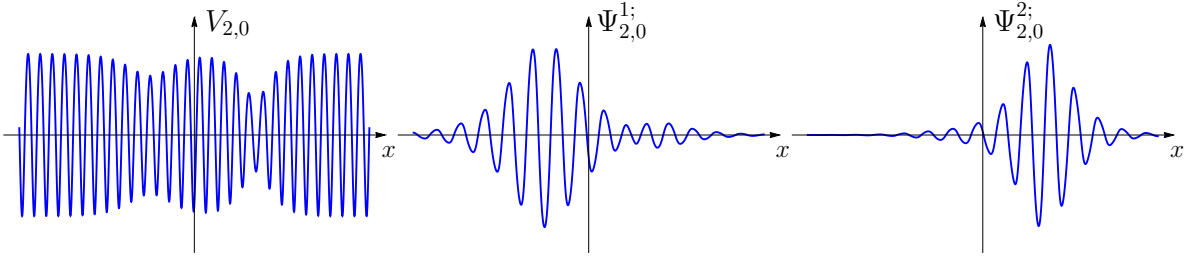}\\
  \caption{Each of the two pulse type  bound states of the system $H_{2,0}$
 is localised  in one of the two periodicity defects of the potential $V_{2,0}$, 
which are showing up as compression modulations.
 The states also reveal a small tunnelling (asymmetry) 
 in the direction of the other deformation.
}\label{fig7}
\end{figure}
The amplitude of these 
oscillating functions tends  exponentially to zero
in both limits $x\rightarrow\pm \infty$,
which confirms their bound state nature, see Figure \ref{fig7}.
The relations (\ref{W2limit+}) and 
(\ref{W2limit-}) tell us  that the Darboux-Crum transformation
generated on the basis of the states 
appearing there on the right hand sides 
produces  a potential translated  in
 $(\beta^+_1+i{\rm {\bf K}}')
+(\beta^+_2+i{\rm {\bf K}}')$. 
Using this fact and taking into account 
the imaginary period $2i{\rm {\bf K}}'$ of $V_{0,0}(x)$
we find that  
$$
V_{2,0}^{-\infty}(x)=\lim_{x\rightarrow-\infty}V_{2,0}(x)=
V_{0,0}(x+\beta^+_1+\beta^+_2),
$$
and, analogously, 
$$
V_{2,0}^{+\infty}(x)=\lim_{x\rightarrow+\infty}V_{2,0}(x)=
V_{0,0}(x-\beta^+_1-\beta^+_2).
$$
 Therefore, 
 similarly to the case of
 soliton defects corresponding to the 
  bound states in the lower forbidden band,
  the two-soliton defect associated with 
  the presence of the two bound states 
  in the intermediate  (upper) prohibited band 
 produces the phase shift  described by
 Eq. (\ref{phase_n}) with $n=2$ and $\beta^-_j$ there changed for $\beta^+_l$, where
 the parameters $\beta^+_1$ and 
 $\beta^+_2$
 obey the condition
(\ref{beta+beta-}).
The  bound states here are described by  infinitely oscillating 
wave functions,
which have an infinite number of zeros 
and exponentially decreasing amplitudes. 
This situation contrasts with
the bound states introduced into the 
lower forbidden band, where the wave 
functions  are also  
exponentially decreasing but
have  finite number of zeros, 
similarly  to the nature of ordinary bound states.

The system 
(\ref{H20}) is also characterized by the Lax-Novikov integral,
which in the present case is the differential operator of order
$7$,
\begin{equation}\label{P3factor20}
	\mathcal{P}_{2,0}=\A_{2,0}\mathcal{P}_{0,0}\A_{2,0}^\dagger\,,
	\qquad
	[\mathcal{P}_{2,0},H_{2,0}]=0\,.
\end{equation}
The second order operators $\A_{2,0}$ and $\A_{2,0}^\dagger$ 
intertwining the  Lam\'e system $H_{0,0}$ with
$H_{2,0}$ have  the form (\ref{Aj}) and 
(\ref{An}) with the functions $\mathcal{F}_+(1)$ and 
$\mathcal{F}_-(2)$ changed here, respectively,  for 
$\Phi_+(1)$ and $\Phi_-(2)$.
They satisfy relations of the form (\ref{HA0nH}) with $n=2$,
where $H_{0,n}$ has to be changed for $H_{2,0}$,
and constants $\varepsilon^-_j$ have to be changed for 
corresponding energy values $\varepsilon^+_l$, $l=1,2$, 
of the non-physical eigenstates from the intermediate 
prohibited band we used in the construction.

Analogously to the discussion presented in the previous subsection,
it is not difficult to show that the infinite period limit 
applied to the system (\ref{H20}) corresponds 
to reflectionless system given by potential
(\ref{V3V1}).

The described procedure of the introduction of the 
periodicity defects with eigenvalues within the intermediate 
prohibited band can be generalised for the  
case of arbitrary even number of the solitons.
This can be done in a systematic way by
choosing linear combinations of the wave functions 
of the form (\ref{Phi+}) and (\ref{Phi-})
with alternating lower indexes $+$ and $-$,
cf. (\ref{W_n}), with the  restriction on the 
parameters $\beta^+$ which generalises that from
(\ref{beta+beta-}),
\begin{equation}\label{beta+beta-int}
0<\beta^+_1<\beta^+_2<\ldots\beta^+_{2\ell}<{\rm{\bf K}} \quad 
\Leftrightarrow\quad 
\mathcal{E}(\beta^+_1)>\mathcal{E}(\beta^+_2) \ldots
>\mathcal{E}(\beta^+_{2\ell})\,.
\end{equation}
In the basis of such a construction
there is the property $\vert \W_{{2\ell,0}}(x)\vert>0$
guaranteed by the choice (\ref{beta+beta-int}),
where $\W_{2\ell,0}(x)$ is the Wronskian of the corresponding 
$2\ell$ non-physical eingestates of the Lam\'e system, 
\begin{equation}\label{W_2n-int}
\W_{2\ell,0}(x)=W(\Phi_{+}(1),
\Phi_{-}(2),\ldots,\Phi_{+}(2\ell-1),\Phi_{-}(2\ell))\,.
\end{equation}
The proof of this property is given in 
 Appendix~\footnote{Like in the 
shortly discussed in Section  \ref{PT1Dar} procedure corresponding to the
reflectionless P\"oschl-Teller system, the defects
also can be introduced  in such a way that 
their associated energies will appear  between the already 
placed discrete energy levels, but the final picture 
will be  described equivalently by the Darboux-Crum transformation
based on the Wronskian (\ref{W_2n-int}).}.
As a generalisation of (\ref{H20}) and (\ref{P3factor20}), 
the Hamiltonian and Lax-Novikov 
integral are  given here by the relations 
\be\label{H2l0}
	H_{2\ell,0}=H_{0,0}-2\frac{d^2}{dx^2}\log \W_{2\ell,0}\,,
\ee
\begin{equation}\label{P2l0}
	\mathcal{P}_{2\ell,0}=\A_{2\ell,0}\mathcal{P}_{0,0}\A_{2\ell,0}^\dagger\,,
	\qquad
	[\mathcal{P}_{2\ell,0},H_{2\ell,0}]=0\,.
\end{equation}
They satisfy the Burchnall-Chaundy relation of the form 
\begin{equation}
	\mathcal{P}_{2\ell,0}^2=H_{2\ell,0}(H_{2\ell,0}-k'^2)(H_{2\ell,0}-1)
	\prod_{l=1}^{2\ell}(H_{2\ell,0}-\varepsilon^{+}_l)^2\,.
\end{equation}
Here $\varepsilon^{+}_l=\mathcal{E}(\beta^+_l)$
are the eigenvalues of the bound states
\begin{equation}
	\Psi^{l;}_{2\ell,0}(x;\beta^+_1,C_1,\ldots,\beta^+_{2\ell},C_{2\ell})=
	\frac{W(\Phi_+(1),\Phi_-(2),
	\ldots,\Phi_{-}(2\ell),
	\Phi_{{(-1)^l}}(l))}
	{\W_{2\ell,0}}\,,\qquad l=1,\ldots, 2\ell\,.
\end{equation}
Other physical and non-physical eigenstates 
of $H_{2\ell,0}$ of eigenvalues $\mathcal{E}(\alpha)$
are given by
\begin{equation}
	\Psi^\alpha_{2\ell,0;\pm}(x;\beta^+_1,C_1,\ldots,\beta^+_{2\ell},C_{2\ell})=
	\frac{W(\Phi_+(1),\Phi_-(2),\ldots,\Phi_{-}(2\ell),
	\Psi^\alpha_\pm)}
	{\W_{2\ell,0}}\,.
\end{equation}

{}From this picture with 
even number $2\ell\geq 2$ of  bound states
in the intermediate forbidden band, one can obtain systems 
which contain odd number $2\ell-1$  of 
discrete energy levels in the 
same prohibited band of the initial one-gap Lam\'e system.  
This can be achieved  by sending anyone of
the $2\ell$ solitons  to infinity.

Let us see how this procedure works in the case 
of the system  (\ref{H20}). 
For the sake of definiteness, we send 
the first  soliton, associated with the higher 
discrete energy level
$\mathcal{E}(\beta^+_1)$, to infinity.
Another case corresponding to the limit associated with the
soliton related to the lower discrete energy level
can be realised in a similar way.
To send the indicated soliton to infinity, 
we take a limit $C_1\rightarrow \infty$. In analogous way,
one can also consider the limit $C_1\rightarrow 0$.

\begin{figure}[!h]
  \centering
  \includegraphics[scale=0.5]{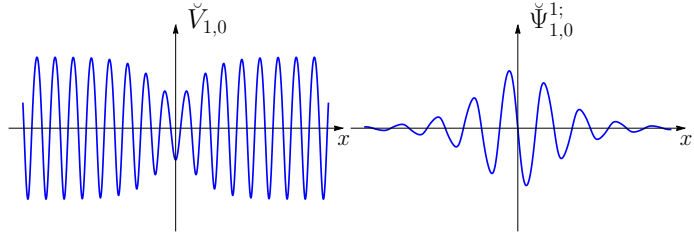}\\
  \caption{Sending one soliton 
  to infinity results in a potential supporting  
  one bound state less.
  System $\breve{H}_{1,0}$ is related with the
  Lam\'e system $H_{0,0}$ by the Darboux-Crum transformation
  of the second order, while it is related with the 
  singular Treibich-Verdier system by the first order 
  Darboux transformation.
  The symmetric state (presented by odd function here) 
  is centered in the soliton 
  deformation of the the potential, and the tunnelling  
  related to the soliton sent to infinity disappears. 
}\label{fig8}
\end{figure}

In the limit $C_1\rightarrow \infty$, 
the potential $V_{2,0}(x)$ given by Eq. (\ref{H20})
transforms into 
\begin{equation}\label{C1inty}
	\lim_{C_1\rightarrow\infty}V_{2,0}(x)\equiv \breve{V}_{1,0}(x;\beta^+_1)=
	V_{0,0}(x)-2\dfrac{d^2}{dx^2}\log
	W({\Psi}_+^{\beta^+_1},{\Phi}_-(2))\,.
\end{equation}
The Hamiltonian 
$\breve{H}_{1,0}(x;\beta^+_1)=-\frac{d^2}{dx^2}+
\breve{V}_{1,0}(x;\beta^+_1)
$
possesses single bound state of energy $\varepsilon^+_2$,
which can be obtained as a 
limit of the bound eigenstate $\Psi^{2;}_{2,0}(x)$ 
of $H_{2,0}$,
\be
	\lim_{C_1\rightarrow\infty}\Psi^{2;}_{2,0}(x)=
	\breve{\Psi}^{1;}_{1,0}(x)\,,
\ee	
see Figure \ref{fig8}.
In correspondence with the results 
of Section \ref{Darbtrans},  the Darboux transformation based 
on the single eigenfunction ${\Psi}_+^{\beta^+_1}(x)$
produces the Treibich-Verdier potential, 
 $V_{0,0}(x)-2\frac{d^2}{dx^2}\log
{\Psi}_+^{\beta^+_1}=V_{0,0}(x+\beta^+_1+i{\rm{\bf K}}')$,
and we can present  (\ref{C1inty}) 
in the equivalent form 
\begin{equation}
	\breve{V}_{1,0}(x)=V_{0,0}(x+
	\beta^+_1+i{\rm{\bf K}}')-2\dfrac{d^2}{dx^2}\left(\log
	\frac{W({\Psi}_+^{\beta^+_1},
	{\Phi}_-(2))}{{\Psi}_+^{\beta^+_1}} \right).
\end{equation}
Function ${W({\Psi}_+^{\beta^+_1},{\Phi}_-(2))}/{\Psi_+^{\beta^+_1}}$ 
appearing in the argument of the logarithm is an eigenfunction
of the system $H_{0,0}(x+\beta^+_1+i{\rm{\bf K}}')$.
The Bloch-like eigenstates of this Hamiltonian 
operator  can be obtained from the corresponding eigenstates
of the Lam\'e system $H_{0,0}(x)$,
 $\Psi_\pm^\alpha(x+\beta^+_1+i{\rm{\bf K}}')
 ={N}_\pm(\alpha)
\breve{\Psi}_\pm^\alpha(x+\beta^+_1)$, where
\be\label{Psibreve}
	\breve{\Psi}_\pm^\alpha(x)=
	\frac{\Theta(x\pm \alpha)}{{\rm H}(x)}e^{\mp x Z(\alpha)}
\ee
and ${N}_\pm(\alpha)=
 \exp\left(\mp i(\frac{\alpha\pi}
 {2{\rm{\bf K}}}+ {\rm{\bf K}}'Z(\alpha)\right)$.
Therefore, we have 
\be\label{WbrevePsi}
	\frac{W({\Psi}_+^{\beta^+_1},{\Psi}_\pm^{\beta^+_2})}{\Psi_+^{\beta^+_1}}=
	C_\pm
	\breve{\Psi}_\pm^{\beta^+_2}(x+\beta^+_1).
\ee
Putting in both sides of the last relation $x=-\beta^+_1$ (or, $x=\mp\beta^+_2$ 
to escape simple poles at both sides),
we define the real nonzero constants $C_\pm$ in (\ref{WbrevePsi}),
\begin{equation}
	C_\pm=\mp 
	\frac{{\rm H}(\beta^+_2\mp\beta^+_1)
	{\rm H}'(0)}{\Theta(\beta^+_1)\Theta(\beta^+_2)}
	\exp\left(\pm\beta^+_1{\rm Z}(\beta^+_2)\right).
\end{equation}
Making a shift $x\rightarrow x-\beta^+_1$ in (\ref{C1inty}),
all this gives us 
\be\label{V10(x)}
	{V}_{1,0}(x)\equiv\breve{V}_{1,0}(x-\beta^+_1)=1+k'^2
	-2\frac{{\rm{\bf E}}}{{\rm{\bf K}}}-
	2\frac{d^2}{dx^2}\log \chi^{\beta^+_2}_{1,0}\,,
\ee
\be\label{chiV10}
	\chi^{\beta^+_2}_{1,0}(x)=\breve{C}_2\Theta(x+\beta^+_2)
	\exp\left(-x{\rm Z}(\beta^+_2)\right)
	+\frac{1}{\breve{C}_2}\Theta(x-\beta^+_2)
	\exp\left(x{\rm Z}(\beta^+_2)\right).
\ee	
Here a real constant $\breve{C}_2$ is given in terms 
of $C_2$ by
$\breve{C}_2=C_2\sqrt{\frac{{\rm H}(\beta^+_2-
\beta^+_1)}{{\rm H}(\beta^+_2+\beta^+_1)}}\exp\left(\beta^+_1{\rm Z}
(\beta^+_2)\right)>0$, and we have taken into account 
the relation $\frac{d^2}{dx^2}\log {\rm H}(x)=
\dn^2 (x+i{\rm{\bf K}}')-\frac{{\rm{\bf E}}}{{\rm{\bf K}}}$.
In the limit $C_1\rightarrow\infty$,
the Wronskian in the denominator of 
the eigenstate (\ref{boundinter1}) of energy $\mathcal{E}(\beta^+_1)$
of the system $H_{2,0}$ 
blows up exponentially, and  this state
disappears. On the other hand,
the state  (\ref{boundinter2}) transforms into the bound state
of energy 
$\mathcal{E}(\beta^+_2)$
of the system $H_{1,0}(x)=-\frac{d^2}{dx^2}+V_{1,0}(x)$,
\be\label{boundV10}
	\Psi^{2;}_{2,0}(x-\beta^+_1)\rightarrow
	\breve{\Psi}_{1,0}^{1;}(x-\beta^+_1)= {\rm const}\,
	\frac{{\rm H}(x)}{\chi^{\beta^+_2}_{1,0}(x)}\,.
\ee

The presence of this bound state in the spectrum of $H_{1,0}(x)$
 is the unique difference in comparison with the spectrum  of the
one-gap Lam\'e system $H_{0,0}(x)$.
The system $H_{1,0}(x)$ is related with 
$H_{0,0}(x)$, however, by the second order Darboux-Crum 
 transformation 
of the form (\ref{C1inty}) with $x$ changed there for
$x-\beta^+_1$.
On the other hand, 
the system $H_{1,0}(x)$ can be related with the 
singular Treibich-Verdier system described by the 
potential $V_{0,0}(x+i{\rm{\bf K}}')$, by the first order 
Darboux transformation  based on the function
$\breve{\Psi}_\pm^\alpha(x-\beta^+_1)$
given by Eq. (\ref{Psibreve}), which is  
eigenfunction of the singular 
$PT$-invariant Hamiltonian operator $H_{0,0}(x+i{\rm{\bf K}}')$.
This picture is analogous to that for the P\"oschl-Teller system
 when we want to  introduce there the 
bound state between the already existing bound state and the 
continuous  part of the spectrum, see Section \ref{PT1Dar}.

In correspondence with 
the described picture,  the system $H_{1,0}(x)$
is characterised by the irreducible  Lax-Novikov integral
\be\label{LN5VT}
	\mathcal{P}_{1,0}(x)=A_\psi \mathcal{P}_{0,0}(x+i{\rm{\bf K}}')
	 A_\psi^\dagger\,,
	\qquad
	\psi=\breve{\Psi}_+^{\beta^+_1}(x-\beta^+_1)\,,
\ee
which is the differential operator of order $5$,
where $\mathcal{P}_{0,0}(x)$ is the 
Lax-Novikov integral (\ref{LN00}) of the Lam\'e system
$H_{0,0}(x)$.
In (\ref{LN5VT}) one can take, equivalently,  
$\psi=\Psi^{\beta^+_1}(x-\beta^+_1+i{\rm{\bf K}}') $. 

Notice a remarkable similarity of the potential 
$V_{1,0}$ given by Eqs. (\ref{V10(x)}) and 
(\ref{chiV10}) with the potential $V_{0,1}$ 
defined by Eqs.   (\ref{V01}) and (\ref{var01}).
The important difference 
of the both potentials is, however,  
that ${\rm Z}(\beta^+_2)$ 
presents in the structure of $V_{1,0}$, while
in the structure of the potential $V_{0,1}$ there appears
${\rm z}(\beta^-_1)$ defined in  Table \ref{T1}.
Unlike the nodeless bound state (\ref{bound1}) of the system
$V_{0,1}$, the bound state (\ref{boundV10}) 
of the system $V_{1,0}$
has infinite number of zeroes at $x_n=2n{\rm{\bf K}}$,
and its amplitude, like that of the wave function  (\ref{bound1}),
decreases exponentially as $x$ goes to $\pm\infty$.

When $x\rightarrow\pm\infty$, Hamiltonian $H_{1,0}(x)$ asymptotically
transforms into
$H_{0,0}(x\mp \beta^+_2)-\mathcal{E}(\beta^+_2)$,
and we get
the phase displacement  
$\Delta\phi(\beta^+_2)=-2\beta^+_2$
generated  by the one-soliton potential defect, which 
supports one bound state within  the
upper prohibited band of the original
one-gap Lam\'e system.

Let us   notice  that one can also introduce an odd number 
of bound states into the gap by taking, instead of (\ref{beta+beta-}),
the set of parameters 
$0=\beta^+_1<\beta^+_2<\ldots\beta^+_{2\ell}<{\rm{\bf K}}$,
or $0<\beta^+_1<\beta^+_2<\ldots\beta^+_{2\ell}={\rm{\bf K}}$.
This assumes the change of the state
$\Phi_{+}(1)$  in Wronskian (\ref{W_2n-int}) for 
$\sn\,x$ in the first case, or  $\Phi_{-}(2\ell)$ for
$\cn\, x$ in the second case.
Such alternatives, however, do not give anything new.
They are reproduced  just  by taking, respectively,  limits 
$\beta^+_1\rightarrow 0$,  or  
$\beta^+_{2\ell}\rightarrow {\rm{\bf K}}$ in  general 
picture presented in this subsection.

\subsection{Bound states in both forbidden bands}

\begin{figure}[!h]
  \centering
  \includegraphics[scale=0.8]{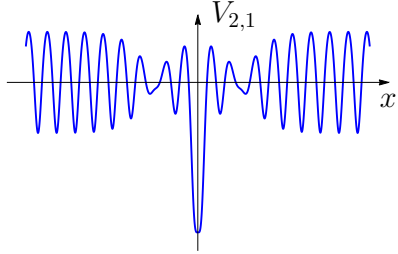}\\
  \caption{Potential supporting two bound states 
  in the gap and one bound state in the lower forbidden band.
  The defects in the form of the two compression 
  modulations and a potential soliton well can be 
  displaced arbitrarily in the periodic background.}\label{fig9}
\end{figure}

One can introduce periodicity defects into
the Lam\'e system by constructing the potentials 
which support bound states in both, lower and upper forbidden
bands. Similarly to the already discussed cases, 
the  construction is based on the property that
the Wronskian 
\begin{equation}
	\W_{2\ell,n}(x)=W\left(\Phi_+(1),\Phi_-(2),\ldots, \Phi_-(2\ell),
	\mathcal{F}_{+}(1),\ldots,\mathcal{F}_{s_n}(n)\right).
\end{equation}
is a nodeless smooth function on all the real line,
see Appendix.
In this way, the most general family of one-gap Hamiltonians 
with $2\ell+n$ defects (solitons) introduced into
the periodic background of Lam\'e potential
$V_{0,0}(x)$
is defined by
\be\label{H2ln}
	H_{2\ell,n}=H_{0,0}-2\frac{d^2}{dx^2}\log \W_{2\ell,n}(x) \,.
\ee
The defects correspond to   
$2\ell$ bound states in  the spectral gap
and $n$ bound states in the lower prohibited band.  
On Figure \ref{fig9} it is shown the form of the potential for
the simplest case $\ell=n=1$.

Each member 
of the family of Hamiltonians 
(\ref{H2ln}) 
possesses a nontrivial integral 
\begin{equation}\label{P2ln}
	\mathcal{P}_{2\ell,n}=\A_{2\ell,n}\mathcal{P}_{0,0}\A_{2\ell,n}^\dagger\,,
	\qquad
	[\mathcal{P}_{2\ell,n},H_{2\ell,n}]=0\,,
\end{equation}
satisfying the relation
\begin{equation}\label{Pgenericln}
	\mathcal{P}_{2\ell,n}^2=H_{2\ell,n}(H_{2\ell,n}-k'^2)(H_{2\ell,n}-1)
	\prod_{l=1}^{2\ell}(H_{2l,n}-\varepsilon^+_l)^2
	\prod_{j=1}^{n}(H_{2\ell,n}-\varepsilon^-_j)^2\,.
\end{equation}
Here $\A_{2\ell,n}$ is the differential operator of order 
$2\ell+n$, which is defined 
by  $\A_{2\ell,n}=A_{2\ell,n}\ldots A_{2\ell,1}\A_{2\ell,0}$,
where
\be
	A_{2\ell,j}=\frac{\W_{2\ell,j}}{\W_{2\ell,j-1}}
	\frac{d}{dx}\frac{\W_{2\ell,j-1}}{\W_{2\ell,j}}\,,\qquad
	j=1,\ldots, n\,.
\ee
The first order differential operator $A_{2\ell,n}$
and its conjugate generate the intertwining relations  
\be
	A_{2\ell,n}H_{2\ell,n-1}=H_{2\ell,n}A_{2\ell,n},\qquad 
	A^\dag_{2\ell,n}H_{2\ell,n}=H_{2\ell,n-1}A^\dag_{2\ell,n}\,,
\ee
and factorise the neighbour Hamiltonians 
$H_{2\ell,n}$ and $H_{2\ell,n-1}$ in the form 
\be
A_{2\ell,n}A^\dag_{2\ell,n}=H_{2\ell,n}-\varepsilon^-_n,\qquad 
A^\dag_{2\ell,n}A_{2\ell,n}=H_{2\ell,n-1}-\varepsilon^-_n\,.
\ee
The $2\ell$ bound states of $H_{2\ell,n}$  of energies 
$\varepsilon^+_l$, $l=1,\ldots, 2\ell$,
within the gap  
are given by 
\begin{equation}
\Psi_{2\ell,n}^{l;}=
	\frac{W(\Phi_+(1),\Phi_-(2),..., \Phi_-(2\ell),\mathcal{F}_{+}(1),
	\ldots,
	\mathcal{F}_{s_n}(n),
	\Phi_{(-1)^l}(l))}
	{\W_{2\ell,n}},
\end{equation}
while  the $n$ 
bound states of energies  $\varepsilon^-_j$,
$j=1,\ldots,n$,  in the lower prohibited band 
have the form
\begin{equation}
	\Psi_{2\ell,n}^{;j}=
	\frac{W(\Phi_+(1),\Phi_-(2),..., \Phi_-(2\ell),
	\mathcal{F}_{+}(1),...,\mathcal{F}_{s_n}(n),
	\mathcal{F}_{-s_j}(j))}
	{\W_{2\ell,n}}\,.
\end{equation}
Here we do not indicate explicitly 
the parameters which define  the functions 
$\Psi_{2\ell,n}^{l;}$ and  $\Psi_{2\ell,n}^{;j}$
being in general of the form 
$\Psi(x;\beta^+_1,C^+_1,\ldots,\beta^+_{2\ell},C^+_{2\ell},
\beta^-_1,C^-_1,\ldots\beta^-_n,C^-_n)$.
Other, physical as well as non-physical,  eigenstates of $H_{2\ell,n}$ 
of eigenvalues $\mathcal{E}(\alpha)$ 
are 
given by
\begin{equation}
	\Psi^\alpha_{2\ell,n;\pm}=
	\frac{W(\Phi_+(1),\Phi_-(2),..., \Phi_-(2\ell),\mathcal{F}_{+}(1),...,\mathcal{F}_{s_n}(n),
	\Psi^\alpha_\pm)}
	{\W_{2\ell,n}}\,.
\end{equation}

It is always possible to eliminate any of the bound states 
from the spectrum taking
the limit $C_r^\pm\rightarrow 0$, or  $C_r^\pm\rightarrow \infty$
for the corresponding parameter. 
In the case if we take such a limit for the parameter   
 $C^+_l$ of the state 
$\Phi_{(-1)^{l+1}}(l)$,  we obtain 
$H_{2\ell,n}(x)\rightarrow 
\breve{H}_{2\ell-1,n}(x;\beta_l^+)$,
where $\breve{H}_{2\ell-1,n}(x;\beta_l^+)$ is the Hamiltonian
of the system with $2\ell-1$ bound states in the gap.
Similarly to the case discussed in the previous subsection,
the $\breve{H}_{2\ell-1,n}(x;\beta_l^+)$
can also be obtained by the Darboux-Crum transformation
of order $2\ell-1+n$ applied to the singular 
Treibich-Verdier system. 
The   Lax-Novikov integral 
$\breve{\mathcal{P}}_{2\ell-1,n}(x;\beta_l^+)$ of 
$\breve{H}_{2\ell-1,n}(x;\beta_l^+)$
appears from (\ref{P2ln})
via the indicated limit through the reduction, 
$\mathcal{P}_{2\ell,n}(x)
\rightarrow \big(\breve{H}_{2\ell-1,n}(x;\beta_l^+)-
\varepsilon^+_l\big)
\breve{\mathcal{P}}_{2\ell-1,n}(x;\beta_l^+)$. 
On the other hand, if we take one of the 
two specified limits 
for the parameter  $C_j^-$,
we obtain the Hamiltonian 
 $\tilde{H}_{2\ell,n-1}(x;\beta_j^-)$,
 which corresponds to the system ${H}_{2\ell,n-1}(x)$ 
 of the form (\ref{H2ln}) with the 
 displaced argument, $x\rightarrow x+\beta^-_j$. The initial parameters $\beta^-_i$
 with $i=j+1,\ldots,n$ transform into the
 parameters $\beta^-_{i}$, $i=j,\ldots,n-1$,
 of the resulting system,
and  the same happens with 
the corresponding parameters $C^-_i$. 
Moreover,  all parameters $C^\pm$ undergo rescaling, 
 $C^+_l\rightarrow c^+_l(\beta^+_l,\beta^-_j)C^+_l$, 
 $l=1,\ldots,2\ell$, $C^-_{i}
 \rightarrow c^-_{i}(\beta^-_{i},\beta^-_j)C^-_{i}$,
 $i=1,\ldots,n-1$, where $c^+_l>0$ and $c^-_{i}>0$
 are some functions of the indicated arguments,
 whose explicit form we do not write down 
 explicitly here.
 
Notice that in the most general case of  
one-gap
quantum system 
$H_{2\ell-m,n}=-\frac{d^2}{dx^2}+
V_{2\ell-m,n}(x)$  supporting 
$2\ell-m+n\geq 1$ bound states,   
relation  (\ref{phase_n})  is generalised 
for
 \be\label{Delphigen}
\Delta\phi=-2\sum_{j=1}^n\beta^-_j -2
\sum_{l=1}^{2\ell-m}\beta^+_l\,,
\ee
where 
$n\geq 0$, $2\ell -m\geq 0$, $m=0,1$, and 
the 
omission of the corresponding sum  is assumed 
when
$n=0$, or $\ell=0$.
This  is the  net phase displacement 
between $x=+\infty$ and $x=-\infty$ periodic asymptotics 
of the  potential $V_{2\ell-m,n}(x)$,
which is the  one-gap Lam\'e potential 
$V_{0,0}(x)$
perturbed by $n\geq 0 $ soliton defects 
of the potential well type and $2\ell-m\geq 0$,
 periodicity defects 
of the compression modulations nature.

In conclusion of this  Section, 
let us note that  the notion of 
Hill's discriminant (Lyapunov function)
is defined for a Schr\"odinger equation 
with periodic potential, 
and reflects coherently the properties of the eigenstates
under the shift of the quantum system for its period
\cite{solitons,Hill}. 
The Darboux-Crum transformations
which do not violate the periodicity of the potential,
produce isospectral systems, and  
do not change the corresponding discriminants \cite{CJP,Rosu}. 
The systems we constructed here, are  \emph{almost isospectral} 
to the one-gap Lam\'e system. 
Their potentials are not periodic  functions, and so, 
the Hill's discriminant can not be defined for them 
in a usual way. It can be considered only in the regions 
$x\rightarrow -\infty$ and $x\rightarrow
+\infty$, where the periodicity (with a relative phase 
displacement defect)
 is restored asymptotically.  At the same time, it is necessary 
 to bare in mind that the Lyapunov function reflects the 
 stability properties of the points in the spectrum: 
 for periodic quantum systems, two linearly independent 
 Bloch-Floquet  states 
 correspond to all the points inside the allowed bands, 
 while the edge points are treated as non-stable 
 because there one of the two solutions  is unbounded
 \cite{Hill}.
 Since the periodicity defects we constructed introduce into the spectrum
 of Lam\'e system only the  discrete energy values
 corresponding to non-degenerate bound states,
 one can say that they  do not change the properties 
 of stability of the spectrum of the initial system.

\section{Exotic supersymmetry}

According to the analysis presented above,
any pair of the Hamiltonians
$H_{2\ell_1-m_1,n_1}$ and 
$H_{2\ell_2-m_2,n_2}$, where $m_{1,2}=0,1$,
can be  related by means of the two pairs
of intertwining operators. One pair of mutually conjugate 
operators intertwines the Hamiltonians directly.
Another pair has  higher differential order, and 
does the same  job via a virtual periodic 
one-gap system. The operators of the second pair
involve in their structure 
the Lax-Novikov integral of the Lam\'e system
$H_{0,0}$, or of its analog corresponding
to the singular on the real line 
Treibich-Verdier one-gap system. 
Each of the subsystems in the pair ($H_{2\ell_1-m_1,n_1}$, 
$H_{2\ell_2-m_2,n_2}$)  is also characterised by its 
proper Lax-Novikov  integral.
As a result, if we consider the extended system 
given by the matrix $2\times 2$ Schr\"odinger operator
composed from the pair of the indicated Hamiltonians,
it will be described not just by  
the $\mathcal{N}=2$ linear or nonlinear 
supersymmetry as it would be expected 
for the ordinary pair of 
Darboux(-Crum) related 
quantum mechanical systems. Instead,
as in the case of non-periodic reflectionless systems, 
it will be characterised 
by an exotic  nonlinear $\mathcal{N}=4$ supersymmetric 
structure that involves  the  two nontrivial bosonic generators 
composed from the Lax-Novikov integrals of the subsystems.

{}From the perspective of physical applications the most 
interesting case corresponds to  the pairs of the Schr\"odinger 
Hamiltonians, which  
can be related by the mutually conjugate first-order
Darboux intertwiners  alongside with  the pair of higher 
order intertwiners. 
It is this case that we consider in this section 
in detail.

We start from the general discussion of the 
picture corresponding to a basic case, from which other 
cases can be obtained via certain limiting procedures.
Then we illustrate this by considering the simplest 
examples, which
reveal all the peculiarities of the exotic
supersymmetric structure. 

\subsection{Exotic supersymmetry with the first-order supercharges: 
generic picture}

The first-order differential operators  $A_{2\ell,n}$ and 
$A_{2\ell,n}^\dagger$
 intertwine the
Hamiltonians $H_{2\ell,n-1}$ and $H_{2\ell,n}$,
\be\label{AHDar}
	A_{2\ell,n}H_{2\ell,n-1}=H_{2\ell,n}A_{2\ell,n},
	\qquad
	H_{2\ell,n-1}A_{2\ell,n}^\dagger=A_{2\ell,n}^\dagger H_{2\ell,n}\,,
\ee
and factorise them,
\be
	A^\dag_{2\ell,n}A_{2\ell,n}=H_{2\ell,n-1}-\varepsilon_n^-,
	\qquad A_{2\ell,n}A^\dag_{2\ell,n}=H_{2\ell,n}-\varepsilon_n^-\,,
\ee
where $\varepsilon_n^-=\mathcal{E}(\beta^-_n+i{\rm {\bf K}}')$.
These relations allow us to consider the 
extended system described by 
the  Hamiltonian
\be\label{HextSUSY}
\cH_{2\ell,n}=\left(
                      \begin{array}{cc}
                        H_{2\ell,n-1} & 0 \\
                        0 & H_{2\ell,n} \\
                      \end{array}
                    \right)\,,
\ee
and by the pair of matrix operators 
\be\label{Saint}
S^1_{2\ell,n}=\left(
                      \begin{array}{cc}
                        0 & A^\dag_{2\ell,n} \\
                       A_{2\ell,n} & 0 \\
                      \end{array}
                    \right),\qquad S^2_{2\ell,n}=i\sigma_3S^1_{2\ell,n}\,.
\ee
Taking the trivial integral $\Gamma=\sigma_3$ as a 
$\Z_2$-grading operator, we identify $\cH_{2\ell,n}$ as 
the bosonic
operator, $[\Gamma, \cH_{2\ell,n}]=0$, and 
$S^a_{2\ell,n}$, $a=1,2$, as the fermionic ones, 
$\{\Gamma, S^a_{2\ell,n}\}=0$. 
They  generate a  superalgebra of $\mathcal{N}=2$
supersymmetric quantum mechanics
\be\label{SUSYN=2}
	[\cH_{2\ell,n},S^a_{2\ell,n}]=0,
	\qquad \{S^a_{2\ell,n},S^b_{2\ell,n}\}=
	2\delta^{ab}(\cH_{2\ell,n}-\varepsilon_n^-)\,.
\ee
By the redifinition of the Hamiltonian  via an additive shift, 
$\cH_{2\ell,n}-\varepsilon_n^- \rightarrow 
\cH_{2\ell,n}$, one can
transform  
(\ref{SUSYN=2}) into the standard form 
of $\mathcal{N}=2$ superalgebra describing the 
system with the 
zero energy of the non-degenerate ground 
state appearing in the spectrum of  
the `lower' subsystem 
of the extended matrix system. 
Since the subsystems $H_{2\ell,n-1}$ and 
$H_{2\ell,n}$
 possess  the nontrivial
Lax-Novikov integrals being differential operators
of orders  $4\ell+2n+1$ and $4\ell+2n+3$, 
the extended system 
(\ref{HextSUSY}) possesses also two 
nontrivial  bosonic integrals which we define 
in the form
\be\label{PbosInt}
P^1_{2\ell,n}=\left(
                      \begin{array}{cc}
                        (H_{2\ell,n-1}-\varepsilon_n^-)\mathcal{P}_{2\ell,n-1} & 0 \\
                        0 & \mathcal{P}_{2\ell,n} \\
                      \end{array}
                    \right), 
                    \qquad P^2_{2\ell,n}=\sigma_3P^1_{2\ell,n}\,.
\ee
We introduced here the additional factor 
in the upper component whereby the upper and lower components
of these integrals are 
operators of the same differential order.  
The commutation relations 
\be
	 [\cH_{2\ell,n},P^a_{2\ell,n}]=0\,,
 	\qquad [P^a_{2\ell,n}, P^b_{2\ell,n}]=0\,,\qquad
	[P^1_{2\ell,n},S^a_{2\ell,n}]=0
\ee
extend the superalgebraic relations (\ref{SUSYN=2}),
and show that the integral $P^1_{2\ell,n}$
is the bosonic central charge. 
On the other hand,   the nontrivial commutator 
$[P^2_{2\ell,n},S^a_{2\ell,n}]$ 
generates the second pair of the fermionic supercharges 
$Q^a_{2\ell,n}$, 
which are the matrix differential operators of the order 
 $2(2\ell+n+1)$. As we shall see,
 the anti-commutator of $Q^a_{2\ell,n}$ with $Q^b_{2\ell,n}$
 produces
 a polynomial in matrix Hamiltonian
 $\cH_{2\ell,n}$, while the anti-commutator  of 
 $Q^a_{2\ell,n}$ with $S^b_{2\ell,n}$ generates 
 the central charge $P^1_{2\ell,n}$. 
 The second bosonic integral $P^2_{2\ell,n}$
 generates finally a kind of a rotation between the 
 supercharges $S^a_{2\ell,n}$ and $Q^a_{2\ell,n}$.

Taking in (\ref{HextSUSY}) the limit $C^+_{l}\rightarrow \infty$ or
$C^+_{l}\rightarrow 0$ with $l$ chosen from the set 
$1,\ldots,2\ell$, we obtain another extended system 
\be\label{HextSUSYbreve}
\breve{\cH}_{2\ell-1,n}=\left(
                      \begin{array}{cc}
                        \breve{H}_{2\ell-1,n-1} & 0 \\
                        0 & \breve{H}_{2\ell-1,n} \\
                      \end{array}
                    \right)\,.
\ee
As we saw, the application of the limits $C^+_{l}\rightarrow \infty$ or
$C^+_{l}\rightarrow 0$ to the corresponding Lax-Novikov 
integrals of the subsystems produces the reducible
operators.
The irreducible nonsingular 
Lax-Novikov integrals of $\breve{H}_{2\ell-1,n-1}$ 
and $\breve{H}_{2\ell-1,n}$ 
have orders $4\ell+2n-1$ and $4\ell+2n+1$,  
and include in their structure the Lax-Novikov 
integral of the singular Treibich-Verdier one-gap system.
The bosonic integrals $\breve{P}^a_{2\ell-1,n}$
of the extended matrix system (\ref{HextSUSYbreve})
are constructed   from $\breve{\mathcal{P}}_{2\ell-1,n-1}$ 
and $\breve{\mathcal{P}}_{2\ell-1,n}$  like in (\ref{PbosInt}).
Again, $\breve{P}^1_{2\ell-1,n}$ will play the role 
of the central charge of the nonlinear superalgebra,
while the commutator
 $[\breve{P}^2_{2\ell-1,n},\breve{S}^a_{2\ell-1,n}]$
 will generate the second pair of the supercharges 
$ \breve{Q}^a_{2\ell-1,n}$.  
The exotic superalgebra of the system (\ref{HextSUSYbreve})
will have as a result a form similar to that for the system
(\ref{HextSUSY}).

Let us change  index $n$ for $n+1$ in
(\ref{HextSUSY}), 
and take one of the two limits 
\be
	\lim_{C_{n+1}^-\rightarrow 0,\infty} H_{2\ell,n+1}(x)= 
	\tilde{H}_{2\ell,n}(x;\mp\beta_{n+1}^-)\,,
\ee
where the upper and lower sign on the r.h.s. corresponds, respectively,
to the $0$ and $\infty$ cases. 
In such a limit 
 we get the extended system described by the Hamiltonian 
\be\label{cHtilde}
\tilde{\cH}_{2\ell,n}=\left(
                      \begin{array}{cc}
                        H_{2\ell,n} & 0 \\
                        0 & \tilde{H}_{2\ell,n} \\
                      \end{array}
                    \right),
\ee
where $ \tilde{H}_{2\ell,n}$ corresponds to one of the indicated limits,
$\tilde{H}_{2\ell,n}(x;\mp\beta_{n+1}^-)$.
Here we have used the definition of the functions (\ref{Fsi+-}),
and have taken into account that for the function  
(\ref{Fdef}) the identity 
$F(-x;\beta^-)=F(x;-\beta^-)$ is valid.
The initial subsystems 
$H_{2\ell,n}$ and  $H_{2\ell,n+1}$  in (\ref{HextSUSY})
with $n$ changed for $n+1$
are related by the first order intertwining operators 
$A_{2\ell,n+1}$ and $A^\dagger_{2\ell,n+1}$. 
Then the pair  of
$H_{2\ell,n}(x)$ and  $\tilde{H}_{2\ell,n}(x;\mp\beta_{n+1}^-)$
in (\ref{cHtilde})
is related by the first order intertwining operators 
\be\label{X2ln}
	X_{2\ell,n}(x;\mp\beta_{n+1}^-)
	\equiv\lim_{C_{n+1}^-\rightarrow 0,\infty} A_{2\ell,n+1}=
	\frac{\hat{\W}_{2\ell,n}(F(x;\mp\beta_{n+1}^-))}{\W_{2\ell,n}}
	\frac{d}{dx}\frac{\W_{2\ell,n}}{\hat{\W}_{2\ell,n}(F(x;\mp\beta_{n+1}^-))}
\ee
and $X^\dagger_{2\ell,n}(x;\mp \beta_{n+1}^-)$,
where 
$\hat{\W}_{2\ell,n}(f(x))\equiv W(\Phi_+(1), \ldots,
 \Phi_-(2\ell), \mathcal{F}_+(1) 
 \ldots \mathcal{F}_{s_{n}}(n), f(x) )$.
 The subsystems in (\ref{cHtilde}) are completely isospectral,
 and the exotic  supersymmetry 
 in this case has a structure similar to that of
 the system (\ref{HextSUSY}). However, 
unlike  (\ref{HextSUSY}), the  
system (\ref{cHtilde}) is characterised by the 
spontaneously broken exotic supersymmetry,
and this fact, as we shall see,
 is properly reflected by the `fine structure' of the
 nonlinear superalgebra. 
 
 Another interesting case which could be mentioned 
 corresponds to the limit 
\be
	\lim_{\beta_{n+1}^-\rightarrow \beta_{n}^-} 
	H_{2\ell,n+1}=\lim_{\beta_{n+1}^-\rightarrow \beta_{n}^-} 
	\tilde{H}_{2\ell,n}(x;\mp\beta_{n+1}^-)=H_{2\ell,n-1}\,.
\ee
However, if we apply such a limit to the system (\ref{HextSUSY}) 
with index $n$ changed
for $n+1$, we obtain just a system  of the form 
$\cH_{2\ell,n}$ but with the permutted upper and lower 
corresponding Hamiltonians.

\subsection{Unbroken exotic supersymmetry}

Consider  now the simplest case of the extended systems  
(\ref{HextSUSY}) with $\ell=0$, $n=1$. 
Besides the first order operators $A_{0,1}$ and $A^\dagger_{0,1}$, 
the pair of Hamiltonians $H_{0,0}$ and  $H_{0,1}$
are intertwined by the differential operators of  order $4$,
$B_{0,1}=A_{0,1}\mathcal{P}_{0,0}(x)$ and $B^\dagger_{0,1}$.
The systems $H_{0,0}$ and  $H_{0,1}$
are also characterized by the 
 Lax-Novikov integrals 
$\mathcal{P}_{0,0}(x)$ and 
$\mathcal{P}_{0,1}(x)=A_{0,1}\mathcal{P}_{0,0}(x)A^\dag_{0,1}$. 
Besides the integrals of the form (\ref{Saint}) and 
(\ref{PbosInt}), the extended matrix system 
is characterised also by the pair of the supercharges
\be\label{Qaint}
Q^1_{0,1}=\left(
                      \begin{array}{cc}
                        0 & B^\dag_{0,1} \\
                       B_{0,1} & 0 \\
                      \end{array}
                    \right),\qquad Q^2_{0,1}=i\sigma_3Q^1_{0,1}\,.
\ee
The fermionic integrals $S^{a}_{0,1}$ and 
$Q^{a}_{0,1}$, and the bosonic integrals 
$P^{a}_{0,1}$ together with the Hamiltonian 
$\mathcal{H}_{0,1}$ generate the following 
nonlinear superalgebra, 
\be\label{susyexact1}
	\{S^{a},S^{b}\}=
	2\delta^{ab}(\mathcal{H}-\varepsilon^-_1)\,, \qquad
	\{Q^{a},Q^{b}\}=
	2\delta^{ab}(\mathcal{H}-\varepsilon^-_1)C_3(\mathcal{H})\,,
\ee
\be\label{susyexact2}
	\{S^{a},Q^{b}\}=
	2\delta^{ab}P^{1}\,,
\ee
\be\label{susyexact3}
	[P^{2},S^{a}]=-2i\epsilon^{ab}(\mathcal{H}-\varepsilon^-_1)Q^{b}\,,\qquad
	[P^{2},Q^{a}]=-2i\epsilon^{ab}(\mathcal{H}-\varepsilon^-_1)C_3(\mathcal{H})
	S^{b}\,,
\ee
\be\label{susyexact4}
	[P^{1},Q^{a}]=0,\quad
	[P^{1},S^{a}]=0\,,
\ee
where $C_3(\mathcal{H})=\mathcal{H}(\mathcal{H}-k'^2)(\mathcal{H}-1)$,
$\epsilon^{ab}$ is the antisymmetric tensor, $\epsilon^{12}=1$, and 
for the sake of simplicity we omitted the lower indexes.
The unique non-degenerate state with energy 
$\mathcal{E}=\varepsilon^-_1$ 
appearing in the spectrum of
subsystem  $H_{0,1}$
is annihilated by the shifted 
Hamiltonian $\mathcal{H}-\varepsilon^-_1$ and by 
all the integrals $S^{a}$, $Q^{a}$ and $P^{a}$.
This means that the exotic supersymmetry 
of the extended Schr\"odinger system 
is unbroken. The doubly degenerate energy values
corresponding to the edges of the allowed bands
of the subsystems are the zeros of the 
third order polynomial appearing in the 
superalgebra structure: $C_3(\mathcal{E})=0$ for
$\mathcal{E}=0,\,k'^2,\, 1$. This reflects the property
that the corresponding edge states of the subsystems
are detected by the fourth-order supercharges $Q^a$
as well as by the bosonic integrals $P^a$: all these 
operators annihilate them. One can also show 
that the physical eigenstates $\Psi^\alpha_\pm$  and 
$A_{0,1}\Psi^\alpha_\pm$ of the `upper' and 
`lower' subsystems  inside their valence and conduction
bands possessing the quasi-momentum of the
opposite sign (they correspond to the different 
lower indexes of the Bloch states) are  distinguished by the 
bosonic integrals $P^a$. 

\begin{figure}[!h]
  \centering
  \includegraphics[scale=0.7]{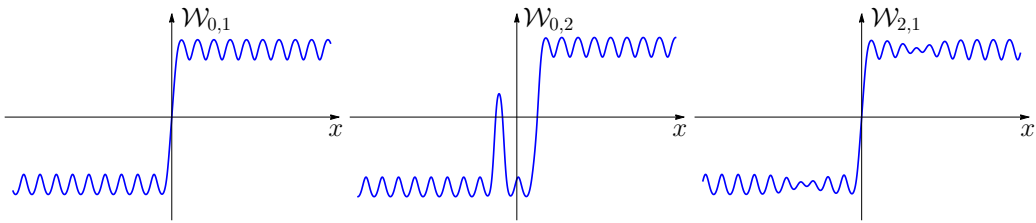}\\
  \caption{Topological superpotentials in the form of 
  the kink
  which incorporate one bound state into the spectrum. 
  On the left, it is shown  the superpotential which relates the systems 
 $H_{0,0}$ with  $H_{0,1}$. 
 The superpotential in the center corresponds to the 
 pair  of the systems $H_{0,1}$ and $H_{0,2}$, while that 
 on the right corresponds to the pair of $H_{2,0}$ and 
  $H_{2,1}$.
}\label{fig10}
\end{figure}

The second relation $[P^1,S^a]=0$ from (\ref{susyexact4})
can be rewritten as a nonlinear differential equation 
for the superpotential $\mathcal{W}_{0,1}(x)$ shown on Figure
\ref{fig10}, see Eq. (\ref{WjOmega}).  
This 
corresponds here to the first equation of the stationary mKdV 
hierarchy, which  can be associated with the extended system 
with one non-periodic soliton defect introduced into the 
one-gap Lam\'e system. At the same time, the equation
$[\mathcal{H},P_{1}]=0$ can be presented in the 
form of the nonlinear differential equations of the third order
for the potentials $V_\pm(x)
\equiv \cW_{0,1}^2\pm \cW_{0,1}'+\varepsilon^-_1$.
These equations correspond to the first equation 
of the stationary KdV hierarchy, which can be associated 
with the one-gap Lam\'e system itself  and with 
its deformation $V_-(x)$ produced by the one-soliton defect
introduced into the periodic background of the one-gap Lam\'e system.

The generic case of the extended systems
(\ref{HextSUSY})  and (\ref{HextSUSYbreve})
is described by the exotic nonlinear superalgebras of
the same form. The unique difference is that the third order
polynomial $C_3(\mathcal{H})$ appearing here  will be 
changed  for the structure 
polynomials of the form (\ref{Pgenericln}), which are 
associated with the square of the corresponding 
Lax-Novikov integrals.

\subsection{Spontaneously broken exotic supersymmetry}

The case of the  spontaneously
broken exotic supersymmetry realised  in the one-gap systems
with the non-periodicity defects can be illustrated by the extended 
system with the mutually displaced one-gap Lam\'e systems
 $H_{0,0}(x)$ and  
 $\tilde{H}_{0,0}(x;\beta^-)=H_{0,0}(x+\beta^-)$. 
 Though such systems are periodic, all the principle features of 
 the structure of the exotic supersymmetry
 we observe in this case appear also in the extended systems composed from
the  completely isospectral  systems with soliton defects.
  
\begin{figure}[!h]
  \centering
  \includegraphics[scale=0.7]{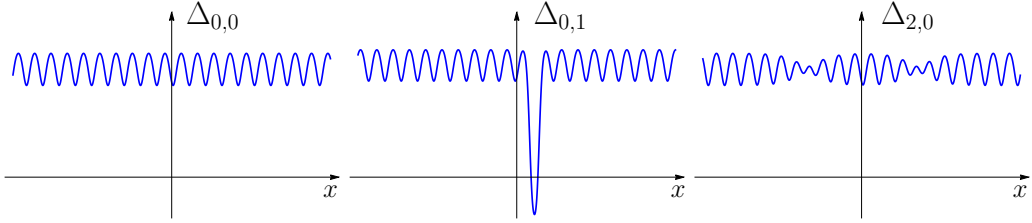}\\
  \caption{Nontopological superpotentials 
  which allow to displace the periodic 
  potential network of the Lam\'e system as
  well as the non-periodic defects in it. 
 The non-trivial displacements of the  defects correspond to 
 a nonlinear interaction between the soliton defects
themselves,  and to their interaction with the periodic background
 (see Figure \ref{fig4}). 
 According to Figure \ref{fig10}, these superpotentials 
 are obtained by sending the  kink,
 and the associated ground state of the Hamiltonian
 (\ref{HextSUSY}), to minus infinity,
 that generates the supersymmetry breaking.
The shown superpotentials 
 relate the following isospectral pairs:
  $H_{0,0}$  and $\tilde{H}_{0,0}$ (on the left), 
  $H_{0,1}$ and  $\tilde{H}_{0,1}$ (in the center),
  $H_{2,0}$  and  $\tilde{H}_{2,0}$  (on the right).}\label{fig11}
\end{figure}

The isospectral Hamiltonians  $H_{0,0}(x)$ and  $H_{0,0}(x+\beta^-)$
are connected by the first order differential operator 
\be\label{X00}
	X_{0,0}(x;\beta^-)=F(x;\beta^-)\frac{d}{dx}\frac{1}{F(x;\beta^-)}
	= \frac{d}{dx}+\Delta_{0,0}(x;\beta^-)\,,
\ee
and by its Hermitian conjugate operator,
where  
\be\label{Delta00}
\Delta_{0,0}(x,\beta^-)=
{\rm {Z}}(x)-{\rm {Z}}(x+\beta^-)+{\rm z}(\beta^-)
\ee
is the  superpotential shown on Figure \ref{fig11}.
To simplify notations, in what follows  in this subsection 
we  omit lower indices  in Hamiltonians, 
 intertwining operators and corresponding
 Lax-Novikov integrals,
and put $\beta^-=\beta$.
Recall that  $0<\beta<{\rm {\bf K}}$.

The operator (\ref{X00}) and its conjugate  
factorise the Hamiltonians,
\be\label{X00X00}
	X^\dag(x;\beta)X(x;\beta)=
	H(x)-\varepsilon(\beta),\quad
	X(x;\beta)X^\dag(x;\beta)=
	H(x+\beta)-\varepsilon(\beta)\,,
\ee
and  intertwine them, 
\be\label{HX00H}
X(x;\beta)H(x)=
H(x+\beta)X(x;\beta)\,,
\qquad 
	X^\dagger(x;\beta)H(x+\beta)=
	H(x)X^\dagger(x;\beta)\,,
\ee
where $\varepsilon(\beta)=\mathcal{E}(\beta+i{\rm{\bf K}}')=-{\rm cs}^2\beta$.
These first  order intertwining operators 
are related by
$X^\dag(x;\beta)=-X(x+\beta;-\beta)$,
that follows from the identity 
$1/F(x;\beta)=F(x+\beta;-\beta)\exp(-\beta{\rm z}(\beta))$,
and corresponds according to (\ref{X2ln}) to the limit $C^-\rightarrow \infty$
of the first order operator $A_{0,1}$. 
In this limit the topologically nontrivial superpotential
$\mathcal{W}_{0,1}$ transforms into the topologically trivial
superpotential $\Delta_{0,0}$, see Figures \ref{fig10} and
\ref{fig11}. 
One can construct the second intertwiner 
being the differential operator of the order $2$ 
by taking a composition of the two first-order intertwiners 
(\ref{X00}), 
\be\label{Y00}
	\mathcal{G}(x;\beta',\beta)=
	X(x+\beta';\beta-\beta')X(x;\beta')\,,
\ee
$\mathcal{G}(x;\beta',\beta)H(x)=
H(x+\beta)\mathcal{G}(x;\beta,\beta')$,
where we assume that $\beta'\neq \beta$.
The first  factor on the r.h.s. 
in (\ref{Y00})  intertwine  the 
$H(x)$ with the Hamiltonian of the virtual system $H(x+\beta')$ 
and then this is intertwined by the second factor with 
$H(x+\beta)$. Notice also that 
$\mathcal{G}^\dagger(x;\beta',\beta)=
\mathcal{G}(x+\beta; \beta'-\beta,-\beta)$.

One could think here that 
in this way 
intertwining operators of the higher order 
$n>2$
can be constructed,
but this is impossible because of the identity 
\cite{PAN,AP2}
\be\label{YYgauge}
\mathcal{G}(x;\beta',\beta)=
\mathcal{G}(x;\beta'',\beta)
+G(\beta,\beta',\beta'')X(x;\beta)\,,
\ee
 from where it follows that the third-order differential 
operator 
\be
	X(x+\beta';\beta-\beta')\mathcal{G}(x;\beta'',\beta')=
	-(H(x+\beta)-\varepsilon(\beta'-\beta))X(x;\beta)
	+G(\beta',\beta'',\beta)\mathcal{G}(x;\beta',\beta),
\ee
which intertwines $H(x)$ and $H(x+\beta)$,
reduces effectively to the first- and second-order  intertwining 
operators 
$X(x;\beta)$ and $\mathcal{G}(x;\beta',\beta)$. 
Here we used the notations 
$G(\beta,\beta',\beta'')\equiv g(\beta,-\beta')-g(\beta,-\beta'')$,
\begin{equation}\label{gcs}
    g(\beta,\beta')\equiv \ns\,\beta\,\ns\,\beta'\,\ns\,(\beta+\beta')
    \left(1-\cn\,\beta\,\cn\,\beta'\,\cn\, (\beta+\beta')\right)\,.
\end{equation}

The  relation (\ref{YYgauge}) reflects effectively 
a kind of the `gauge' nature of the parameter 
$\beta'$, which appears in the structure of
$\mathcal{G}(x;\beta',\beta)$  and 
is associated with a virtual system $H(x+\beta')$.
On the other hand, from the same relation 
and definition (\ref{Y00}) one finds
that the second order operator 
\be\label{Y00true}
	{Y}(x;\beta)=
	X(x+\beta';\beta-\beta')X(x;\beta')
	-g(\beta,-\beta')X(x;\beta)
\ee
is invariant under the change $\beta'\rightarrow \beta''$.
Thus, being a certain  linear combination  of (\ref{X00}) 
and (\ref{Y00}), ${Y}(x;\beta)$ is the `gauge-invariant' 
second order intertwining operator, 
$Y(x;\beta)H(x)=
H(x+\beta)Y(x;\beta)$,
which does not depend on the value of the
virtual parameter  in spite of its appearance on the r.h.s.
in (\ref{Y00true}). The conjugate operator
acts in the opposite direction, and similarly to 
the first order intertwining operator,
we have 
$Y^\dagger(x;\beta)=
Y(x+\beta;-\beta)$.

One can represent  ${Y}(x;\beta)$ 
in the explicitly $\beta'$-independent form  
in terms of the superpotential 
(\ref{Delta00}) and parameter $\beta$.
However, we do not need here such an  expression,
and will use the representation (\ref{Y00true}).

{}From  the properties of $X(x;\beta)$ and 
${Y}(x;\beta)$ it follows that the third order operators
$X^\dagger(x;\beta){Y}(x;\beta)$
and $Y^\dagger(x;\beta){X}(x;\beta)$
 reduce, up to the additive constants, to the 
third order Lax-Novikov integral $\mathcal{P}(x)=\mathcal{P}_{0,0}(x)$
given by Eq. (\ref{LN00}), and to $\mathcal{P}(x+\beta)$,
respectively. Namely, we have 
\be\label{XYP}
	{X}^\dagger(x;\beta)Y(x;\beta)=
	-i\mathcal{P}(x)-\mathcal{N}_0(\beta)\,,\qquad
	{X}(x;\beta)Y^\dagger(x;\beta)=
	i\mathcal{P}(x+\beta)-\mathcal{N}_0(\beta)\,,
\ee
and the pair of identity relations, which can be obtained from 
(\ref{XYP}) by the Hermitian conjugation.
The $\beta$-dependent constant $\mathcal{N}_0(\beta)$ 
 is given by~\footnote{Notice here that for the limit case $\beta={\rm {\bf K}}$,
$\mathcal{N}_0({\rm {\bf K}})=0$. Then for the choice 
$\beta'={\rm {\bf K}}+i{\rm {\bf K}}'$ the coefficient $g$
in (\ref{Y00true}) turns into zero,
and the Hermitian conjugate form of  the first relation in
(\ref{XYP})  corresponds to factorisation (\ref{P3factor}).
Another choice, for instance,   
$\beta'=i{\rm {\bf K}}'$, gives a factorisation 
$i\mathcal{P}_{0,0}(x)=A_{1/\sn\,x}A_{{\sn\,x}/{\dn\,x}}A_{\dn\,x}$. } 
 \be\label{N0}
 	\mathcal{N}_0(\beta)=\dn\,\beta \,\cn\,\beta{\rm ns}^3\beta=
	\frac{1}{2}\frac{d}{d\beta}\,\varepsilon(\beta)\,.
\ee

Similarly to (\ref{HX00H}),
the second order intertwining operators
generate the second-order polynomial in the
isospectral Hamiltonians,
 \be\label{YYH}
	Y^\dagger(x;\beta)Y(x;\beta)=
	\mathcal{N}_2\left(H(x),\beta\right)\,,\qquad
	Y(x;\beta)Y^\dagger(x;\beta)=
	\mathcal{N}_2\left(H(x+\beta),\beta\right)\,,
\ee
where 
\be\label{N2Hb}
	\mathcal{N}_2\left(H(x),\beta\right)=
	H^2(x)+c_1(\beta)H(x)+c_2(\beta)\,,
\ee
\be\label{c1c0b}
	c_1(\beta)=-k'^2-{\rm ns}^2\beta=\varepsilon(\beta)-
	1-k'^2\,,\qquad
		c_2(\beta)=\dn^2\beta\,{\rm ns}^4\beta=
		\left(
		\varepsilon(\beta)-1\right)\left(
		\varepsilon(\beta)-k'^2\right)\,.
\ee
Finally, for the products of the intertwining operators 
with the Lax-Novikov integral we obtain
\be\label{XP00}
	-iX(x;\beta)\mathcal{P}(x)=
	 \mathcal{N}_1(H(x+\beta),\beta)\, Y(x;\beta)
	 +\mathcal{N}_0(\beta)\,X(x;\beta)
\ee
\be\label{P00Xdag}
	i\mathcal{P}(x)X^\dagger(x;\beta)=
	\mathcal{N}_1(H(x),\beta)\, Y^\dagger(x;\beta)
	 +\mathcal{N}_0(\beta)\,X^\dagger(x;\beta)\,,
\ee
\be\label{YP00}
	iY(x;\beta)\mathcal{P}(x)=
	 \mathcal{N}_2(H(x+\beta),\beta)\, X(x;\beta)
	 +\mathcal{N}_0(\beta)\,Y(x;\beta)\,,
\ee
\be\label{P00Ydag}
	-i\mathcal{P}(x)Y^\dagger(x;\beta)=
	\mathcal{N}_2(H(x),\beta)\, X^\dagger(x;\beta)
	 +\mathcal{N}_0(\beta)\,Y^\dagger(x;\beta)\,,
\ee
and four other relations given by the Hermitian conjugation. 
Here we introduced the notation
\be\label{N1Hb}
	\mathcal{N}_1(H(x),\beta)=H(x)-\varepsilon(\beta)\,.
\ee
The operators $X(x;\beta)$ and 
$Y(x;\beta)$, and their conjugate 
ones intertwine 
the Lax-Novikov integrals $\mathcal{P}(x)$ and 
$\mathcal{P}(x+\beta)$
exactly in the same way as they do this 
with the corresponding Hamiltonians.

Now  we are in a position to identify the 
superalgebra of the extended  Schr\"odinger system 
 $\tilde{\cH}={\rm diag}\,
(H(x),H(x+\beta))$,
which corresponds to (\ref{cHtilde}) with $\ell=n=0$
and lower component $\tilde{H}_{0,0}(x;\beta^-_1)$.
This extended system is characterised 
by  the two pairs of the 
fermion integrals 
$\tilde{S}^a(x;\beta)$ and  $\tilde{Q}^a(x;\beta',\beta)$,
constructed from the first-, $X^\dagger(x;\beta)$,
$X(x;\beta)$, and 
second-order,  $Y^\dagger(x;\beta)$,
$Y(x;\beta)$, intertwining operators
in the form similar to that in (\ref{Saint}), 
and by the two boson integrals 
$\tilde{P}^1=
{\rm diag}\,(\mathcal{P}(x),\mathcal{P}(x+\beta))$
and  $\tilde{P}^2=\sigma_3\tilde{P}^1$.
These $2\times 2$ matrix operators 
generate the following exotic nonlinear 
$\mathcal{N}=4$ superalgebra:
\be\label{susytilde1}
\{\tilde{S}^{a},\tilde{S}^b\}=
2\delta^{ab}\mathcal{N}_1(\tilde{\mathcal{H}},\beta)\,,\qquad
\{\tilde{Q}^{a},\tilde{Q}^b\}=
2\delta^{ab}\mathcal{N}_2(\tilde{\mathcal{H}},\beta)\,,
\ee
\be\label{susytilde2}
\{\tilde{S}^{a},\tilde{Q}^{b}\}=-2\epsilon^{ab}\tilde{P}^1
-2\delta^{ab}\mathcal{N}_0(\beta)\,,
\ee
\be\label{susytilde3}
[\tilde{P}^{2},\tilde{S}^{a}]=
-2i\mathcal{N}_1(\tilde{\mathcal{H}},\beta)\tilde{Q}^{a}
-2i\mathcal{N}_0(\beta)\tilde{S}^{a}\,,\qquad
[\tilde{P}^{2},\tilde{Q}^{a}]=
2i\mathcal{N}_2(\tilde{\mathcal{H}},\beta)\tilde{S}^{a}
+2i\mathcal{N}_0(\beta)\tilde{Q}^{a}\,,
\ee
\be\label{susytilde4}
[\tilde{P}^{1},\tilde{Q}^{a}]=0\,,\qquad
[\tilde{P}^{1},\tilde{S}^{a}]=0\,,
\ee
where $\mathcal{N}_1(\tilde{\mathcal{H}},\beta)$
and $\mathcal{N}_2(\tilde{\mathcal{H}},\beta)$
are defined as above with the operator argument
$H(x)$ changed for 
$\tilde{\mathcal{H}}$. The matrix 
Hamiltonian operator $\tilde{\mathcal{H}}$ 
plays here, as well as in the superalgebra we considered
in the previous subsection, the role
of the central element.  
Note  that  the 
constants appearing in the structure 
of $\mathcal{N}_1(\tilde{\mathcal{H}},\beta)$
and $\mathcal{N}_2(\tilde{\mathcal{H}},\beta)$ 
correspond to the energies of the 
doubly degenerate states of the system 
at the edges of the allowed bands: 
$\mathcal{E}=0,\, k'^2,\, 1$.

The  sub-superalgebra 
generated by the supercharges 
$\tilde{S}^a$ and by 
the 
Hamiltonian
 $\tilde{\mathcal{H}}$ 
 with  $0<\beta<{\rm {\bf K}}$
corresponds to the case 
of the  spontaneously broken  linear (Lie)
 $\mathcal{N}=2$ supersymmetry.
 The first-order 
 supercharges do not annihilate the 
 two ground states $\Psi^t_+\equiv(\dn x,0)$
 and $\Psi^t_-\equiv (0,\dn (x+\beta))$
 being eigenstates of zero energy of the extended system.
 This is obvious from the first relation 
 from (\ref{susytilde1}) and Eq. (\ref{N1Hb}).
 The quantity 
 $-\varepsilon(\beta)={\rm cs}^2\beta>0$
 defines here the scale of supersymmetry breaking.
{}The second relation from 
(\ref{susytilde1}) and Eqs. 
(\ref{N2Hb}) and (\ref{c1c0b}) show  
that the second-order supercharges $\tilde{Q}^a$
also do not  annihilate these states.
These edge states, however,   as well as the 
edge states of energies $k'^2$ and $1$,
 which correspond to the two other 
doubly degenerate energy levels of 
 $\tilde{\mathcal{H}}$,
 are zero modes of the bosonic 
 generators $\tilde{P}^a$.
 
 The  limit case $\beta={\rm {\bf K}}$
 corresponding to $\varepsilon=0$
 is special here. At $\beta={\rm {\bf K}}$,
 the coefficient  $\mathcal{N}_0$
 turns into zero, and 
 the indicated two ground states 
are zero modes of the first-order supercharges.
The structure 
 of the non-linear superalgebra 
 (\ref{susytilde1})--(\ref{susytilde4})
 essentially simplifies  because of
 disappearance of the three terms in
 Eqs.  (\ref{susytilde2}) and  (\ref{susytilde3}).
 In this case the second-order supercharges
 $\tilde{Q}^{a}$ annihilate the doubly degenerate
 states at the  edges of the valence and conduction bands
 of energies $k'^2$ and $1$.
 Since 
 the second-order supercharges  $\tilde{Q}^{a}$
 do not annihilate the 
 degenerate pair of the ground states  
 in this case either, 
 the extended system $\tilde{\mathcal{H}}$
 with $\beta={\rm {\bf K}}$
  is characterised 
 by the partially broken exotic nonlinear 
$\mathcal{N}=4$  supersymmetry. 

Notice that though at $\beta={\rm {\bf K}}$, 
the sub-supersymmetry 
$\mathcal{N}=2$ 
generated by $\tilde{\mathcal{H}}$ and 
$\tilde{S}^{a}$ is unbroken,
the subsystems $H(x)$ and $H(x+\beta)$ are completely isospectral,
and  the superextended system is characterised by zero 
Witten index \cite{Witten}. 
This is a characteristic peculiarity  of the quantum 
  supersymmetric systems composed from 
 the  periodic completely isospectral pairs,
  which was 
  noted for the first time by 
  Braden and Macfarlane  \cite{BraMac}
  for the particular case of the pair of 
  one-gap  periodic  
  Lam\'e  systems shifted mutually
  for the half-period $\beta={\rm {\bf K}}$,
  and later was discussed in a more broad
  context of ``self-isospectrality" 
  by   Dunne and Feinberg \cite{DunFei}.
  In  the framework  of 
  the non-linear ``tri-supersymmetric" structure
it was analysed then in  \cite{CJNP,CJP}.
  
 In the context of the breaking of 
 the exotic supersymmetry
 it is worth to notice that, generally speaking,  
 the second-order supercharges are not defined uniquely here.
 Instead of $\tilde{Q}^{a}$, one can take 
 linear combinations of $\tilde{Q}^{a}$
 and $\tilde{S}^{a}$,
 for instance $\hat{Q}{}^a=\tilde{Q}^{a}
 +\gamma \tilde{S}^{a}$, where $\gamma$ is a real
 constant.  The particular choice 
 $\gamma=\dn\, \beta /\sn\,\beta\cn\,\beta$
 gives then the supercharges $\hat{Q}{}^a$,
 which  satisfy the anti-commutation relations
  $\{\hat{Q}{}^a,\hat{Q}{}^b\}=2\delta^{ab}
  \tilde{\mathcal{H}}(\tilde{\mathcal{H}}+\varrho(\beta))$,
  where $\varrho(\beta)=k'^2{\rm sc}^2\beta$.
 {}Hence, for $\beta\neq {\rm {\bf K}}$, 
 the supercharges $\hat{Q}{}^a$
annihilate the ground states of zero energy
 of the system   $\tilde{\mathcal{H}}$
 (while other states from their kernels correspond to 
 non-physical eigenstates of $\tilde{\mathcal{H}}$). 
 In this case the exotic supersymmetry generated 
 by  $\tilde{S}^{a}$,  $\hat{Q}{}^a$,  $\tilde{P}^{a}$
 and $\tilde{\mathcal{H}}$ should be interpreted 
 as partially  broken. However, 
 the second-order supercharges 
 $\hat{Q}{}^a$, unlike $\tilde{Q}^{a}$,
 are not defined for the limit case
 $\beta={\rm {\bf K}}$. 
 The supercharges  
 $\hat{Q}{}^a$ with the indicated choice of the 
 parameter $\gamma$ correspond 
 to the the second-order intertwining 
 generators (\ref{Y00}) with 
 $\beta'={\rm {\bf K}}$.

As in the case of the unbroken exotic supersymmetry
we considered in the previous subsection,
the Lax-Novikov matrix integral 
$\tilde{P}^1$  plays here the role of the 
bosonic central  charge,
and the second 
relation in (\ref{susytilde4})
corresponds to the stationary equation of the mKdV hierarchy 
for the topologically trivial superpotential 
 $\Delta_{0,0}(x,\beta)$. 
The relation  
$[\tilde{\mathcal{H}},\tilde{P}^{1}]=0$
corresponds to the pair of stationary 
equations of the KdV hierarchy 
for the functions
$V_\pm(x)=
\Delta_{0,0}(x,\beta)^2\pm\Delta_{0,0}'(x,\beta)+\varepsilon(\beta)$,
which represent the potentials of the corresponding 
mutually shifted Schr\"odinger systems.

The superalgebra 
(\ref{susytilde1})--(\ref{susytilde4}) in comparison 
with that
of the unbroken exotic supersymmetry case 
(\ref{susyexact1})--(\ref{susyexact4})
contains the
terms  with the coefficient $\mathcal{N}_0(\beta)$
in (\ref{susytilde2}), (\ref{susytilde3}),
which are absent in (\ref{susyexact2}), (\ref{susyexact3}). 
There are also other obvious differences in these two
forms of superalgebras, which reflect properly the
unbroken and  spontaneously  broken  character  
of the 
exotic supersymmetries, and different 
topological nature of the corresponding
superpotentials.  At the formal level,
some of these differences are associated 
with a nontrivial limit procedure applied 
to the fourth-order intertwining operators 
$B_{0,1}=A_{0,1}\mathcal{P}_{0,1}(x)$
and $B^\dagger_{0,1}$, in terms of which
the fourth-order supercharges $Q^a$ 
were constructed in the previous subsection.
In correspondence with the limit
(\ref{X2ln}), we have $B_{0,1}\rightarrow
X(x;\beta)\mathcal{P}(x)$, 
$\mathcal{P}(x)=\mathcal{P}_{0,0}(x)$.
But according to the relation
(\ref{XP00}), the fourth-order intertwining 
operator we  obtain in the limit is reducible,
and, finally, instead  of the fourth-order
intertwining operators, here we have  
the second-order operators
$Y(x;\beta)$ and $Y^\dagger(x;\beta)$,
which intertwine the completely isospectral 
pair of the Schr\"odinger systems 
$H(x)=H_{0,0}(x)$ and $H(x+\beta)=H_{0,0}(x+\beta)$.

\section{Discussion and outlook}

To conclude, we 
summarise shortly the results, and  
point out further possible research directions.
\vskip0.05cm

We showed how by applying 
the Darboux-Crum 
transformations to 
the quantum one-gap Lam\'e system,
an  arbitrary countable number of  bound states  
can be introduced
into the  forbidden bands of its spectrum.
These states are trapped  by 
localised perturbations of the periodic 
potential background of the initial system.
The nature of the perturbations 
depends on whether they support 
discrete energy levels 
in the lower forbidden band, or
in the finite gap separating the allowed
valence and conduction bands. 
In the first case the perturbations
have a nature of the smooth 
soliton potential 
\emph{wells} superimposed on the 
background of the Lam\'e system, 
while the discrete energy levels  
in the gap are supported by 
compression \emph{modulations}  of the 
periodic  background. 
Though both types of perturbations 
have a soliton nature, to distinguish, 
we identify them here 
as the $W$-type and
$M$-type defects, respectively.
The nature of  the 
bound states  is essentially different   in these 
two cases. The $n\geq 1$ bound states trapped 
by the $W$-type defects are described
by modulated by the background 
wave functions with finite number $0\leq j\leq n-1$ of 
nodes on the real line. 
In contrast, the bound states supported by 
the $M$-type defects have 
infinite number of nodes, and represent 
oscillating trapped pulses. 

The obtained  non-periodic systems  
are reflectionless: their physical states 
inside the valence and conduction bands
are described by the  
Darboux-Crum transformed Bloch-states
of the Lam\'e system, just  like 
the scattering states of quantum systems 
with multi-soliton potentials are given by
a Darboux-Crum transformation of 
free particle plane waves.
Similarly to the multi-soliton reflectionless  potentials, which 
exponentially tend to a constant value corresponding 
to the free particle case, here the 
asymptotics of  the perturbed potentials 
corresponds to the periodic one-gap Lam\'e potential.
We showed that the net phase displacement (defect) between
$x=+\infty$ and $x=-\infty$ periodic asymptotics 
of the potential are given by a simple sum of the same 
parameters which 
determine, via the elliptic $\dn^2$ -parametrisation,
the discrete energy levels.

The procedure 
for introducing the $W$- and the $M$-type 
periodicity defects has some important differences.
In the first case the order $n$ of the 
Darboux-Crum transformation 
corresponds exactly to the number of the
introduced  bound states.
In the second case the same is true 
when the number of discrete 
energy values  is even. The 
odd number of the discrete energy levels  in the 
gap is obtained by sending one of the already 
introduced $2\ell$ $M$-type defects to infinity.
The resulting potential with $2\ell-1$ $M$-defects
is related to the initial Lam\'e system 
by $2\ell$-th order Darboux-Crum transformation.
At the same time, it can be related 
by the Darboux-Crum transformation
of order $2\ell-1$ with a singular one-gap
Treibich-Verdier system
obtained by a displacement of the regular Lam\'e system 
for one of its two complex half-periods.
The indicated complex  displacement can itself be generated 
by the first-order Darboux transformation. This 
explains the existence of  two
alternative Darboux-Crum transformations whose orders
differ by one.

 The procedure described in this article
 allows us to construct 
the irreducible Lax-Novikov integrals 
of motion 
for the perturbed systems $H_{2\ell-m,n}$ via the 
Darboux-Crum dressing of the Lax-Novikov 
integral of the initial periodic Lam\'e system
$H_{0,0}$. This is similar, again, to the situation with
the transparent quantum systems described  by multi-soliton 
potentials, for which the Lax-Novikov integrals
are the Darboux-Crum dressed  form of the 
momentum operator of the free particle.  
The Lax-Novikov integrals here
are differential operators
of order $2(n+2\ell-m)+3$ for the system
with $n\geq 0$ $W$-type and $2\ell-m\geq 0$, $m=0,1$,
$M$-type defects.
The condition of conservation of these 
integrals generates  
a nonlinear differential equation of 
order $2(n+2\ell-m)+3$ for the potential
$V_{2\ell-m,n}(x)$.
This ordinary nonlinear differential equation 
of odd order in the highest derivative 
 belongs to  the 
stationary KdV hierarchy. 
 
For an extended system composed from
an arbitrary pair
of the Hamiltonians $H_{2\ell_1-m_1,n_1}$
and  $H_{2\ell_2-m_2,n_2}$, 
which possess 
$n_i\geq 0$, $i=1,2$, discrete energy levels 
in the lower forbidden band 
and $2\ell_i-m_i\geq 0$,
$m_i=0,1$, bound states in the gap, 
the presence of the Lax-Novikov integrals has 
an essential consequence. 
The whole system is now described 
not just by an $\mathcal{N}=2$ linear or nonlinear 
supersymmetry as would be expected in the case of 
a Darboux-Crum related pair of ordinary, 
non-transparent,  or not periodic finite-gap,
quantum Hamiltonians.
Instead, such a system is characterised 
by an exotic nonlinear $\mathcal{N}=4$ 
supersymmetry that, besides 
two pairs of the fermion supercharges
of odd and even differential orders, 
 involves two bosonic 
generators composed from the Lax-Novikov
integrals of the subsystems.
We investigated in more detail the most interesting, 
from the point of view of physical applications, 
case, when
two of the four fermionic supercharges 
are matrix differential operators of order 
one.  
In this case, one of the matrix Lax-Novikov 
bosonic integrals
plays a role of central charge of a nonlinear 
superalgebra, and its commutativity 
with  first order supercharges 
generates a higher order differential equation 
for the superpotential which  
belongs to the stationary mKdV hierarchy.
The second bosonic integral generates rotations
between the pair of  first order  supercharges 
and the pair of higher order supercharges. 

When the spectra of Schr\"odinger 
superpartners are different 
only in the lowest discrete energy
level present in one of the two subsystems, 
that corresponds to the almost isospectral case, 
the superpotential has a topologically 
nontrivial modulated crystalline kink-type nature.
This case is described by an unbroken exotic nonlinear 
$\mathcal{N} = 4$ supersymmetry, in which
the ground state is annihilated by all four 
supercharges and two bosonic integrals. 
On the other hand, in the completely 
isospectral case, 
the pair of Schr\"odinger Hamiltonians
is characterised by  a superpotential
of a topologically trivial, modulated kink-antikink
type nature.
Such pairs  can be obtained from the pairs 
of almost isospectral case
just by sending the
$W$-type defect associated with
the lowest energy discrete value 
to infinity.  The completely isospectral pairs 
are described by a spontaneously broken 
exotic nonlinear $\mathcal{N}=4$ 
supersymmetry. Unlike the unbroken supersymmetry 
case, in such systems the two 
states corresponding to the lowest 
doubly degenerate energy value are 
annihilated (in a generic case) only by the 
bosonic  Lax-Novikov integrals.

When one of the two first-order supercharges 
is reinterpreted as the  matrix Hamiltonian
operator, 
we arrive at the Bogoliubov-de Gennes system,
in which the superpotential  
will play the role of a scalar Dirac 
potential.
The results presented  here allow us then,
particularly,  
to obtain new types of self-consistent condensates 
and associate with them 
new solutions for the Gross-Neveu model,
which correspond to the kink
and kink-antikink type configurations in the 
crystalline background. We are going to consider this
problem elsewhere.

It is worth noticing that Dirac Hamiltonians with 
scalar potential appear, in different physical context, 
in description of the low-energy charge carriers 
in graphene and related carbon nanostructures. 
This fact opens potential applications of the results 
in physics of condensed matter systems, 
following the ideas of \cite{JNPtun,twisted1,twisted2}.

The discussed constructions can be 
generalised to the case of
the PT-symmetric one-gap potentials.
To achieve this, it is sufficient to  
apply the complex shift  considered in Section
\ref{Darbtrans} to the described 
Hermitian systems with periodicity defects. 
Such systems have an immediate application  
 in the context of the PT-symmetric quantum mechanics
and optics.

An interesting development of the presented  results is 
to `reconstruct' the time dependence
for defects in a periodic background of the 
one-gap Lam\'e system in correspondence with
dynamics illustrated, as an example, by Figure \ref{fig4}. 
This would provide us a new class of solutions
for the KdV and mKdV equations. 
At the same time, it is natural to consider the
generalisation of the construction to the case 
of quantum $n$-gap systems with $n > 1$.
One can also wonder if,  somehow, 
both $W$-type and $M$-type defects are the result of
``shrinking"  bands from a more generic 
finite-gap Hamiltonian, under some special limit.

Finally, it would also be very interesting to look for the  
$(1+1)$D field theories, in which nontrivial 
solutions are  controlled by stability operator
of the Schr\"odinger type  \cite{AloJuan}
with the potentials of the nature considered here.

\vskip0.2cm

{\bf Acknowledgements}. 
The work of MP has been partially 
supported by FONDECYT Grant No.
1130017. MP thanks Salamanca University 
and Nuclear Physics Institute 
of the ASCR, 
where a part of this work was done, for hospitality.
AA acknowledges the CONICYT scholarship 21120826
and financial support of
Direcci\'on de Postgrado and Vicerrectoria Acad\'emica
of the USACH.
He thanks Salamanca University for the kind hospitality.
FC wishes to thank the warm hospitality of Nuclear Physics Institute 
of the ASCR. FC is partially supported through Fondecyt grant 11121651, 
Conicyt grant 79112034 and by the Alexander von 
Humboldt Foundation. CECs is funded by the 
Chilean Government through the Centers of 
Excellence Base Financing Program of Conicyt.
VJ was supported by the project RVO61389005 
of the NPI ASCR.

\vskip1cm

\section{Appendix}

We show here that 
the family of Hamiltonians 
\be
	H_{2\ell,n}=H_{0,0}-2\frac{d^2}{dx^2}\big(\log W\left(\Phi_+(1),\Phi_-(2),...,
	\Phi_-(2\ell),\mathcal{F}_{+}(1),...,\mathcal{F}_{s_n}(n)\right) \big)
\ee
is given in terms of the \emph{non-singular} potentials, 
which correspond to the soliton defects 
introduced into the periodic background of the 
one-gap Lam\'e system.
To achieve this, 
we demonstrate successively that 
the Wronskians appearing in the structure of 
 $H_{0,n}$, $H_{2\ell,0}$  and, finally,   $H_{2\ell,n}$ are 
 nodeless on the real line.
 The notations we employ are explained in the main text.

\subsection{Lower prohibited band}
In order to show that the potential of 
$H_{0,n}$ is regular, i.e. has no zeros on 
the real line, we will demonstrate that
\be\label{Wf}
	(-1)^{\frac{n(n+1)}{2}}W(\mathcal{F}_{+}(1),\ldots,
	\mathcal{F}_{s_{n+1}}(n+1))>0\,.
\ee
First, we  define 
the two sets of functions,
\be\label{fi}
	f_n(x)\equiv(-1)^{n}\frac{W(\mathcal{F}_{+}(1),\ldots,
	\mathcal{F}_{s_{n+1}}(n+1))}
	{W(\mathcal{F}_{+}(1),\ldots,\mathcal{F}_{s_{n}}(n))}\,,
\ee
and 
\be\label{gin}
	g_n(x)\equiv(-1)^{n}\frac{W(\mathcal{F}_{+}(1),\ldots,
	\mathcal{F}_{s_{n}}(n),\mathcal{F}_{s_{n+2}}(n+2))}
	{W(\mathcal{F}_{+}(1),\ldots,\mathcal{F}_{s_{n}}(n))}\,,	
\ee
which are non-physical eigenstates of $H_{0,n}$
with eigenvalues $\varepsilon^-_{n+1}$  and
$\varepsilon^-_{n+2}$, respectively. 
We will check below that $f_n(x)>0$, while  $g_n(x)$ 
has only one zero. 

In correspondence with the definition $\W_{0,0}=1$ introduced 
in Eq. (\ref{WjOmega}),
for $n=0$  we have   
$f_0=\mathcal{F}_{+}(1)>0$, and 
$g_0=\mathcal{F}_{-}(2)$.
The second function (plotted  for a particular case with
$C_2=1$
in Figure \ref{fig3}) has one zero, which we denote by $x_0$.
Thus, we have 
$g_0(x)>0$ for $x<x_0$  and $g_0(x)<0$  for $x>x_0$.

For  the case $n=1$, we also define the  functions
\be
	f(x)=W(\mathcal{F}_{+}(1),\mathcal{F}_{-}(2)),\qquad
	g(x)=W(\mathcal{F}_{+}(1),\mathcal{F}_{+}(3))\,,
\ee
which appear in the numerators of (\ref{fi}) and (\ref{gin}).
Taking into account that $\mathcal{F}$ are  solutions of the stationary 
 Schr\"odinger equation, it is straightforward to check
 that
\be
	f'(x)=\left(\varepsilon^-_1 - \varepsilon^-_2 
	\right)\mathcal{F}_{+}(1)\mathcal{F}_{-}(2)\,,
\ee
\be
	g'(x)=\left(\varepsilon^-_1 - \varepsilon^-_3 \right)
	\mathcal{F}_{+}(1)\mathcal{F}_{+}(3)\,.
\ee
As  $\varepsilon^-_2 <\varepsilon^-_1<0$, 
we observe that 
  ${\rm sign}\,(f'(x))={\rm sign}\,(\mathcal{F}_{-}(2))$.
  Then  
\be
	f(x_0)=
	\mathcal{F}_{+}(x_0;\beta^-_1,C_1)\mathcal{F}'_{-}(x_0;\beta^-_2,C_2)
\ee
since 
 $\mathcal{F}_{-}(x_0;\beta^-_2,C_2)=0$. 
 {}From the  Schr\"odinger equation we have also 
 $\mathcal{F}'_{-}(x_0;\beta^-_2,C_2)\neq0$, 
 and from the definition (\ref{Fsi+-}) it follows that 
$\mathcal{F}'_{-}(x_0;\beta^-_2,C_2)<0$.
We have then $f(x_0)<0$, and hence,  
${\rm sign}\,(f'(x))={\rm sign}\,(\mathcal{F}_{-}(2))$.
Thus,
the function $f(x)$ increases monotonically from  
$f(-\infty)=-\infty$, it takes a maximum negative value 
$f(x_0)<0$ at $x=x_0$, and then decreases
 again monotonically 
to $f(\infty)=-\infty$. This means that  $f(x)<0$
and, as a consequence,
\be
f_1(x)=-\frac{W(\mathcal{F}_{+}(1),\mathcal{F}_{-}(2))}
{\mathcal{F}_{+}(1)}>0
\ee
for all  $x$.

The derivative $g'(x)$ takes positive values
and grows up  exponentially for $x\rightarrow\pm\infty$.
Therefore, $g(x)$ passes through zero only once 
at some point $x_1$. The function
\be
g_1(x)=-\frac{W(\mathcal{F}_{+}(1),\mathcal{F}_{+}(3))}
{\mathcal{F}_{+}(1)}
 \ee
has then only one zero at this point $x_1$,
and takes positive and negative values for  
 $x<x_1$ and  $x>x_1$, respectively.
 So, we see that the non-physical 
 eigenstates $f_0$ and $f_1$ of $H_{0,0}$ and 
 $H_{0,1}$, respectively,   have no zeros,
 while their  eigenfunctions  $g_0$ and $g_1$ 
 have one zero, where their slope is  negative.

We extend now this result by induction for 
arbitrary $n$ by showing that  
$f_n(x)>0$ while  $g_n$ has only one zero $x_n$, and that  
$g_{n}(x)>0$  and $g_n(x)<0$ for $x<x_n$ and $x>x_n$, 
respectively, and so, $g'_{n}(x_n)<0$.

By 
using the Darboux-Crum construction, we can check that functions 
$f_n(x)$  and $g_n(x)$  are  non-physical eigenstates 
of the Schr\"odinger operator 
\be
	H_{0,n}=H_{0,0}-2\frac{d^2}{dx^2}\log W(\mathcal{F}_{+}(1),\ldots,
	\mathcal{F}_{s_n}(n))
\ee
with eigenvalues 
$\varepsilon^-_{n+1}$ 
and  $\varepsilon^-_{n+2}$.
For $n+1$ we have
\be\label{fn+1}
	f_{n+1}(x)=(-1)^{n+1}\frac{
	W(\mathcal{F}_{+}(1),\ldots,\mathcal{F}_{s_{n+2}}(n+2))}
	{W(\mathcal{F}_{+}(1),\ldots,\mathcal{F}_{s_{n+1}}(n+1))}
	=-\frac{W\left(f_n,g_n\right)}{f_n}\,,
\ee
\be
	W'\left(f_n,g_n\right)=
	\left( \varepsilon^-_{n+1} - \varepsilon^-_{n+2}
	\right)f_n\, g_n\,,
\ee
from where we obtain that 
${\rm sign}\,W'\left(f_n,g_n\right)={\rm sign}\,g_n(x)$.
The zero  $x_n$ of  ${g_n}$  corresponds therefore to the 
maximum of 
$W\left(f_n,g_n\right)$,
\be
	W\left(f_n,g_n\right)(x_n)=
	{g'_n}(x_n)f_n(x_n)<0\,.
\ee
Since  ${\rm sign}\,W'\left(f_n,g_n\right)={\rm sign}\,g_n(x)$, 
the function $-W\left(f_n,g_n\right)$ decreases for $x<x_n$,
and increases for $x>x_n$,  and then $-W\left(f_n,g_n\right)(x_n)>0$ 
for all $x$.
{}From Eq. (\ref{fn+1}) we conclude  that
 $f_{n+1}(x)>0$ for all $x$.

Let us change   $\beta^-_{n+1}$ by  $\beta^-_{n+3}$
in the numerator  of the function 
$f_n(x)$ in (\ref{fi}), 
and redefine  the resulting function as $h_n(x)$.
This function takes positive values, $h_n(x)>0$, 
and we obtain the following relations
\be
	g_{n+1}=(-1)^{n+1}\frac{W(\mathcal{F}_{+}(1),\ldots,
	\mathcal{F}_{s_{n+1}}(n+1),\mathcal{F}_{s_{n+3}}(n+3))}
	{W(\mathcal{F}_{+}(1),\ldots,\mathcal{F}_{s_{n+1}}(n+1))}=
	-\frac{W\left(f_n,h_n\right)}{f_n}\,,
\ee
\be
	W'\left(f_n,h_n\right)=
	\left(\varepsilon^-_{n+1} - \varepsilon^-_{n+3}
	\right)f_nh_n>0\,.
\ee
Consequently,  $W\left(f_n,h_n\right)$ 
increases exponentially from $-\infty$ to $+\infty$ 
passing through one zero, which we call
 $x_{n+1}$. 
Since $f_n(x)>0$ is a regular function,  
and $g_{n+1}$ has only one zero at
$x_{n+1}$, we find that 
 $g_{n+1}(x)>0$ for $x<x_{n+1}$ and $g_{n+1}(x)<0$  for 
 $x>x_{n+1}$.

{}Finally, from the definition (\ref{fi}) of $f_n(x)$ 
we obtain 
\be
	f_nf_{n-1}\ldots f_1\mathcal{F}_{+}(1)=
	\big((-1)^{\sum ^n_{i=1}i }\big)W(\mathcal{F}_{+}(1),
	\ldots,\mathcal{F}_{s_{n+1}}(n+1))\,,
\ee
and since 
\be
	f_nf_{n-1}...,f_1\mathcal{F}_{+}(1)>0\,,
\ee
we demonstrate the necessary 
relation (\ref{Wf}).

\subsection{Upper prohibited band}

In order to show that $H_{2\ell,0}$ is 
non-singular  on the whole real line,
we show that the Wronskian  is a regular nodeless function 
 $W(\Phi_+(1),\Phi_-(2),...,\Phi_-(2\ell))$, where the functions 
 $\Phi_+(2l-1)$  and $\Phi_-(2l)$, $l=1,2\ldots$  correspond 
 to a generalisation of those defined in   (\ref{Phi+}),
 (\ref{Phi-}) for $\ell=1$.
 
Before, we have shown that 
 $W(\Phi_+(1),\Phi_-(2))<0$ by choosing  
 parameters $0<\beta^+_1<\beta^+_2<{\rm{\bf K}}$.
 This condition means  that 
  $1> \varepsilon^+_1 > \varepsilon^+_2 >k'^2$
 for the eigenvalues of the non-physical
 eigenstates $\Phi_+(1)$ and $\Phi_-(2)$ inside the 
 intermediate forbidden band of $H_{0,0}$.
 
To demonstrate the validity of the formulated statement for the next 
case $\ell=2$, we define 
an eigenstate of the one-gap Lam\'e system 
with the displaced argument,
 $x\rightarrow x+\beta^+_3+i{\rm{\bf K}}'$, in the 
 following form
\be
	\breve{\Phi}[1](x,\beta^+_3)=
	\frac{W(\Psi_+^{\beta^+_3}(x),\Phi_+(1))}
	{\Psi_+^{\beta^+_3}(x)}\,.
\ee
This state has infinite number of poles at the zeros of 
${\Psi_+^{\beta^+_3}(x)}$. Between each pair of poles 
$\breve{\Phi}[1](x,\beta^+_3)$
does not change the sign and takes nonzero values.
Its sign is inverted in the neighbour regions separated  by poles.
{}From the theorem on
zeros, the linearly independent state
\be\label{phi2x}
\breve{\Phi}[2](x;\beta^+_3)=\frac{W(\Psi_+^{\beta^+_3}(x),\Phi_-(2))}
{\Psi_+^{\beta^+_3}(x)}
\ee
has also and infinite number of poles, 
but between each pair of poles 
it possesses one zero, which we denote  as $x_i$. 
The  function (\ref{phi2x})  preserves the sign when 
the argument  passes through any pole. 

Now, it is necessary to show that
$W(\breve{\Phi}[1],\breve{\Phi}[2])$ does not have zeros. 
For this we redefine the function $\breve{\Phi}[2]$
up to a sign is such a way that 
its derivative in some $x_{i_0}$ will be positive.
In the same way, we also redefine, 
up to a global sign, the function   
 $\breve{\Phi}[1](x)$ to have  $\breve{\Phi}[1](x_{i_0})<0$.
Thus, we obtain that
\be
	W(\breve{\Phi}[1],\breve{\Phi}[2])(x_i)
	=\breve{\Phi}[1](x_i)\breve{\Phi}'[2](x_i)<0\,,
\ee
while
\be
	W'(\breve{\Phi}[1],\breve{\Phi}[2])=
	(\varepsilon^+_1 - \varepsilon^+_2)
	\breve{\Phi}[1]\breve{\Phi}[2]\,.
\ee
The function
 $W(\breve{\Phi}[1],\breve{\Phi}[2])$ 
has a local extremum at each  $x_i$, 
and its derivative is positive for 
 $x<x_i$  till  a pole, and  is negative for 
 $x>x_i$ till the next  pole
since $x_i$ is a local maximum  of $W(\breve{\Phi}[1],\breve{\Phi}[2])(x)$.
{}From here, 
we conclude that 
$W(\breve{\Phi}[1],\breve{\Phi}[2])(x)$ does not have zeros,
and, hence,  is of one sign. 

Because of the identity 
\be
	W(\Phi_+(1),\Phi_-(2),\Psi_+^{\beta^+_3}(x))=
	\Psi_+^{\beta^+_3}(x)
	W(\breve{\Phi}[1],\breve{\Phi}[2])(x;\beta^+_3),
\ee
the Wronskian $W(\Phi_+(1),\Phi_-(2),\Psi_+^{\beta^+_3}(\pm x))$ 
has exactly the same zeros as 
 $\Psi_\pm^{\beta^+_3}(x)$.
Note that we have
$W(\Phi_+(-x;\beta^+_1,1/C_1),\Phi_-(-x;\beta^+_2,1/C_2)
,\Psi_+^{\beta^+_3}(-x))=-W(-\Phi_+(1),\Phi_-(2),\Psi_+^{\beta^+_3}(-x))
=-W(\Phi_+(1),\Phi_-(2),-\Psi_+^{\beta^+_3}(-x))$.
Using the Wronskian properties, it is easy to see that
$W(a,b)(x)=-W(a,b)(-x)$ and   $W(a,b,c)(x)=-W(a,b,c)(-x)$, but   
$W(a,b,c,d)(x)=W(a,b,c,d)(-x)$.
Taking in account the above relations, we can
write
\be
	{\rm sign}\,W(\breve{\Phi}[1],\breve{\Phi}[2])
	(x;\beta^+_3)={\rm sign}\,W(\breve{\Phi}[1],
	\breve{\Phi}[2])(-x;\beta^+_3)\,.
\ee

Thus, the zeros of the non-physical states of
$H_{2,0}$,
\be
	\frac{W(\Phi_+(1),\Phi_-(2),\Phi_+(3))}{W(\Phi_+(1),
	\Phi_-(2))} \quad {\rm and}\quad \frac{W(\Phi_+(1),\Phi_-(2),
	\Phi_-(4))}{W(\Phi_+(1),\Phi_-(2))}\,,
\ee
are within the intervals 
 $\mathcal{I}_n^+(\beta^+_3)$ and  $\mathcal{I}_n^-(\beta^+_4)$,
 respectively,  see Eq. (\ref{defI+I-}), where 
$\mathcal{I}_n^+(3)\bigcap \mathcal{I}_n^-(4)=\emptyset$. 
As a consequence of the theorem on zeros, 
their zeros are alternated.

Next we can check that under the condition
$0<\beta^+_1<\beta^+_2
 <\beta^+_3<\beta^+_4<{\rm {\bf K}}$,  
the Wronskian 
\be
	W\left(\frac{W(\Phi_+(1),\Phi_-(2),
	\Phi_+(3))}{W(\Phi_+(1),\Phi_-(2))},
	\frac{W(\Phi_+(1),\Phi_-(2),\Phi_-(4))}{W(\Phi_+(1),
	\Phi_-(2))}\right)=
	\frac{W(\Phi_+(1),\Phi_-(2),\Phi_+(3),
	\Phi_-(4))}{W(\Phi_+(1),\Phi_-(2))}
\ee
does not have zeros neither the function
$W(\Phi_+(1),\Phi_-(2),\Phi_+(3),\Phi_-(4))$.

This result can be generalised for the case of Wronskian of
 $2\ell$ states, 
 $W(\Phi_+(1),\Psi_-(2),\ldots,\Phi_{+}(2\ell-1),\Phi_{-}(2\ell))$,
 under the condition 
 $0<\beta^+_1<\beta^+_2<\ldots<\beta^+_{2\ell}<{\rm {\bf K}}$.

Using the identity 
\begin{eqnarray}
	&&W(\Phi_+(1),\ldots,\Phi_{-}(2\ell),
	\Psi_+^{\beta^+_{2\ell+1}}(x))\nonumber\\
	&& =W(\Phi_+(1),\ldots,\Phi_-(2\ell-2),\Psi_+^{\beta^+_{2\ell+1}}(x))
	W(\breve{\Phi}[1,\ldots,2\ell-1],\breve{\Phi}[1,\ldots,2\ell-2,2\ell])\,,
\end{eqnarray}
we have 
\bea
\Psi_+^{\beta^+_{2\ell+1}}(x)W(\breve{\Phi}[1],\breve{\Phi}[2])\times
W(\breve{\Phi}[1,2,3],\breve{\Phi}[1,2,4])\times\ldots
	\times&&\nonumber\\
 \times W(\breve{\Phi}[1,\ldots,2\ell-2,2\ell-1],
	\breve{\Phi}[1,\ldots,2\ell-2,2\ell])&&\nonumber\\
	=W(\Phi_+(1),\ldots,\Phi_-(2\ell),\Psi_+^{\beta^+_{2\ell+1}}(x))\,,&&
\eea
where 
\be
	\breve{\Phi}[1,\ldots,l,l+r](x,\beta^+)=
	\frac{W(\Psi^{\beta^+}_+(x),\Phi_+(1),\ldots,
	\Phi_-(2l),\Phi_{s_{2l+r}}(2l+r))}{W(\Psi^{\beta^+}_+(x),\Phi_+(1),\ldots,
	\Phi_-(2l))}\,,	
\ee
and $r=1,2$, $l=0,1,\ldots$. 
Having in mind all previous demonstrations,
it is clear that 
\be
	\left\vert W(\breve{\Phi}[1,\ldots,2l-2,2l-1],
	\breve{\Phi}[1,\ldots,2l-2,2l])\right\vert>0\,,
\ee
and  the  functions
\be
	\frac{W(\Phi_+(1),\ldots,\Phi_{-}(2\ell),\Phi_+(2\ell+1))}
	{W(\Phi_+(1),\ldots,\Phi_-(2\ell))}\quad {\rm and}\quad
	\frac{W(\Phi_+(1),\ldots,.\Phi_-(2\ell),
	\Phi_-(2\ell+2))}
	{W(\Phi_+(1),\ldots,\Phi_-(2\ell))}
\ee
 have alternating zeros in the intervals  $\mathcal{I}^+_n(\beta^+_{2\ell+1})$ and
 $\mathcal{I}^-_n(\beta^+_{2\ell+2})$, respectively.
Then
 \be
   W\left( \frac{W(\Phi_+(1),\ldots,\Phi_-(2\ell),\Phi_+
   (2\ell+1))}{W(\Phi_+(1),\ldots,\Phi_-(2\ell))},
	\frac{W(\Phi_+(1),\ldots,\Phi_-(2\ell),\Phi_-(2\ell+2))}{
	W(\Phi_+(1),\ldots,\Phi_-(2\ell))}\right)
\ee
\be
	=\frac{W(\Phi_+(1),\ldots,\Phi_-(2\ell+2))}{
	W(\Phi_+(1),\ldots,\Phi_-(2\ell))}
 \nonumber
 \ee
is regular and have no zeros,
which means that 
$W(\Phi_+(1),\ldots,\Phi_-(2\ell+2))$ is non-singuar and nodeless
if and only if  $W(\Phi_+(1),\ldots,\Phi_-(2\ell))$ 
is regular and has no zeros.

Besides, if the potentials of the systems
$H_{2\ell,0}$ are non-singular for all real $x$, by taking limits 
$C_l\rightarrow \infty$, or  $C_l\rightarrow 0$, the 
regularity is preserved 
and we get a regular
Hamiltonians $H_{2\ell-1,0}$ 
with  $2\ell-1$ states in the gap of the Lam\'e system. 

\subsection{Mixed case}

Finally, using the all previous demonstrations, we show  that 
the most general Hamiltonian
\be
	H_{2\ell,n}=H_{0,0}-2\frac{d^2}{dx^2}\big(\log W\left(\Phi_+(1),\Phi_-(2),...,
	\Phi_-(2\ell),\mathcal{F}_{+}(1),...,\mathcal{F}_{s_n}(n)\right) \big)
\ee
has also a nonsingular potential. To this aim, we define 
\be
	F_{2\ell}(x;\beta^-)=\frac{W(\Phi_+(1),...,\Phi_-(2\ell),
	F(x;\beta^-))}{W(\Phi_+(1),...,\Phi_-(2\ell))}\,,
\ee
which is a non-physical eigenstate
of $H_{2\ell,0}$ with eigenvalue 
$\mathcal{E}(\beta^-+i{\rm{\bf K}}')$. 
Using the Wronskian identity
\be
	W(\tilde{\Phi}_1,...,\tilde{ \Phi}_l)=
	W(W(F,\Phi_1)/F,...,W(F,\Phi_l)/F)=W(F,\Phi_1,...,\Phi_l)/F
\ee
where  $\tilde{\Phi}=W(F,\Phi)/F$, we obtain 
\bea
	F_{2\ell}(x;\beta^-)&=&\frac{W(\tilde{\Phi}_+(1)
	,\ldots,\tilde{\Phi}_-(2\ell))}{W(\Phi_+(1),\ldots,\Phi_-(2\ell))}F(x;\beta^-)
	\nonumber\\
	&=&G_{2\ell}(x;\beta^-)F(x;\beta^-)\,.
\eea
$\tilde{\Phi}_i$ is the eigenstate 
of the displaced  Lam\'e system  
$H_{0,0}(x+\beta^-)$, with the properties 
similar to those as  $\Phi_i$. 
We have showed that $W(\Phi_+(1),...,\Phi_-(2\ell))$ 
is nodeless and 
takes finite values of
a definite sign. This implies that  
$W(\tilde{\Phi}_+(1),...,\tilde{\Phi}_-(2\ell))$
share the same properties. 
Hence, function 
$G_{2\ell}(x;\beta^-)$ also possesses the same indicated properties. 
Taking into account the properties of the functions inside the Wronskian  under 
the reflection $x\rightarrow -x$, it is not difficult to show that  
${\rm sign}\, G_{2\ell}(x;\beta^-)={\rm sign}\, G_{2\ell}(x;-\beta^-) $. 
Having the identity  $F(-x;\beta^-)=F(x;-\beta)$, we find  that 
\bea
	\mathcal{F}_{2\ell,+}(x;\beta^-)&=&
	\frac{W(\Phi_+(1),\ldots,\Phi_-(2\ell),\mathcal{F}_+
	(\beta^-))}{W(\Phi_+(1),\ldots,\Phi_-(2l))}\\
	&=&CG_{2\ell}(x;\beta^-)F(x;\beta^-)+
	\frac{1}{C}G_{2\ell}(x;-\beta^-)F(-x;\beta^-)\,.
\eea
Since  $G_{2\ell}(x;\pm \beta^-)$ take values of the same sign 
and increase exponentially,
the function 
$\mathcal{F}_{2\ell,+}$ has no zeros. Then,  
\bea
	\mathcal{F}_{2\ell,-}(x;\beta^-)&=&\frac{W(\Phi_+(1),\ldots,\Phi_-(2\ell),	
	\mathcal{F}_-(\beta^-))}{W(\Phi_+(1),\ldots,\Phi_-(2\ell))}\\
	&=&CG_{2\ell}(x;\beta^-)F(x;\beta^-)-	
	\frac{1}{C}G_{2\ell}(x;-\beta^-)F(-x;\beta^-)
\eea
has only one zero. Here, the functions
$\mathcal{F}_{2\ell,\pm}$ 
are linearly independent 
eigenstates of the operator $H_{2\ell,0}$ 
with eigenvalues
$\mathcal{E}(\beta^-+i{\rm {\bf K}}')$, which 
are analogous to the 
eigenfunctions 
$\mathcal{F}_{\pm}$ of the Lam\'e system $H_{0,0}$,
see  (\ref{Fsi+-}).
Using the arguments presented in the first subsection of Appendix, 
one can show that 
\be
	W(\mathcal{F}_{2\ell,+}(1),\ldots,
	\mathcal{F}_{2\ell,s_{n+1}}(n+1))
\ee
has no zeros. 
From the Crum theorem,
\be
	H_{2\ell,n}=H_{2\ell,0}-2\frac{d^2}{dx^2}\log W(\mathcal{F}_{2\ell,+}(1),
	\ldots,\mathcal{F}_{2\ell,s_{n}}(n))=
	H_{0,0}-2\frac{d^2}{dx^2}\log \W_{2\ell,n}\,,
\ee
and it follows that 
\be
 W\left(\Phi_+(1),\Phi_-(2),\ldots,
	\Phi_-(2\ell),\mathcal{F}_{+}(1),\ldots,\mathcal{F}_{s_n}(n)\right)
\ee
is a smooth  and nodeless function.


\end{document}